\documentclass[prr,reprint,superscriptaddress,altaffillsymbol,amsmath,nofootinbib,longbibliography]{revtex4-1}
\usepackage{xcolor}

\pdfoutput=1

\usepackage[percent]{overpic} 
\newlength{\panelheight}
\usepackage{textcomp}
\usepackage{upgreek}

\usepackage{ifthen}
\usepackage{booktabs} 
\usepackage{natbib}

\usepackage{wasysym}
\usepackage{slashed} 
\usepackage{bbm}
\usepackage{mathrsfs}
\usepackage{amssymb}
\usepackage{dsfont}
\usepackage[export]{adjustbox}

\usepackage{booktabs}
\usepackage{placeins}
\usepackage{array}
\usepackage{color}

\definecolor{OurBlue}{rgb}{0.384,0.616,0.784}
\definecolor{OurRed}{rgb}{0.878,0.388,0.212}

\usepackage[
colorlinks=true, 
linkcolor=OurBlue,
citecolor=OurRed,
urlcolor=OurBlue,
]{hyperref}

\allowdisplaybreaks

\newcommand{\beqra}{\begin{eqnarray}}
\newcommand{\eeqra}{\end{eqnarray}}
\newcommand{\beq}{\begin{equation}}
\newcommand{\eeq}{\end{equation}}

\renewcommand{\epsilon}{\varepsilon}

\renewcommand{\bar}{\overline}

\usepackage{orcidlink}

\usepackage[columnwise]{lineno} 

\begin{document}

\title{\boldmath Dynamical multiferroicity in framework materials}

\author{Marek Matas~\orcidlink{0000-0002-8250-8187}}
\thanks{\href{mailto:marek.matas@fjfi.cvut.cz}{marek.matas@fjfi.cvut.cz}}
\affiliation{Faculty of Nuclear Sciences and Physical Engineering, Czech Technical University in Prague, Czech Republic}

\author{Carl P. Romao~\orcidlink{0000-0002-5519-2519}}
\thanks{\href{mailto:carl.romao@cvut.cz}{carl.romao@cvut.cz}}
\affiliation{Faculty of Nuclear Sciences and Physical Engineering, Czech Technical University in Prague, Czech Republic}
\affiliation{Department of Materials, ETH Zurich, Switzerland}

\begin{abstract}

Dynamical multiferroicity, which describes the magnetic fields generated by circularly polarized phonons in materials, is an established mechanism for optical control of magnetism. Here we perform \textit{ab initio} calculations of dynamical multiferroicity in inorganic and organic framework materials, with the goal of identifying materials which enable the generation of large magnetic fields by light. We find the metal--organic framework material Zn(NH$_4$)(formate)$_3$ to have modes with magnetic moments almost twice that of SrTiO$_3$; these modes involve circular motions of NH$_4^+$ hydrogen ions with high gyromagnetic ratios. The complex structure and flexibility of framework materials can allow such angular momentum localization, and also increase the maximum light-induced magnetization permitted by the Lindemann melting criterion. 

\end{abstract}

\maketitle

\section{Introduction}

When phonons (quantized vibrational excitations) involve circular or elliptical motions of atoms, the excitation can carry an associated angular momentum~\cite{zhang2014angular, zhang2015chiral, uedaChiral2023}. The angular momentum of these axial phonons~\cite{juraschekChiral2025} changes their physical properties in several ways, including by giving the phonon a magnetic moment~\cite{juraschek2017dynamical, shabalaAxial2025}. As phonon frequencies are typically several orders of magnitude higher than magnon frequencies, axial phonons play an important role in ultrafast magnetic processes. For example, during ultrafast demagnetization, the phononic system can respond quickly enough to absorb the excess angular momentum~\cite{dornesUltrafast2019, tauchertPolarized2022}. Coherent excitation of axial phonons can create~\cite{basiniTerahertz2024, kahanaLightinduced2024} or switch~\cite{daviesPhononic2024, groefsemaEmergence2026} magnetization on ultrafast time scales. On the other hand, the long lifetimes of acoustic phonons give magnetic axial acoustic phonons potential use as quantum sensors due to their high moment-to-energy ratio~\cite{romaoChiral2023, matasChiral2025}.

The origins of phonon magnetism can be understood through Ampere’s law, as the atoms in a crystal have nuclear and electronic charges, which are moved in circular orbits by an axial phonon, thereby generating a charge current and a magnetic field~\cite{juraschek2017dynamical}. This phenomenon, known as dynamical multiferroicity, produces Maxwellian phonon magnetic moments on the order of the nuclear magneton ($\upmu_\mathrm{n}$)~\cite{juraschek2019orbital}. However, experimental results have demonstrated that axial phonons can create much larger effective magnetic fields, corresponding to effective moments on the order of the Bohr magneton ($\upmu_\mathrm{B}$)~\cite{luo2023large, basiniTerahertz2024, biggsUltrafast2025}. The effective magnetic field is not necessarily Maxwellian; it parametrizes the time reversal symmetry breaking caused by the axial phonon, for example in terms of orbit--lattice or light--matter coupling strength~\cite{chaudhary2023giant, shabalaPhonon2024, merlinUnraveling2024}. However, as the examples of magnetic switching by axial phonons in Al$_2$O$_3$ and SiO$_2$ demonstrate~\cite{daviesPhononic2024}, the small Maxwellian magnetic field can be sufficient for ultrafast control of magnetism even in materials without strong electron--phonon coupling or known mechanisms of large effective magnetic field generation.

The use of axial phonons to create magnetization in otherwise nonmagnetic materials is one element within the broader program of quantum printing, defined as the imprinting of quantum states from phonons and photons onto matter, allowing the generation and control of ferroic order on demand~\cite{aeppliQuantum2025}. For example, coherent driving of an axial phonon using elliptical light can generate combined magnetic and ferroelectric order through nonlinear phononic coupling~\cite{paivaDynamically2025}. In ferroelectric materials, axial phonons couple to the ferroelectric polarization, creating quasiparticles referred to as multiferrons which carry magnetic and electric dipole and multipole moments~\cite{polsLattice2026, tangMultiferroiclike2026}.

The maximum magnetization which can be achieved in a material by coherent excitation of axial phonons can be increased by increasing either the magnitude of the phonon magnetic moment or the number of phonons which can be excited in the material before it begins to melt or decompose. In both of these regards, flexible framework materials (\textit{i.e.}, materials with a mixture of stiff and compliant directions due to topological underconstraint~\cite{romaoAnisotropic2017}) are promising host materials for magnetic axial phonons. Their structures can accommodate large atomic displacements~\cite{coudertResponsive2015}, which has enabled their use in the storage of molecules and ions~\cite{jiangRecent2022}, and they can contain large proportions of light atoms, enhancing the phonon gyromagnetic ratio~\cite{juraschek2019orbital}. These qualities are especially pronounced in the family of metal--organic frameworks, in which metal ions are connected via organic linkers, allowing a very high degree of control over structure and properties.

\section{Dynamical multiferroicity}\label{sec:dmf}

Herein light-induced phonon magnetism in framework materials is studied using the methodology of Refs.~\cite{juraschek2017dynamical, juraschek2019orbital, chaudharyAnomalous2026a}. We extend the previous survey of Ref.~\cite{juraschek2019orbital}, which considered the phonon magnetic moments created by circularly polarized light in crystals in the rock salt and perovskite structure families, to more complex framework structures, and demonstrate the effects of symmetry lowering and structural flexibility on light-induced phonon magnetism.

The procedure for the calculation of the phonon magnetic properties begins with the phonon angular eigenfrequencies ($\omega_n$), eigendisplacements ($\mathbf{u}_{m,n}$), and Born effective charge tensors ($\mathbf{Z}^*_m$), where $n$ indexes the optical phonons at $\Gamma~(\mathbf{q} = 0)$, and $m$ indexes the atoms in the primitive unit cell. These were calculated from first principles as described in the Computational Methods below (Sec.~\ref{sec:methods}). The IR-active modes were identified by the mode effective charge vector ($\mathbf{Z}_n$)~\cite{gonzeDynamical1997, juraschek2019orbital}:
\begin{align}
    \mathbf{Z}_n = \frac{\sum_m \mathbf{Z}^*_m \mathbf{u}_{m,n} \mathcal{M}_m^{-1/2}}{(\sum_m \mathbf{u}_{m,n} \cdot \mathbf{u}_{m,n})^{1/2}}
\end{align}
where $\mathcal{M}_m$ is the atomic mass and $\mathbf{Z}^*_m$ is the Born effective charge tensor, using a criterion of $||\mathbf{Z}_n|| \geq 10^{-10}~\mathrm{e}~\mathrm{emu}^{-1/2}$. 

Pairs of degenerate IR-active modes were then selected with a tolerance of $1\%$ of the frequency, allowing accidentally degenerate pairs of modes to be considered in materials with low symmetry and large numbers of atoms. Circularly polarized superpositions can be created in such cases so long as the phonon lifetimes ($1/\kappa_{a/b}$) are shorter than the dephasing time $2\pi/(\omega_a - \omega_b)$. Therefore, as literature values of the phonon linewidths ($\kappa$) are generally not available for the studied materials, we assume $2\pi\kappa/\omega \leq 1\%$. The ability of the degenerate pairs to be excited using circularly polarized light was ensured by only considering pairs where the cross product of the normalized mode effective charge vectors ($\hat{\mathbf{Z}}_a \times \hat{\mathbf{Z}}_b$) had a magnitude $\geq 0.95$.

The eigendisplacements of degenerate pairs of optically active modes were then transformed to a circularly polarized basis ($\pm$) as~\cite{zhang2015chiral}:
\begin{align}
\mathbf{u}_{m,\pm} = (\mathbf{u}_{m,a} \pm i\mathbf{u}_{m,b})/\sqrt{2}.
\end{align}
The angular momentum of the circularly polarized phonon is then given by~\cite{zhang2015chiral}:
\begin{align}\label{eq:circpol}
\mathbf{L}_{\pm} = \sum_m \mathcal{M}_m \mathbf{u}_{m,\pm} \times \dot{\mathbf{u}}_{m,\pm} = \sum_m \mathcal{M}_m (\mathbf{u}_{m,\alpha} \times \mathbf{u}_{m,\beta})/2,
\end{align}
and its magnetic moment by~\cite{chaudharyAnomalous2026a}:
\begin{align}\label{eq:magmoment}
\mu_{\pm} = \sum_m \mu_{m,\pm} = \sum_m \mathbf{L}_{m,\pm} \gamma_m  = \sum_m \hbar \mathbf{S}_{m,\pm} \mathbf{Z}^*_{m} / {2 \mathcal{M}_m},
\end{align}
where $\mathbf{S}_{m,\pm}$ is the atomic circular polarization vector.

For each material, we model resonant pumping of the optically active mode with circularly polarized THz--IR radiation by solving the equation of motion for the phonon normal mode coordinate $Q_\alpha$ of the components $\alpha \in \{a, b\}$ of the superposition:
\begin{align}\label{eq:Q_eqm}
    \ddot{Q}_\alpha + \kappa_\alpha \dot{Q}_\alpha + \frac{ \partial U}{\partial{Q_\alpha}} = \mathbf{Z}_\alpha \cdot \mathbf{E}(t).
\end{align}
Here $U$ is the phonon potential and $\mathbf{E}(t)$ is the driving electric field, which we model as an ultrashort laser pulse with components $E_\alpha$, peak magnitude $E_0$, and FWHM duration $\tau$~\cite{juraschekSumfrequency2018}:
\begin{align}
    E_\alpha(t) = E_0~e^{2(\tau/\sqrt{8 \ln2})^2}~\mathrm{cos}(\omega_\alpha t+\phi_\alpha),
\end{align}
 where $\phi_a = 0$ and $\phi_b = \pi/2$. The pulse duration and peak magnitude are scaled by the phonon frequency as $\tau/\omega_\alpha = 5 \pi$ and $E_0/\omega_\alpha^2 = 9.5 \times 10^{-20}~\mathrm{V~m^{-1}~s^2}$, yielding experimentally reasonable values for the laser fluence and phonon occupation~\cite{juraschek2019orbital}.

We use the atomic displacements to evaluate the potential for the excitation to melt the sample using the Lindemann criterion~\cite{lindemann1910ueber}.  To obtain a conservative estimate, we solve Eq.~(\ref{eq:Q_eqm}) in the harmonic approximation, \textit{i.e.}, with $\kappa_\alpha = 0$ and $U = \omega_\alpha^2 Q_\alpha^2/2$. We then obtain the root-mean-square atomic displacements ($\mathbf{d}_{m,\pm}$) from the time evolution of the normal mode coordinate:
\begin{align}
    \langle d_{m,\pm}\rangle = \mathrm{RMS}(\mathbf{u}_{m,a}Q_a(t) \pm i\mathbf{u}_{m,b}Q_b(t))/\sqrt{2}.
\end{align}
The Lindemann criterion states that melting occurs when, for any atom $m$, $\langle d_{m,\pm}\rangle > \eta d_{m,0}$, where $\eta$ is the Lindemann coefficient (usually considered to be $10\%$--$15\%$) and $d_{m,0}$ is the equilibrium interatomic distance~\cite{lindemann1910ueber,vopsonGeneralized2020}. In complex framework structures such as MOFs, the melting process can be controlled by the metal--ligand bonds which stabilize the framework, and therefore the Lindemann criterion can be evaluated in terms of the metal--ligand distance \cite{gaillacMelting2018}. We consider this possibility by evaluating the Lindemann criterion at $\eta = 10\%$ for all interatomic distances, and for only the metal--ligand distances. 

The magnetization induced by the pumped phonon mode ($\mathbf{M}_\mathrm{pulse}$) is determined using Eq.~(\ref{eq:magmoment}) and the phonon density ($N_\mathrm{ph}/V_\mathrm{cell}$)~\cite{juraschek2019orbital}:
\begin{align}
    \mathbf{M}_\mathrm{pulse} = \mu_\pm N_\mathrm{ph} /V_\mathrm{cell} = \mu_\pm \omega_\pm (Q_a+Q_b)^2 /2\hbar V_\mathrm{cell}.
\end{align}
The maximum magnetization that can be obtained without melting the sample is then obtained by scaling the magnetization arising from the model pulse using the Lindemann coefficient:
\begin{align}\label{eq:lindemann}
    M_\mathrm{melt} =  \frac{d_{0,m} ||\mathbf{M}_\mathrm{pulse}|| }{\eta \langle d_{m,\pm}\rangle}.
\end{align}

We have studied a number of materials with varying structural complexity, ranging from simple perovskites to complex MOFs. The chosen materials were 
ScF$_3$ (space group $Pm\bar{3}m$)~\cite{grevePronounced2010}, 
BPO$_4$ ($I\bar{4}$)~\cite{bpogrowth}, 
CaZrF$_6$ ($Fm\bar{3}m$)~\cite{hancockLarge2015}, 
$\alpha$-quartz (SiO$_2$, $P3_121$)~\cite{baurSearch2009}, 
SrTiO$_3$ ($I4/mcm$)~\cite{mccallaUnified2016}, 
CuPt(CN)$_6$ ($I4/mmm$)~\cite{chapmanCompositional2006}, 
CuCl(pyrimidine) (CuCl(pyr), $Pma2$)~\cite{natherSynthesis2004}, 
InF$_3$($4,4'$-bipyridine) (InF$_3$(bpy), $I222$) \cite{petrosyants2010organometallic}, 
AgCl(1,10-phenanthroline) (AgCl(phen), $C_2$)~\cite{odokoCatena2004}, 
CsCuCl$_3$ ($P6_122$)~\cite{schlueterRedetermination1966}, 
Sr(\textsc{l}-tartrate) (Sr(tar), $C222_1$)~\cite{appelhansPhase2009}, 
Li$_2$(\textsc{l}-tartrate) (Li$_2$(tar), $P2_1$)~\cite{yeungChiral2013}, 
AlF(benzene-1,4-dicarboxylate) (AlF(bdc), $Imma$) \cite{nanthamatheeContradistinct2015}, 
LiSn$_2$Br$_3$(CN$_2$) ($Fd\bar{3}m$)~\cite{loeberSynthesis2021}, 
Cd(guanidinium)(formate)$_3$ (Cd(Gua)(fmt)$_3$, $Cc$)~\cite{collingsCompositional2016}, 
Cu$_3$(melaminate) (Cu$_3$(mel), $P6/mcc$)~\cite{kallenbachTricopper2021}, 
Li$_2$(2,2-dimethylsuccinate) (Li$_2$(dmsu), $P\bar{1}$) \cite{sainesLayered2012}, 
Zn(NH$_4$)(formate)$_3$ (Zn(NH$_4$)(fmt)$_3$, $P6_3$)~\cite{xuDisorderOrder2010}, and
Zn$_7$(2,2-dimethylsuccinate)$_6$(OH)$_2$ (Zn$_7$(dmsu)$_6$(OH)$_2$, $P\bar{1}$) \cite{sainesLayered2012}. 

\section{Results and discussion}

\begin{figure*}
  \vspace{1em}       
  \centering
  \begin{overpic}[height=\panelheight]{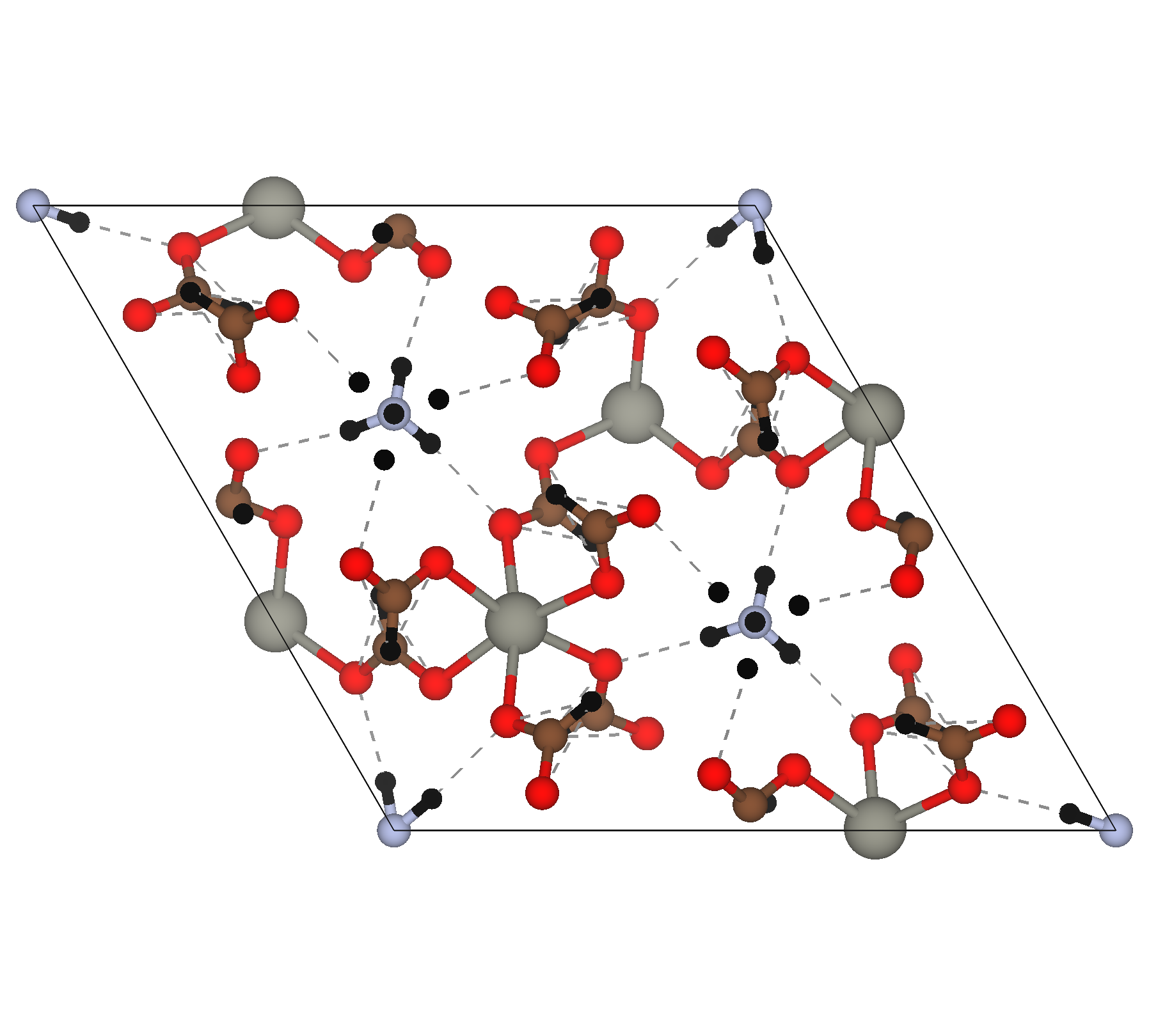}
    \put(3,95){\small\textbf{a)}}
  \end{overpic}
  \hfill
  \begin{overpic}[height=\panelheight]{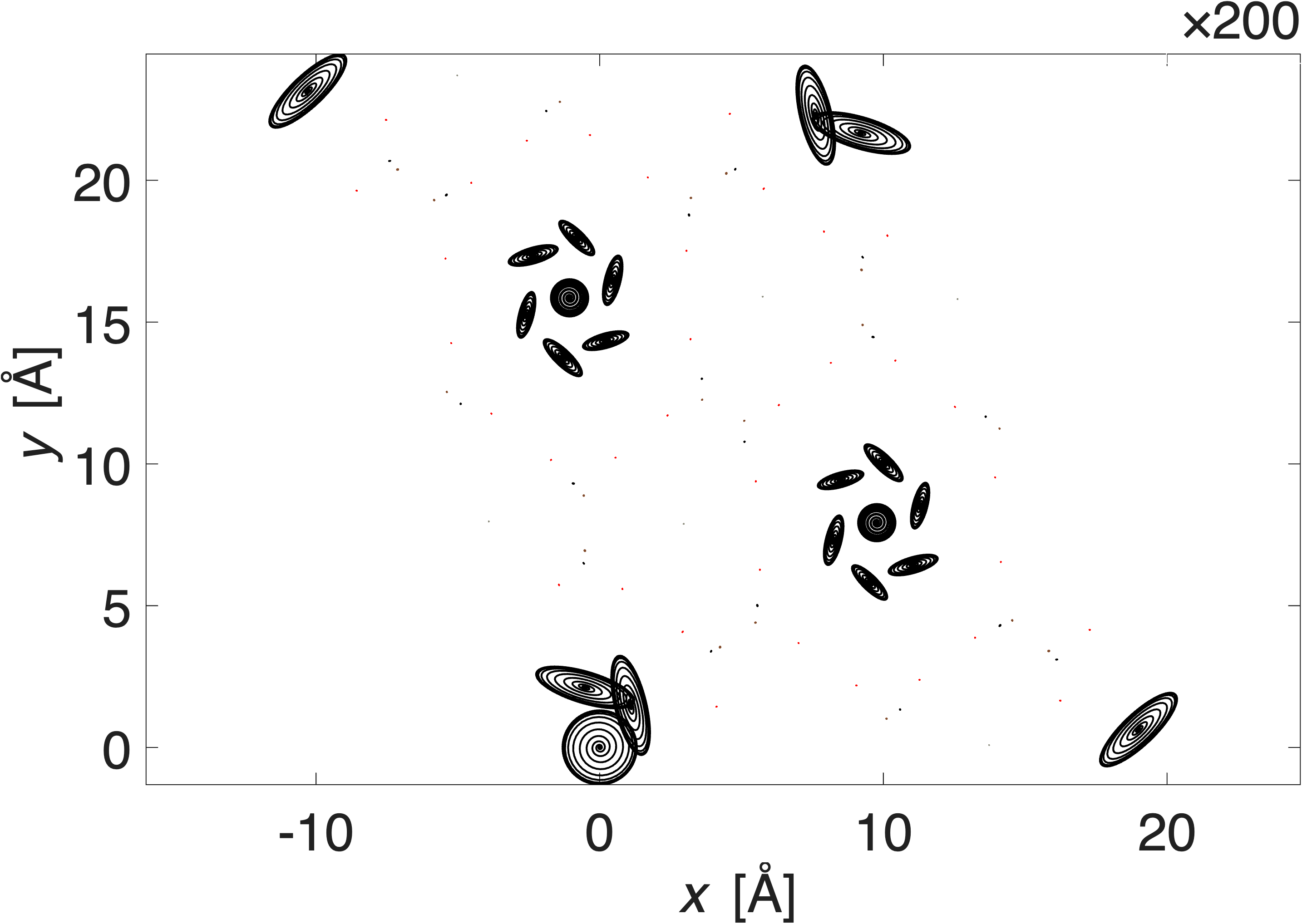}
    \put(3,75){\small\textbf{b)}}
                \put(87,69){\colorbox{white}{\small x141}}
  \end{overpic}
  \hfill
  \begin{overpic}[height=\panelheight]{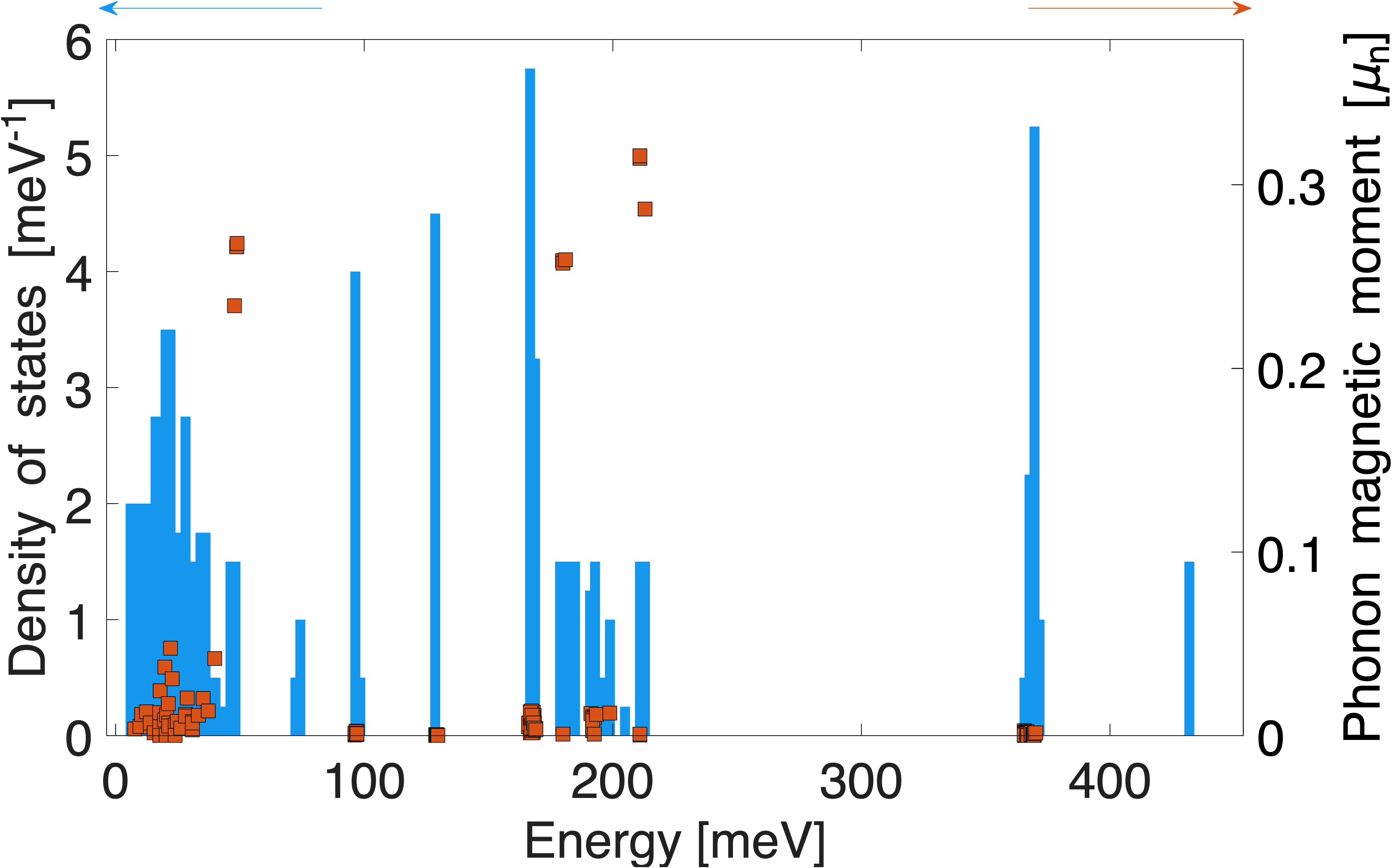}
    \put(3,67){\small\textbf{c)}}
  \end{overpic}
  \caption{Phonon magnetism in Zn(NH$_4$)(fmt)$_3$. \textbf{a)} Crystal structure of the primitive unit cell, with atoms colored by element (Zn gray, O red, N light blue, C brown, H black). \textbf{b)} Projection onto the $xy$ plane of the atomic trajectories in the circularly polarized superposition of the degenerate IR-active mode pair with the largest magnetic moment. Each ellipse traces the motion of one atom; displacements are magnified by the factor indicated in the upper right ($\times 141$) for visibility. The trajectories are shown for the period following a model circularly polarized laser pulse (see Sec.~\ref{sec:dmf} for details). The selected mode at $51$ THz is dominated by the motion of the H atoms within the NH$_4^+$ cation, giving the largest phonon magnetic moment in the survey. \textbf{c)} Phonon density of states (blue, left axis) and phonon magnetic moments (orange squares, right axis, in units of $\upmu_\mathrm{n}$) as a function of phonon energy.}
  \label{fig:znform}
\end{figure*}

\begin{figure*}
  \vspace{1em}       
  \centering
  \begin{overpic}[height=\panelheight]{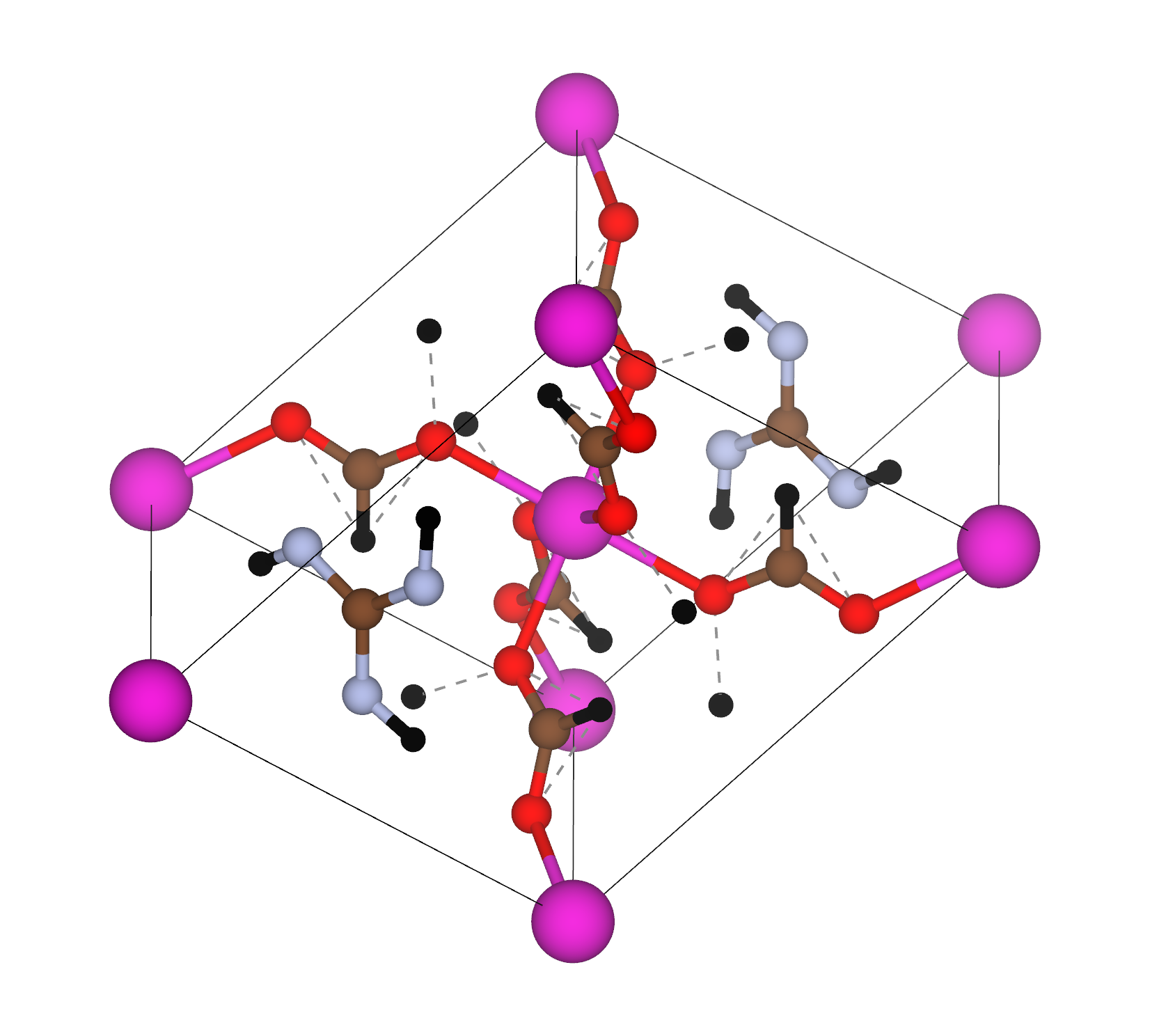}
    \put(3,95){\small\textbf{a)}}
  \end{overpic}
  \hfill
  \begin{overpic}[height=\panelheight]{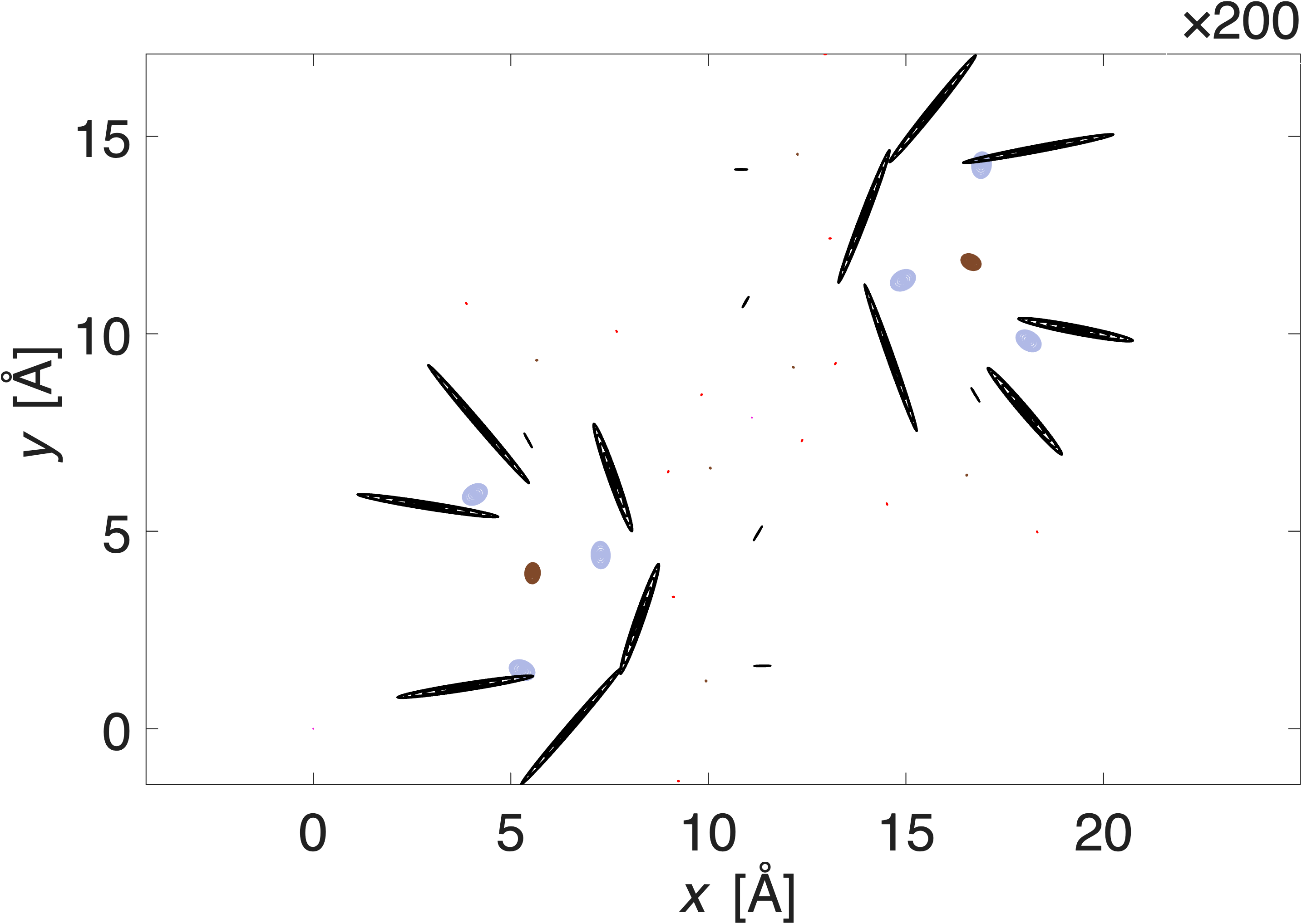}
    \put(3,75){\small\textbf{b)}}
    \put(87,69){\colorbox{white}{\small x141}}
  \end{overpic}
  \hfill
  \begin{overpic}[height=\panelheight]{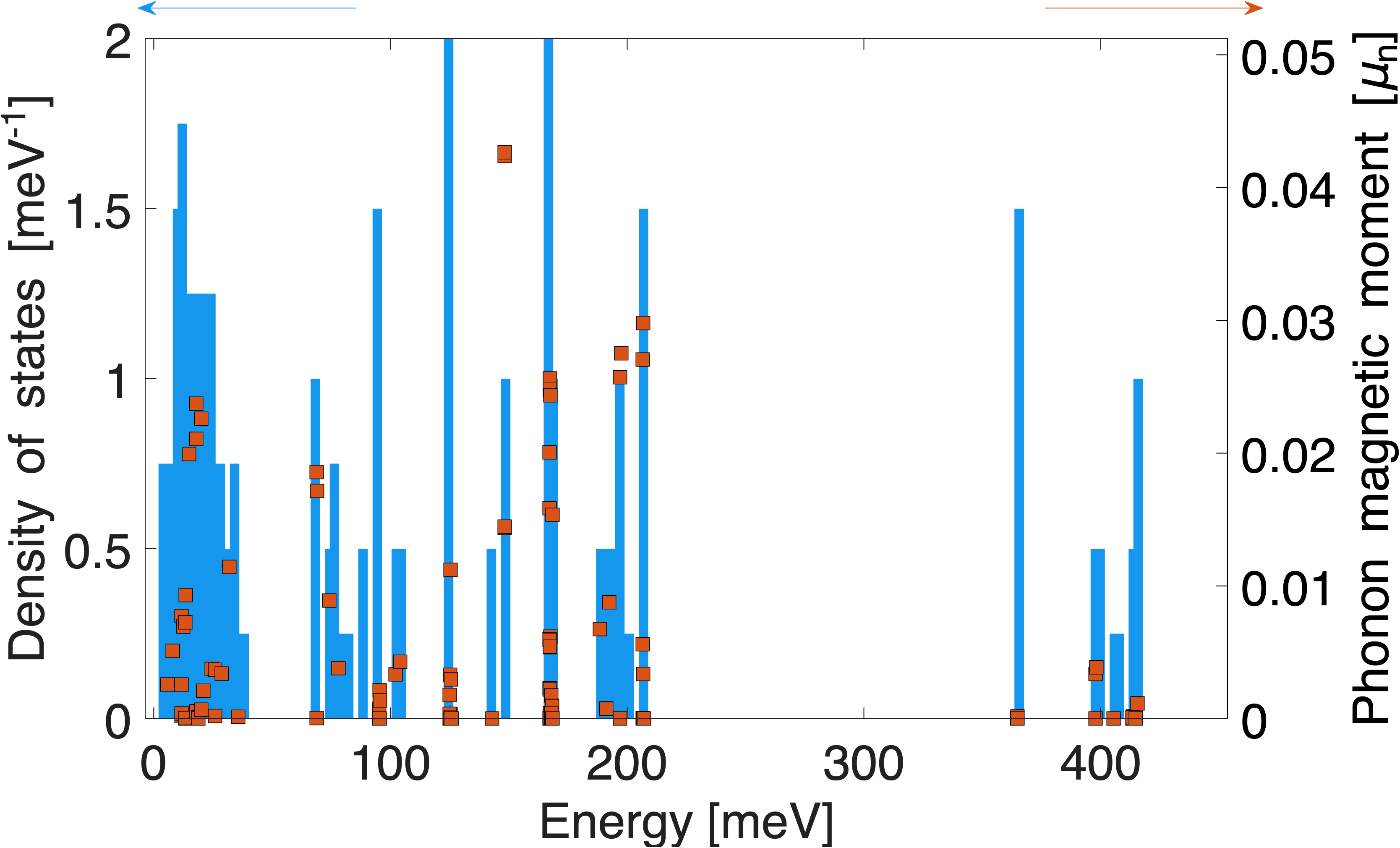}
    \put(3,65){\small\textbf{c)}}
  \end{overpic}  
  \caption{Phonon magnetism in Cd(Gua)(fmt)$_3$. \textbf{a)} Crystal structure of the primitive unit cell, with atoms colored by element (Cd magenta, O red, N light blue, C brown, H black). \textbf{b)} Projection onto the $xy$ plane of the atomic trajectories in the circularly polarized superposition of the degenerate IR-active mode pair with the largest magnetic moment. Each ellipse traces the motion of one atom; displacements are magnified by the factor indicated in the upper right ($\times 141$) for visibility. The trajectories are shown for the period following a model circularly polarized laser pulse (see Sec.~\ref{sec:dmf} for details). Although the guanidinium cation contains H atoms with large effective gyromagnetic ratios, the angular momentum of this mode is not strongly localized on those H atoms, limiting the resulting phonon magnetic moment. \textbf{c)} Phonon density of states (blue, left axis) and phonon magnetic moments (orange squares, right axis, in units of $\upmu_\mathrm{n}$) as a function of phonon energy.}
  \label{fig:cdgform}
\end{figure*}

\begin{figure*}
  \vspace{1em}       
  \centering
  \begin{overpic}[height=\panelheight]{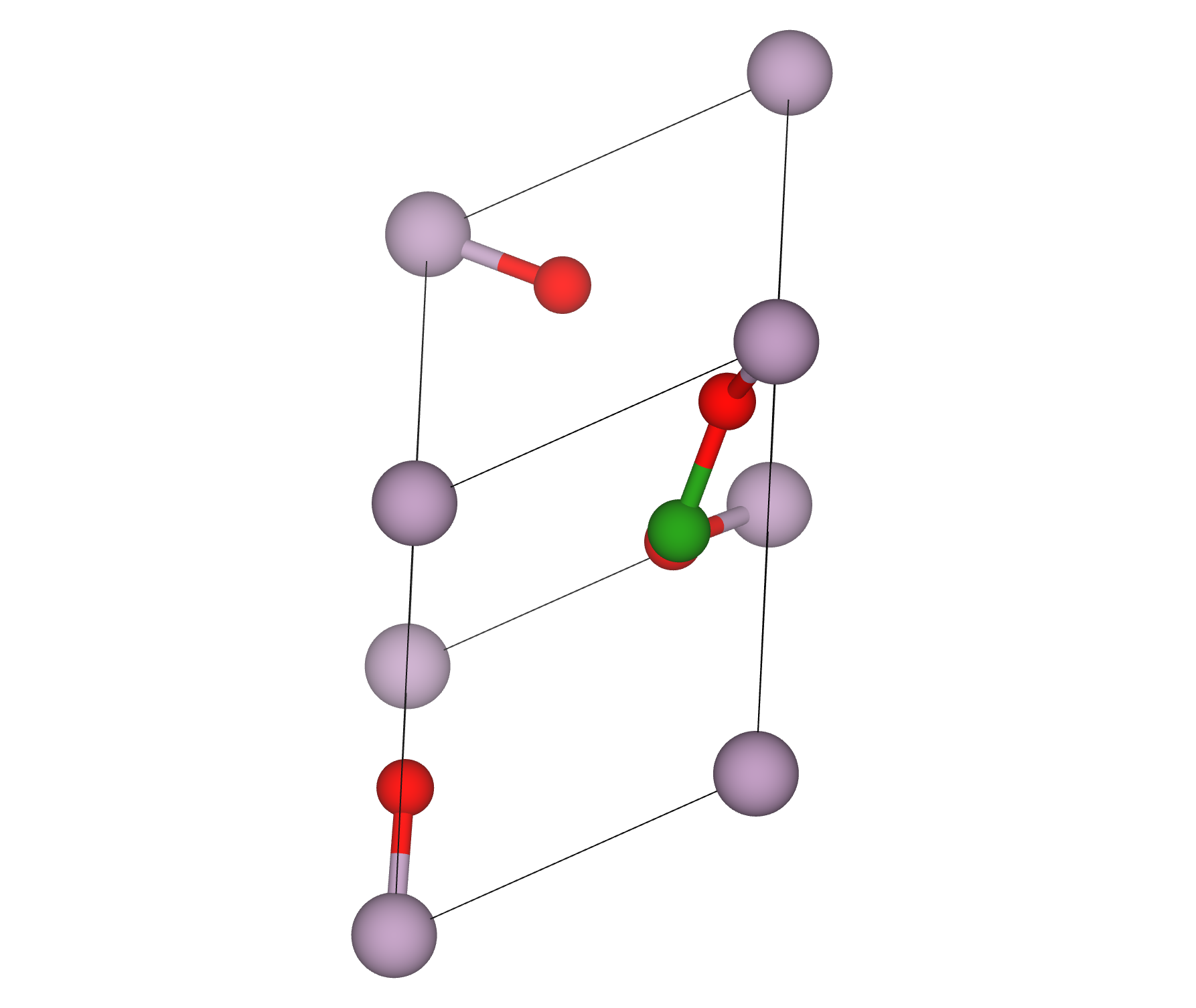}
    \put(3,90){\small\textbf{a)}}
  \end{overpic}
  \hfill
  \begin{overpic}[height=\panelheight]{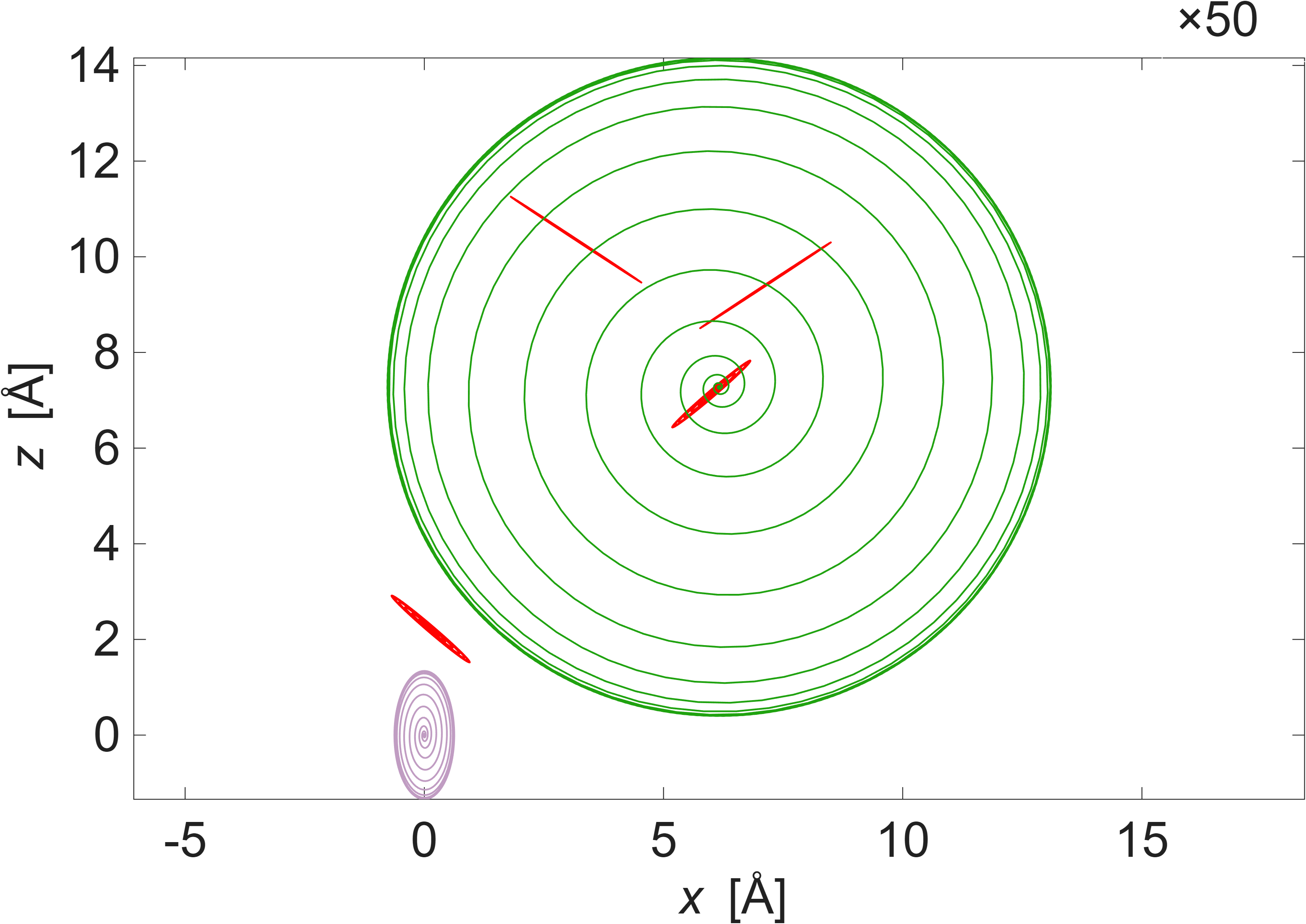}
    \put(3,76){\small\textbf{b)}}
    \put(87,69){\colorbox{white}{\small x50}}
  \end{overpic}
  \hfill
  \begin{overpic}[height=\panelheight]{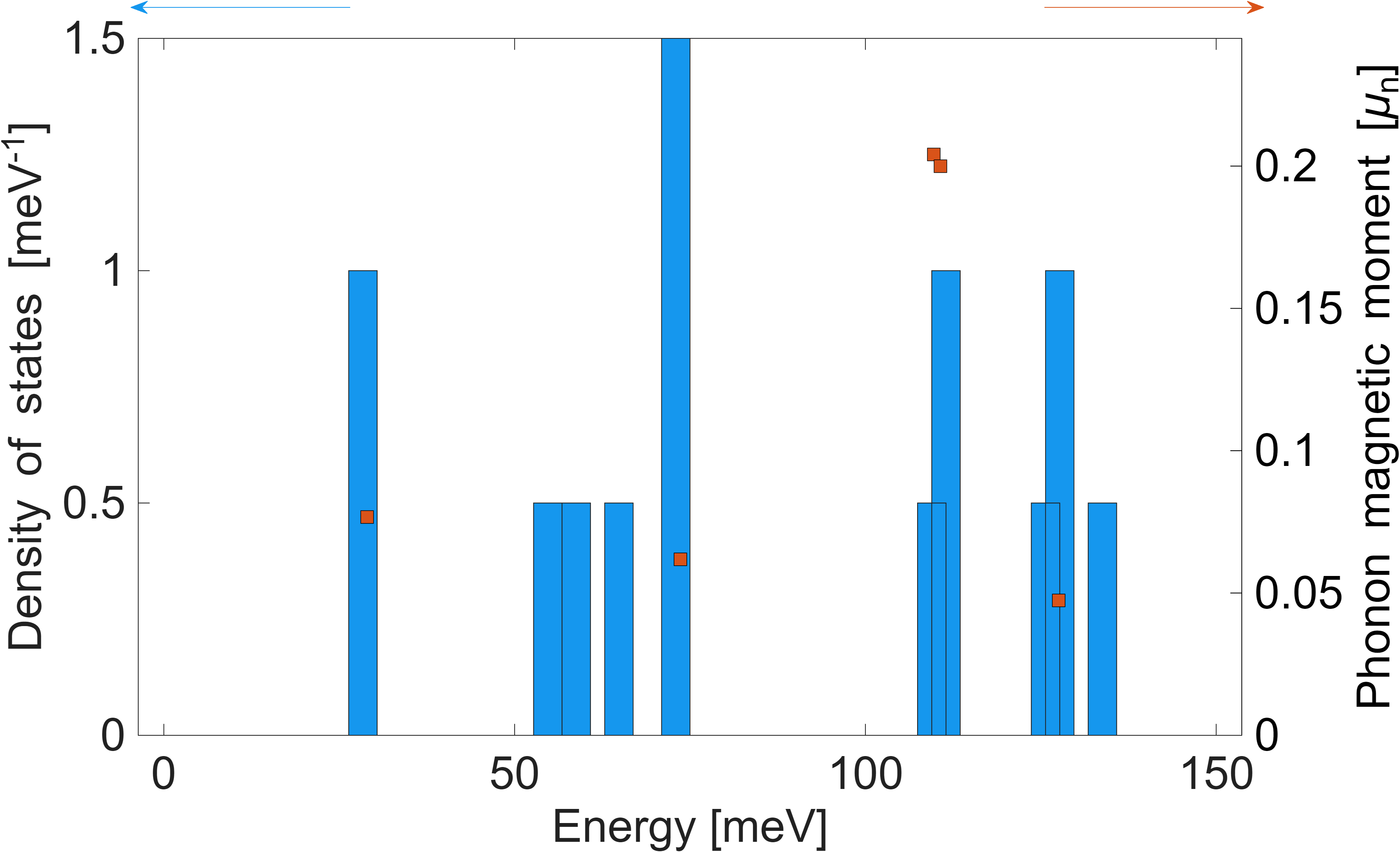}
    \put(3,66){\small\textbf{c)}}
  \end{overpic}
  \caption{Phonon magnetism in BPO$_4$. \textbf{a)} Crystal structure of the primitive unit cell, with atoms colored by element (P purple, O red, B green). \textbf{b)} Projection onto the $xz$ plane of the atomic trajectories in the circularly polarized superposition of the degenerate IR-active mode pair with the largest magnetic moment. Each ellipse traces the motion of one atom; displacements are magnified by the factor indicated in the upper right ($\times 424$) for visibility. The trajectories are shown for the period following a model circularly polarized laser pulse (see Sec.~\ref{sec:dmf} for details). The large magnetic moment is generated by the dominant circular motion of the boron atom; this gives the mode a high degree of angular momentum localization and overall circular polarization. The boron atom is light and possesses a relatively high effective charge ($Z_\mathrm{B}^* \approx -2.75~\mathrm{e}$), enhancing the gyromagnetic ratio. \textbf{c)} Phonon density of states (blue, left axis) and phonon magnetic moments (orange squares, right axis, in units of $\upmu_\mathrm{n}$) as a function of phonon energy.}
  \label{fig:BPO4}
\end{figure*}

\begin{figure*}
  \vspace{1em}       
  \centering
  \begin{overpic}[height=\panelheight]{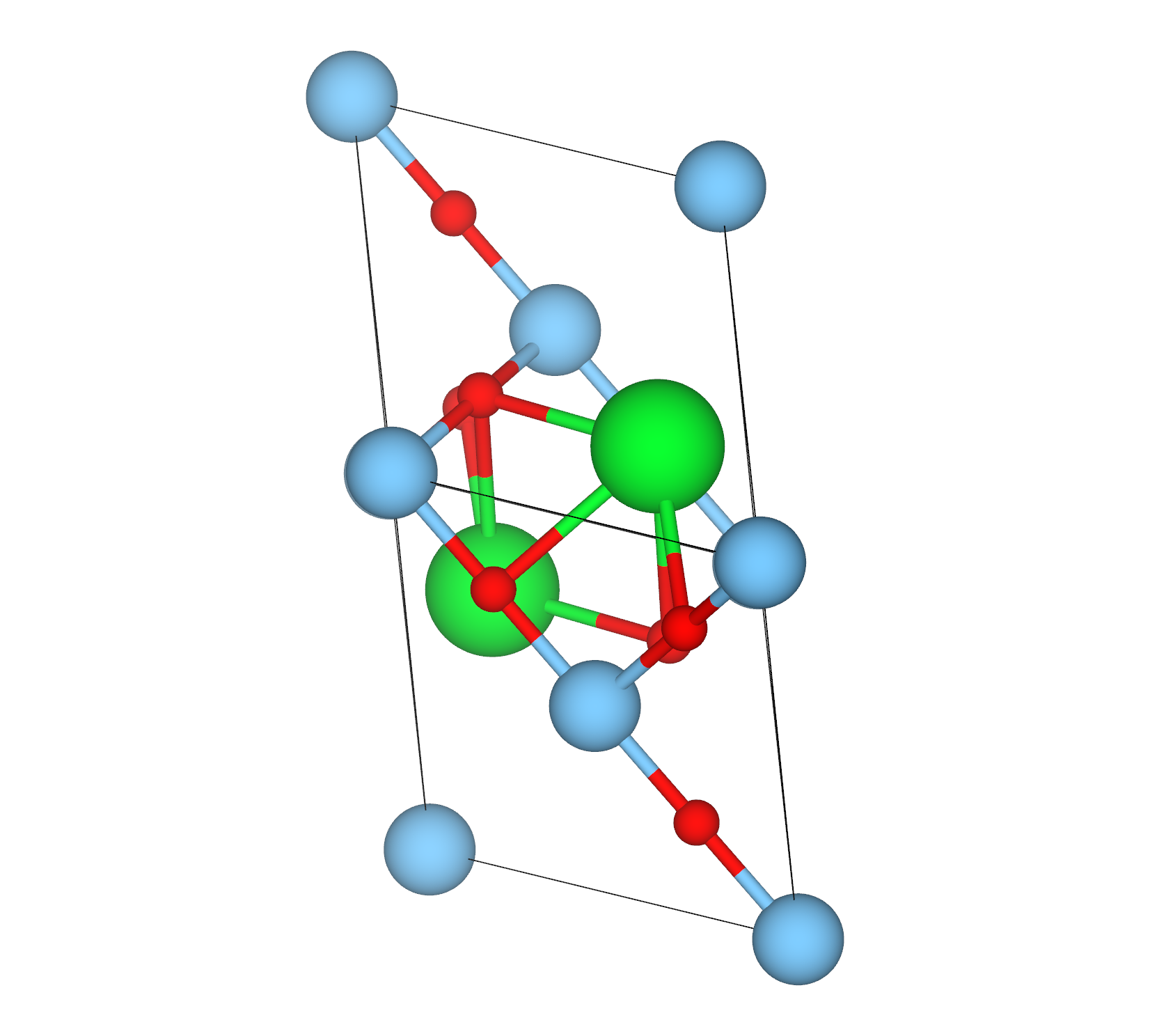}
    \put(3,95){\small\textbf{a)}}
  \end{overpic}
  \hfill
  \begin{overpic}[height=\panelheight]{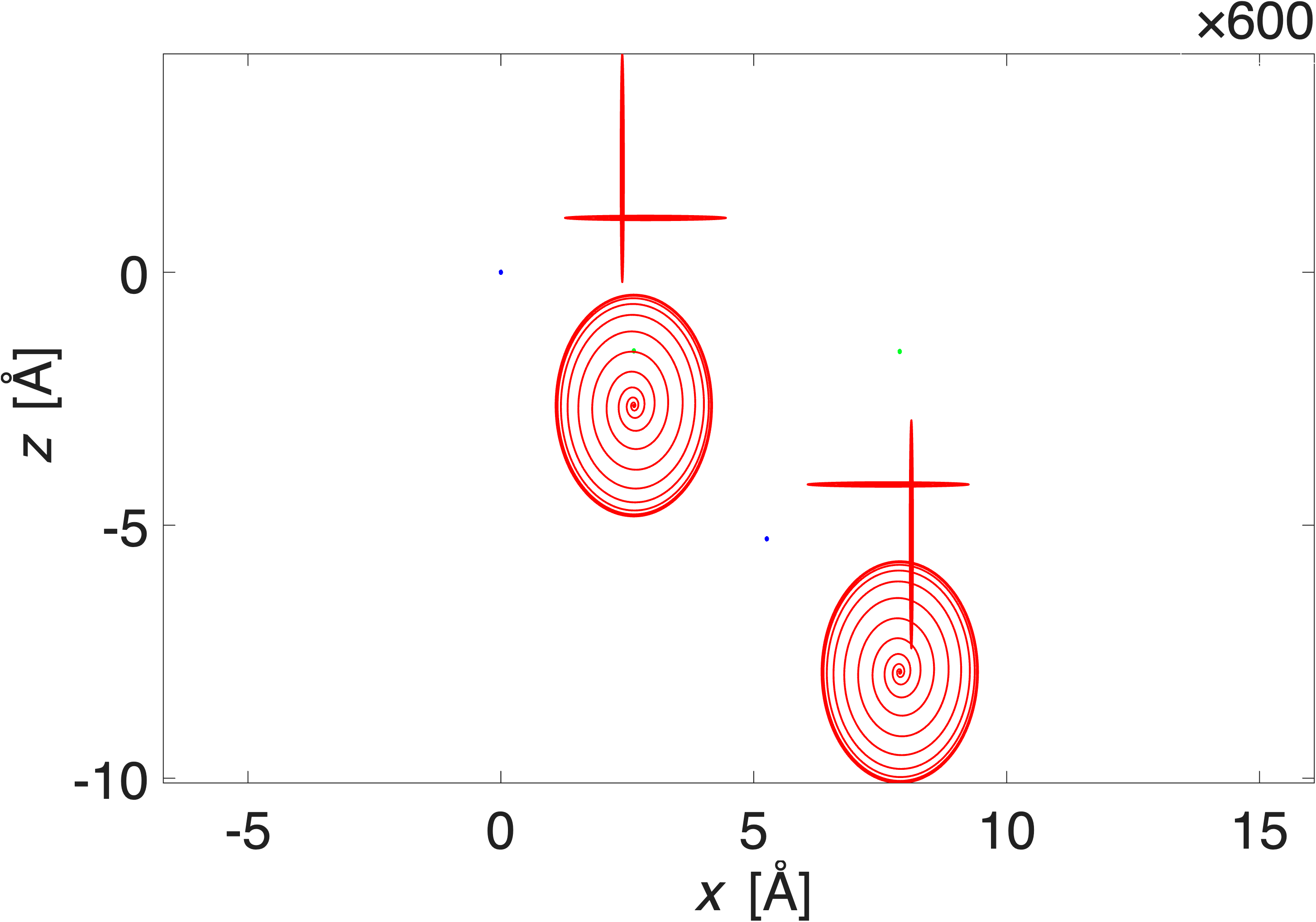}
    \put(3,75){\small\textbf{b)}}
            \put(87,68){\colorbox{white}{\small x424}}

  \end{overpic}
  \hfill
  \begin{overpic}[height=\panelheight]{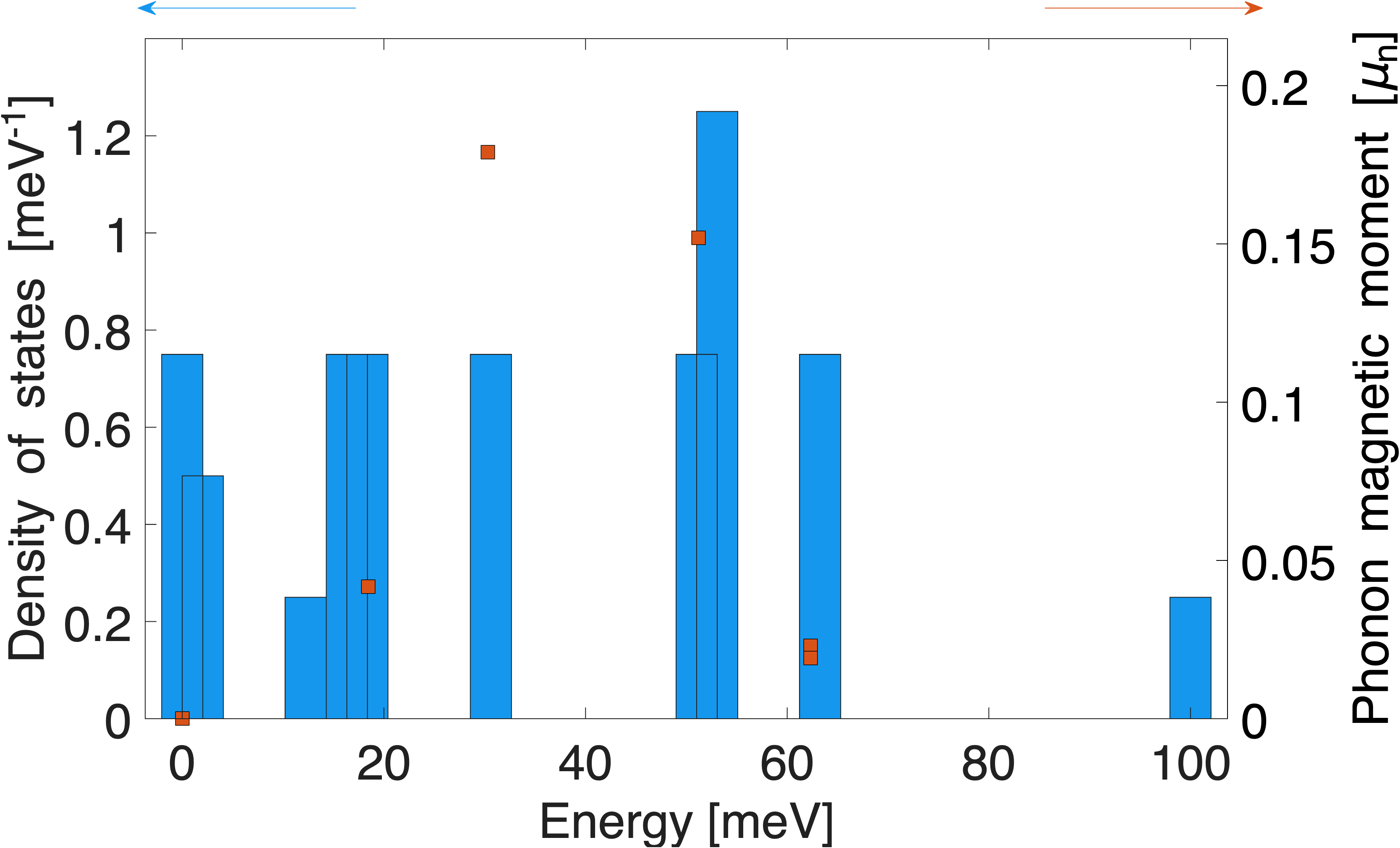}
    \put(3,64){\small\textbf{c)}}
  \end{overpic}
\caption{Phonon magnetism in SrTiO$_3$. \textbf{a)} Crystal structure of the primitive unit cell, with atoms colored by element (Sr green, Ti blue, O red). \textbf{b)} Projection onto the $xz$ plane of the atomic trajectories in the circularly polarized superposition of the degenerate IR-active mode pair with the largest magnetic moment. Each ellipse traces the motion of one atom; displacements are magnified by the factor indicated in the upper right ($\times 424$) for visibility. The trajectories are shown for the period following a model circularly polarized laser pulse (see Sec.~\ref{sec:dmf} for details). The magnetic moment is dominated by oxygen motion: one oxygen sublattice follows an approximately circular trajectory while another is closer to linearly polarized, giving an overall circular polarization near one half. Together with the anomalously large Born effective charge of oxygen in SrTiO$_3$, this produces one of the largest phonon magnetic moments in the dataset. \textbf{c)} Phonon density of states (blue, left axis) and phonon magnetic moments (orange squares, right axis, in units of $\upmu_\mathrm{n}$) as a function of phonon energy.}  \label{fig:STO}
\end{figure*}

\begin{figure*}
  \vspace{1em}       
  \centering
  \begin{overpic}[height=\panelheight]{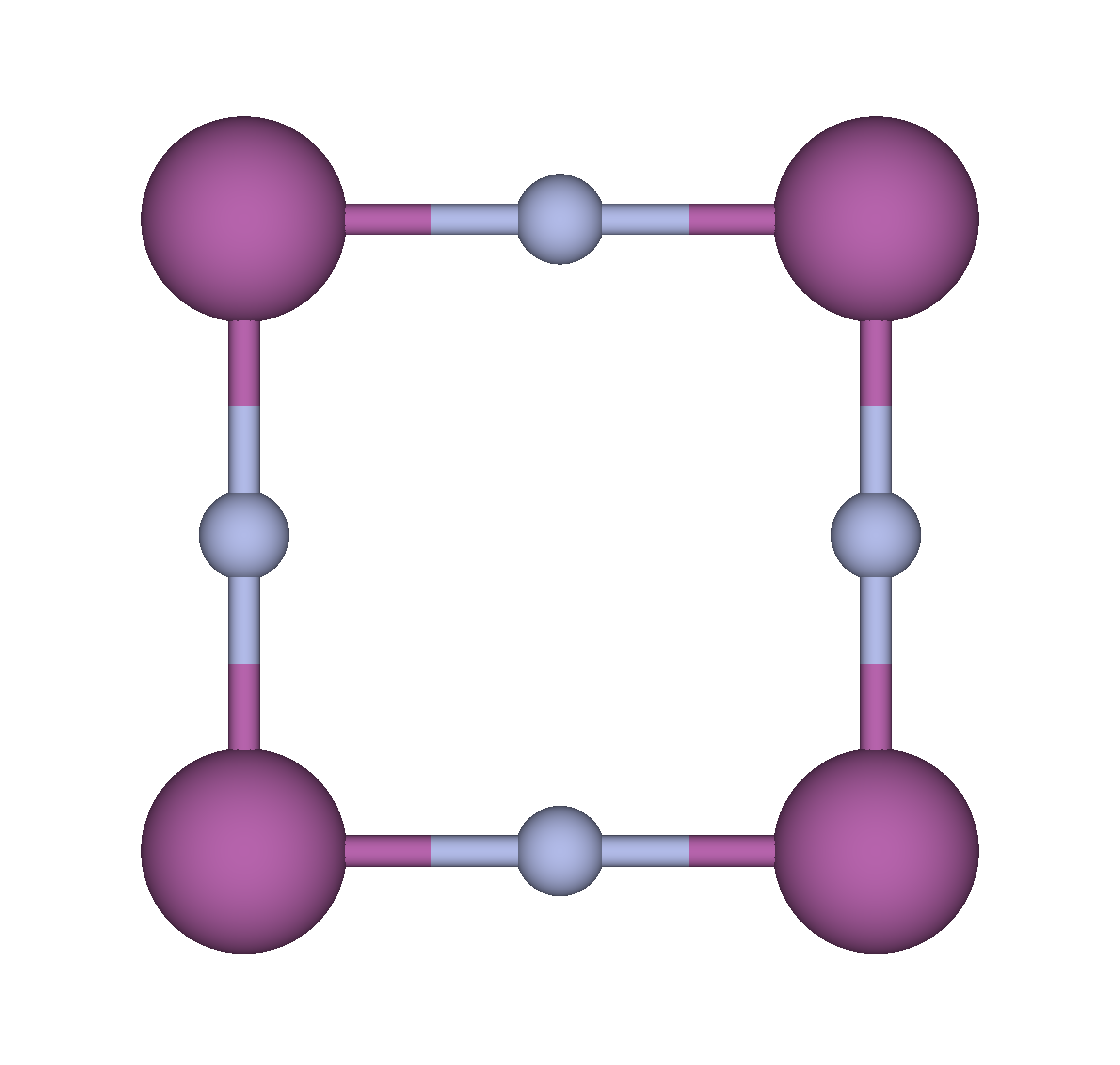}
    \put(3,98){\small\textbf{a)}}
  \end{overpic}
  \hfill
  \begin{overpic}[height=\panelheight]{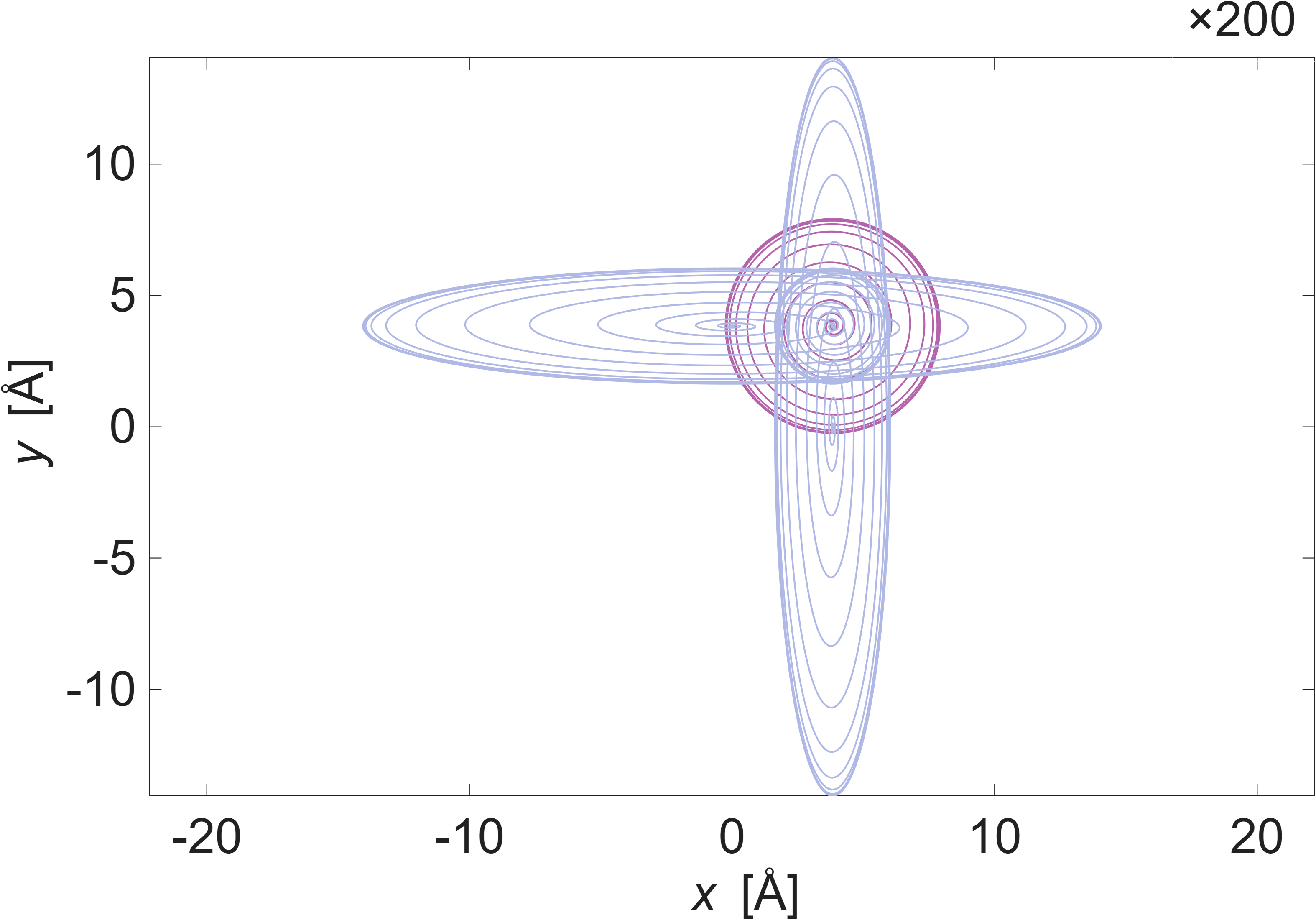}
    \put(3,75){\small\textbf{b)}}
            \put(87,68){\colorbox{white}{\small x200}}

  \end{overpic}
  \hfill
  \begin{overpic}[height=\panelheight]{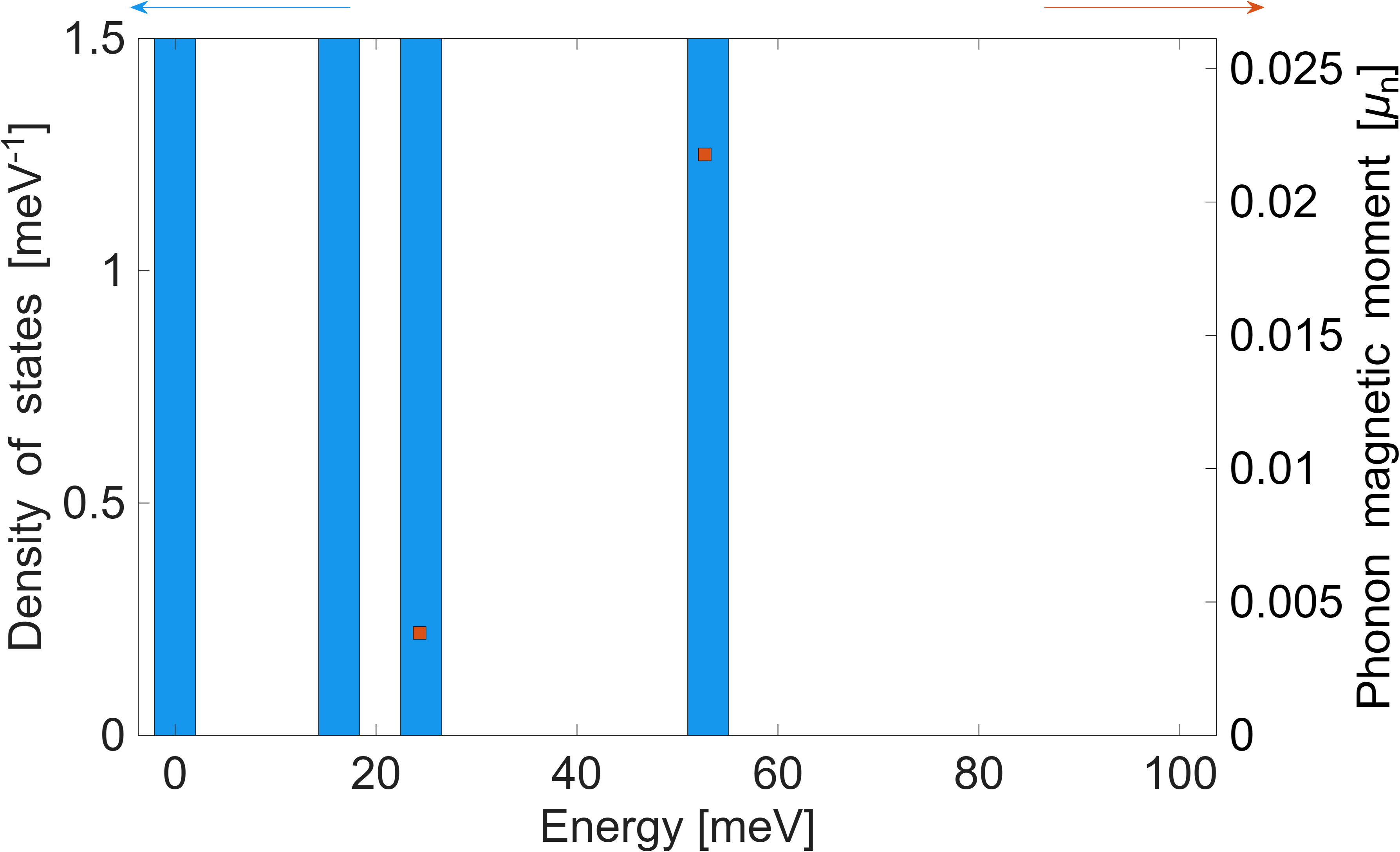}
    \put(3,65){\small\textbf{c)}}
  \end{overpic}
\caption{Phonon magnetism in ScF$_3$. \textbf{a)} Crystal structure of the primitive unit cell, with atoms colored by element (Sc purple, F lilac). \textbf{b)} Projection onto the $xy$ plane of the atomic trajectories in the circularly polarized superposition of the degenerate IR-active mode pair with the largest magnetic moment. Each ellipse traces the motion of one atom; displacements are magnified by the factor indicated in the upper right ($\times 141$) for visibility. The trajectories are shown for the period following a model circularly polarized laser pulse (see Sec.~\ref{sec:dmf} for details). In contrast to SrTiO$_3$, the higher cubic symmetry mixes the Sc and F displacements, preventing strong localization of angular momentum onto a single sublattice and reducing the phonon magnetic moment. \textbf{c)} Phonon density of states (blue, left axis) and phonon magnetic moments (orange squares, right axis, in units of $\upmu_\mathrm{n}$) as a function of phonon energy.}
\label{fig:scf3}
\end{figure*}

\begin{figure*}
  \vspace{1em}       
  \centering
  \begin{overpic}[height=\panelheight]{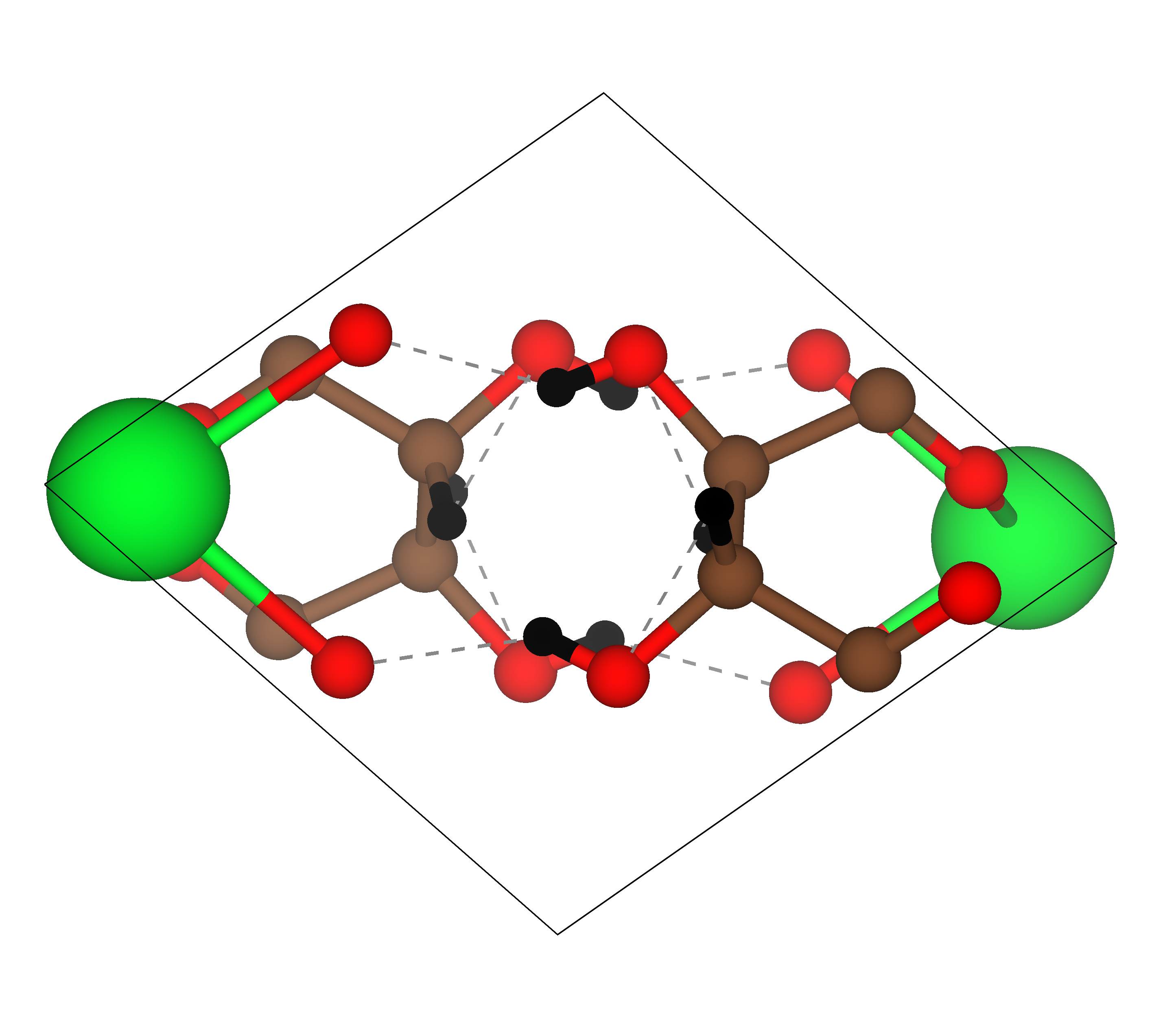}
    \put(3,92){\small\textbf{a)}}
  \end{overpic}
  \hfill
  \begin{overpic}[height=\panelheight]{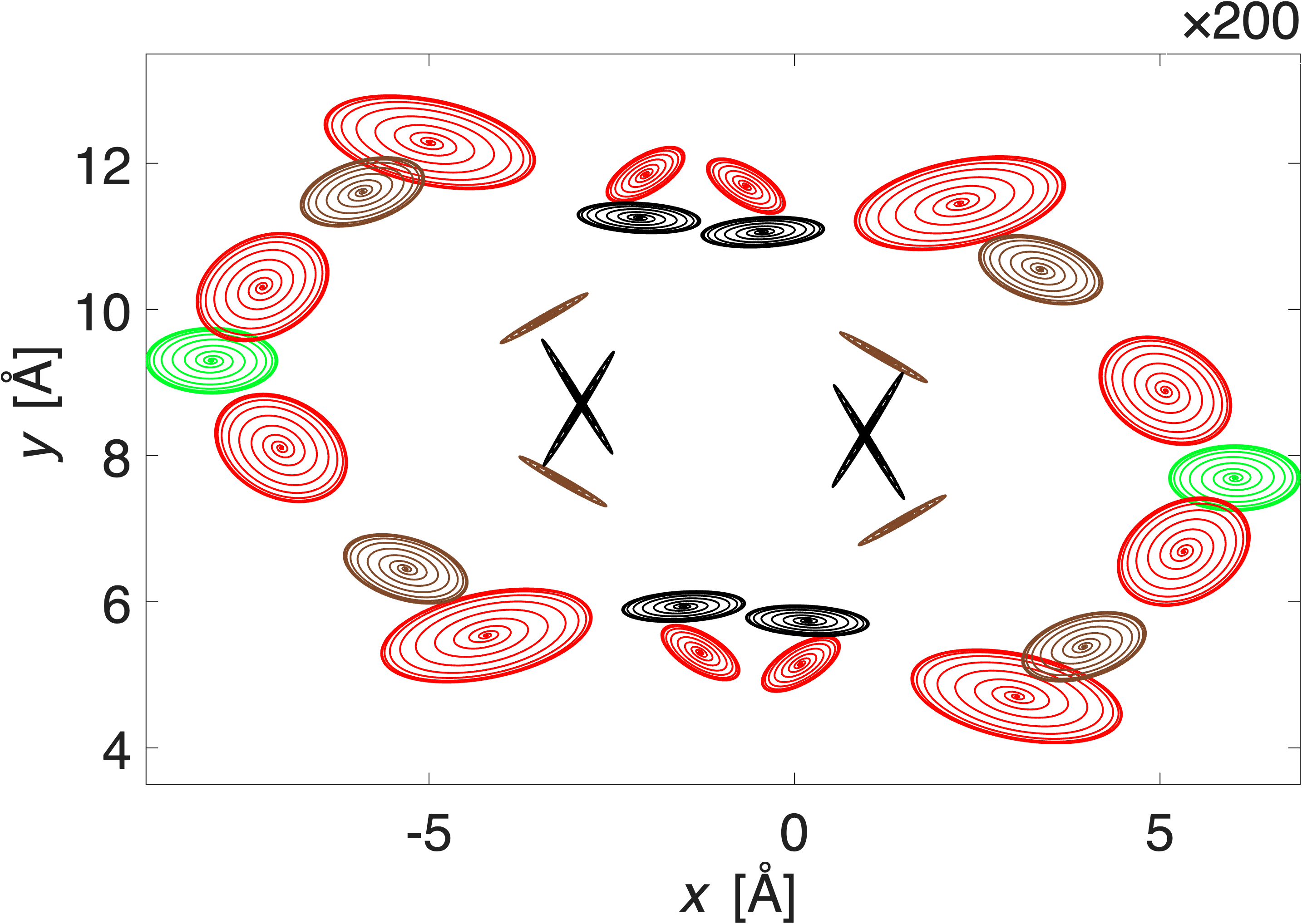}
    \put(3,75){\small\textbf{b)}}
    \put(87,69){\colorbox{white}{\small x141}}
  \end{overpic}
  \hfill
  \begin{overpic}[height=\panelheight]{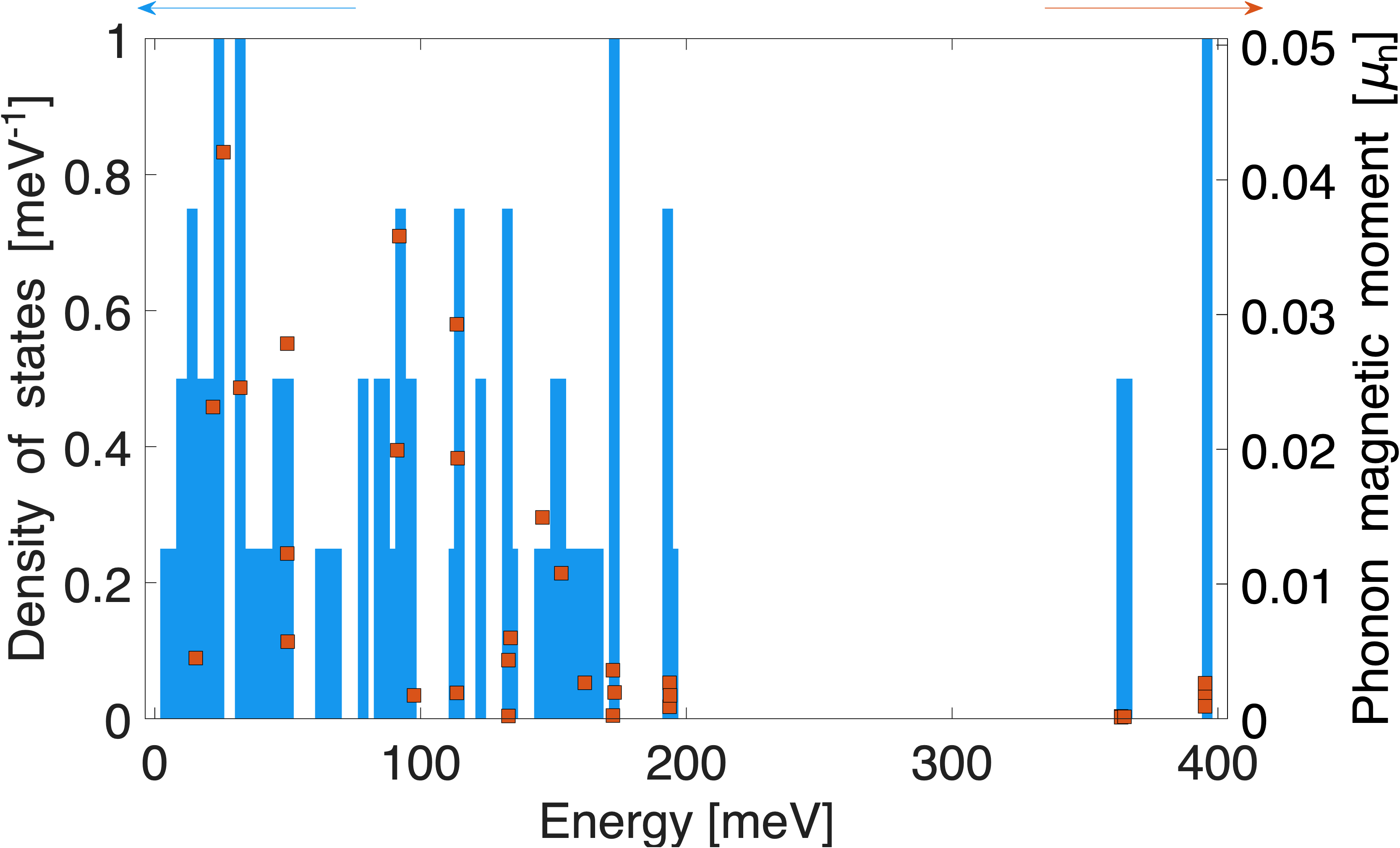}
    \put(3,64){\small\textbf{c)}}
  \end{overpic}
\caption{Phonon magnetism in Sr(tar). \textbf{a)} Crystal structure of the primitive unit cell, with atoms colored by element (Sr green, O red, C brown, H black). \textbf{b)} Projection onto the $xy$ plane of the atomic trajectories in the circularly polarized superposition of the degenerate IR-active mode pair with the largest magnetic moment. Each ellipse traces the motion of one atom; displacements are magnified by the factor indicated in the upper right ($\times 141$) for visibility. The trajectories are shown for the period following a model circularly polarized laser pulse (see Sec.~\ref{sec:dmf} for details). Although several atoms undergo circular motion, pairs of oxygen sublattices rotate in opposite directions, strongly suppressing the net circular polarization and therefore the phonon magnetic moment despite the large effective gyromagnetic ratio. \textbf{c)} Phonon density of states (blue, left axis) and phonon magnetic moments (orange squares, right axis, in units of $\upmu_\mathrm{n}$) as a function of phonon energy.}  \label{fig:cugson}
\end{figure*}

\begin{figure*}
  \vspace{1em}       
  \centering
  \begin{overpic}[height=\panelheight]{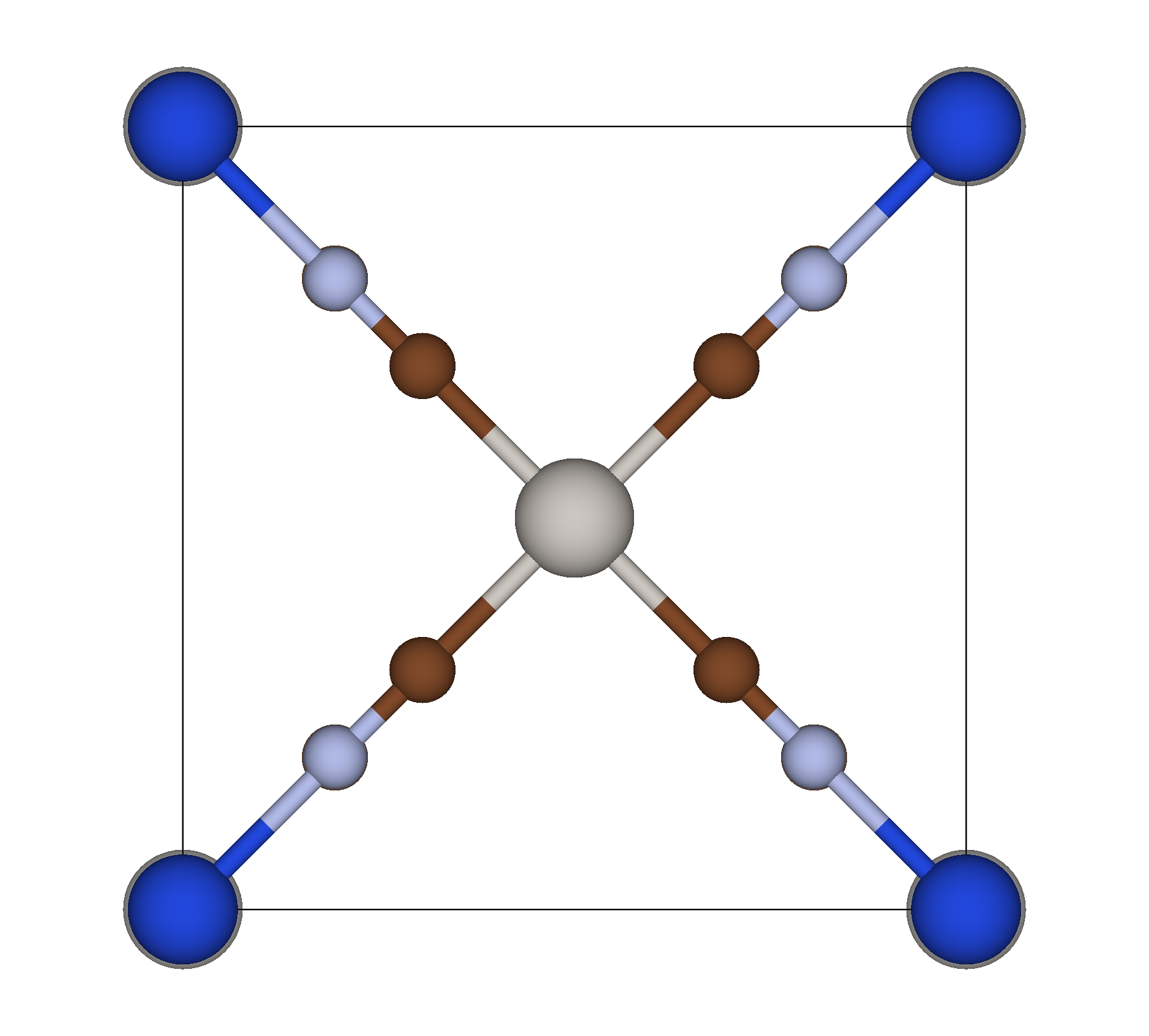}
    \put(3,95){\small\textbf{a)}}
  \end{overpic}
  \hfill
  \begin{overpic}[height=\panelheight]{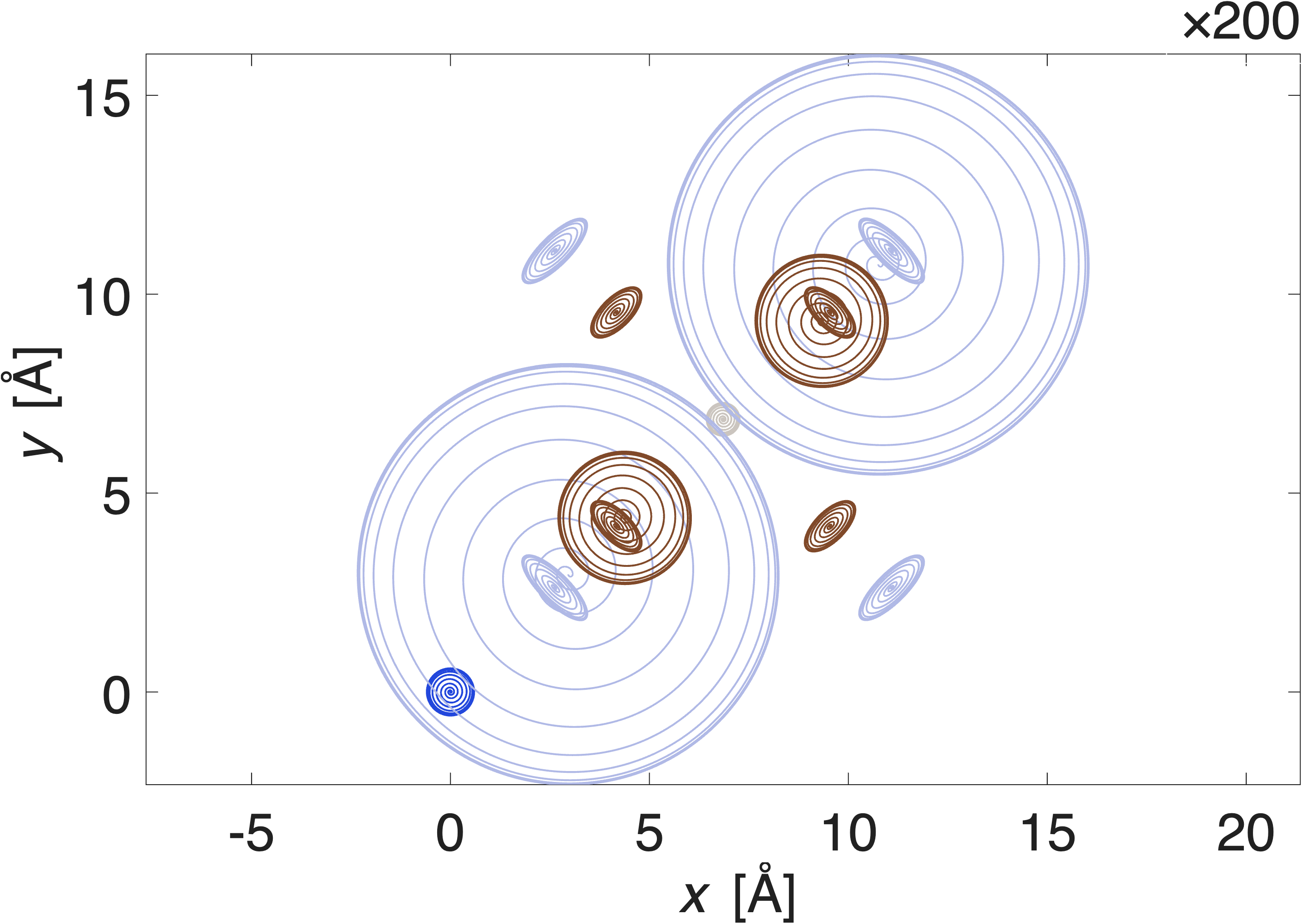}
    \put(3,75){\small\textbf{b)}}
    \put(87,69){\colorbox{white}{\small x141}}

  \end{overpic}
  \hfill
  \begin{overpic}[height=\panelheight]{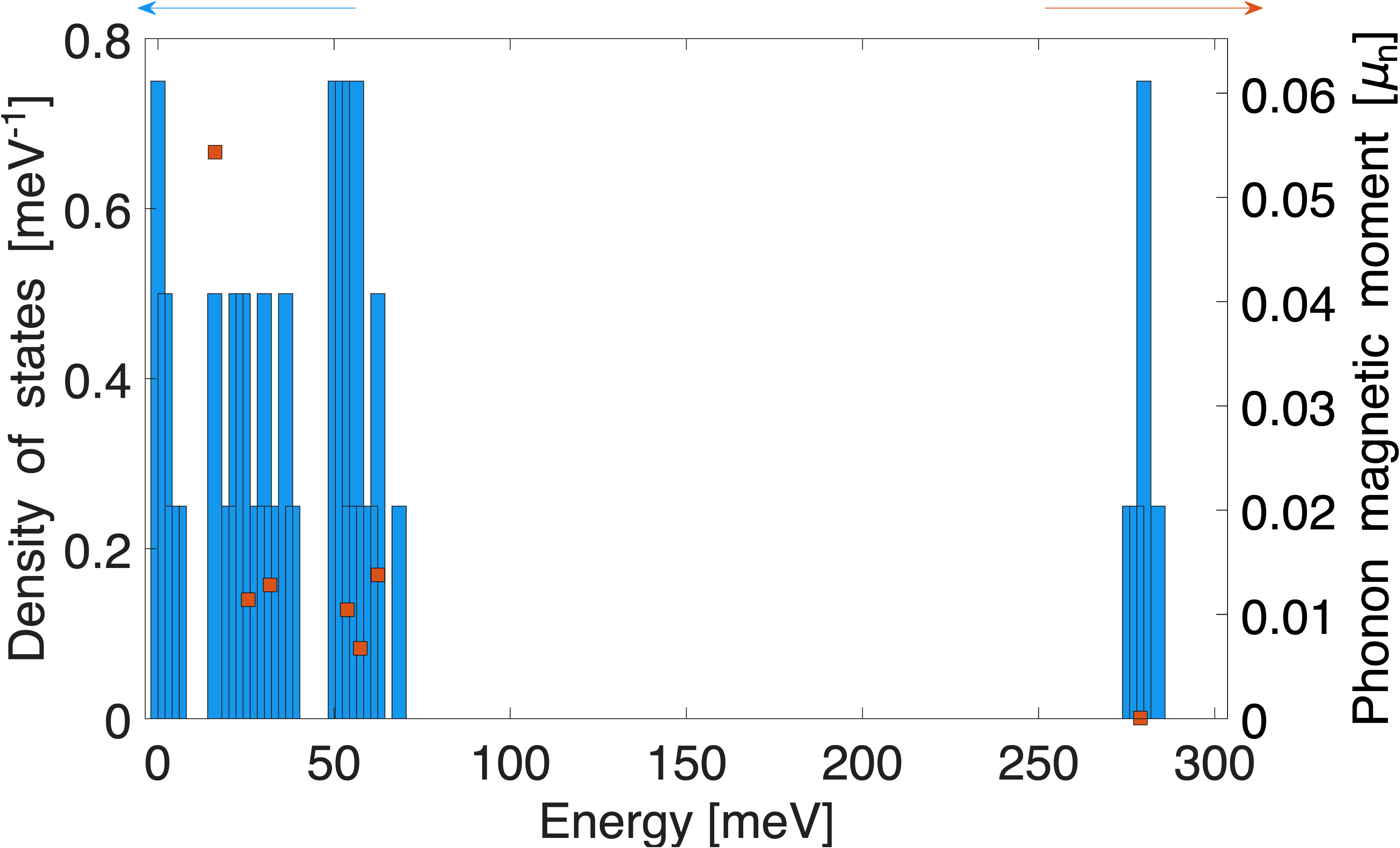}
    \put(3,63){\small\textbf{c)}}
  \end{overpic}
  \caption{Phonon magnetism in CuPt(CN)$_6$. \textbf{a)} Crystal structure of the primitive unit cell, with atoms colored by element (Cu blue, Pt gray, N light blue, C brown, H black). \textbf{b)} Projection onto the $xy$ plane of the atomic trajectories following the model pulse. \textbf{c)} Phonon density of states and phonon magnetic moments as a function of energy.}
  \label{fig:cuptcn6}
\end{figure*}

\begin{figure*}
  \vspace{1em}       
  \centering
  \begin{overpic}[height=\panelheight]{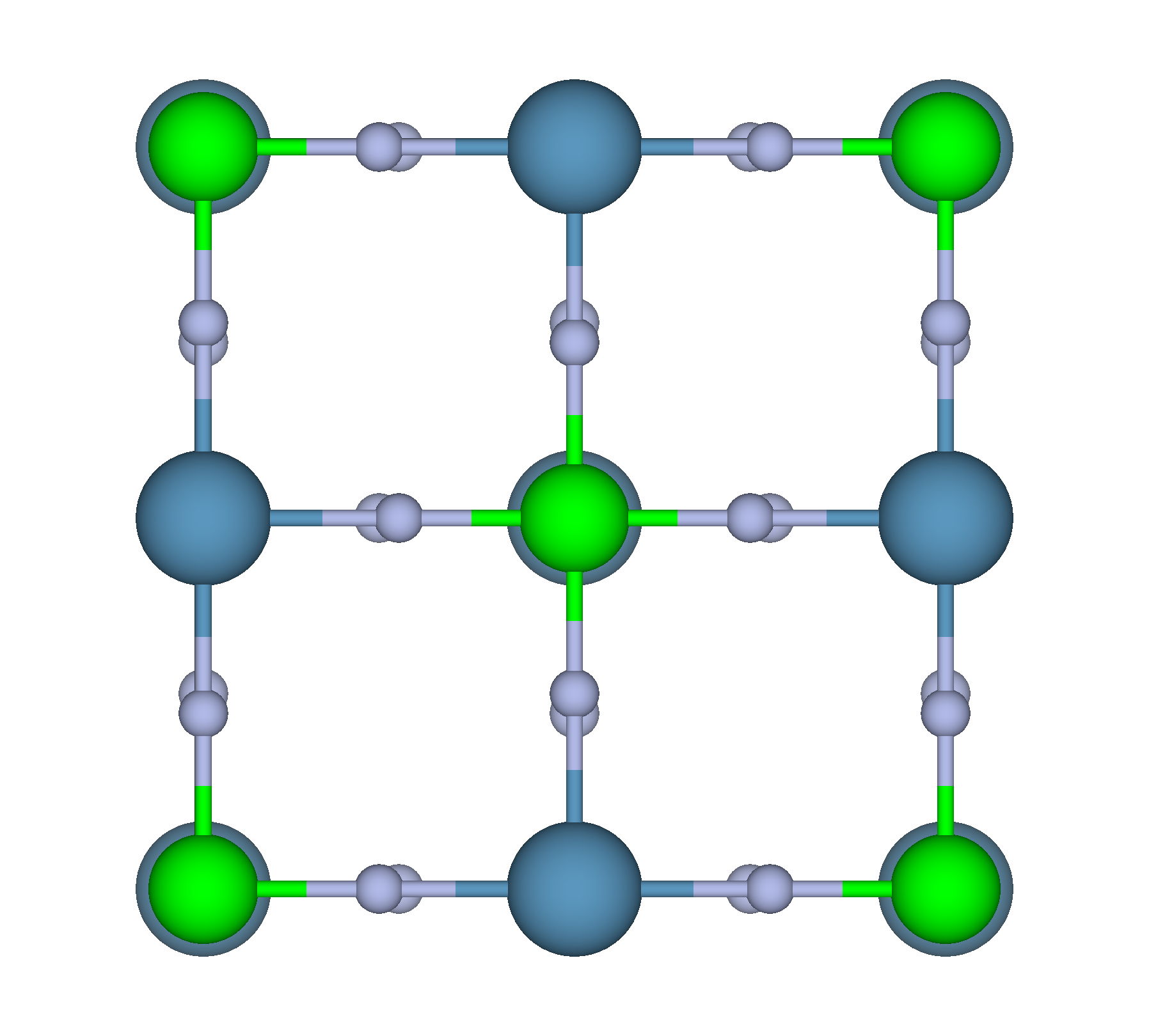}
    \put(3,98){\small\textbf{a)}}
  \end{overpic}
  \hfill
  \begin{overpic}[height=\panelheight]{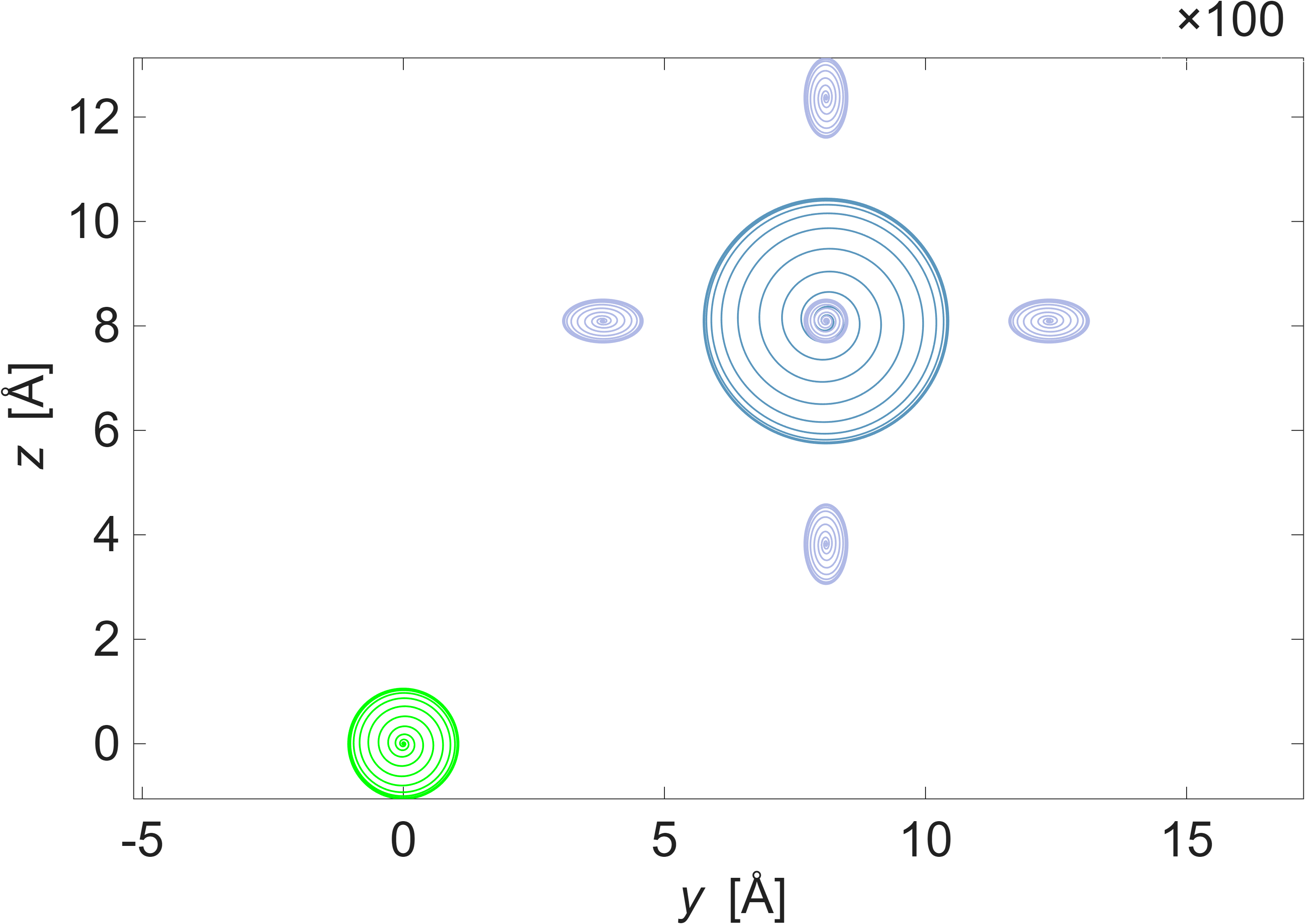}
    \put(3,75){\small\textbf{b)}}
    \put(87,69){\colorbox{white}{\small x100}}
  \end{overpic}
  \hfill
  \begin{overpic}[height=\panelheight]{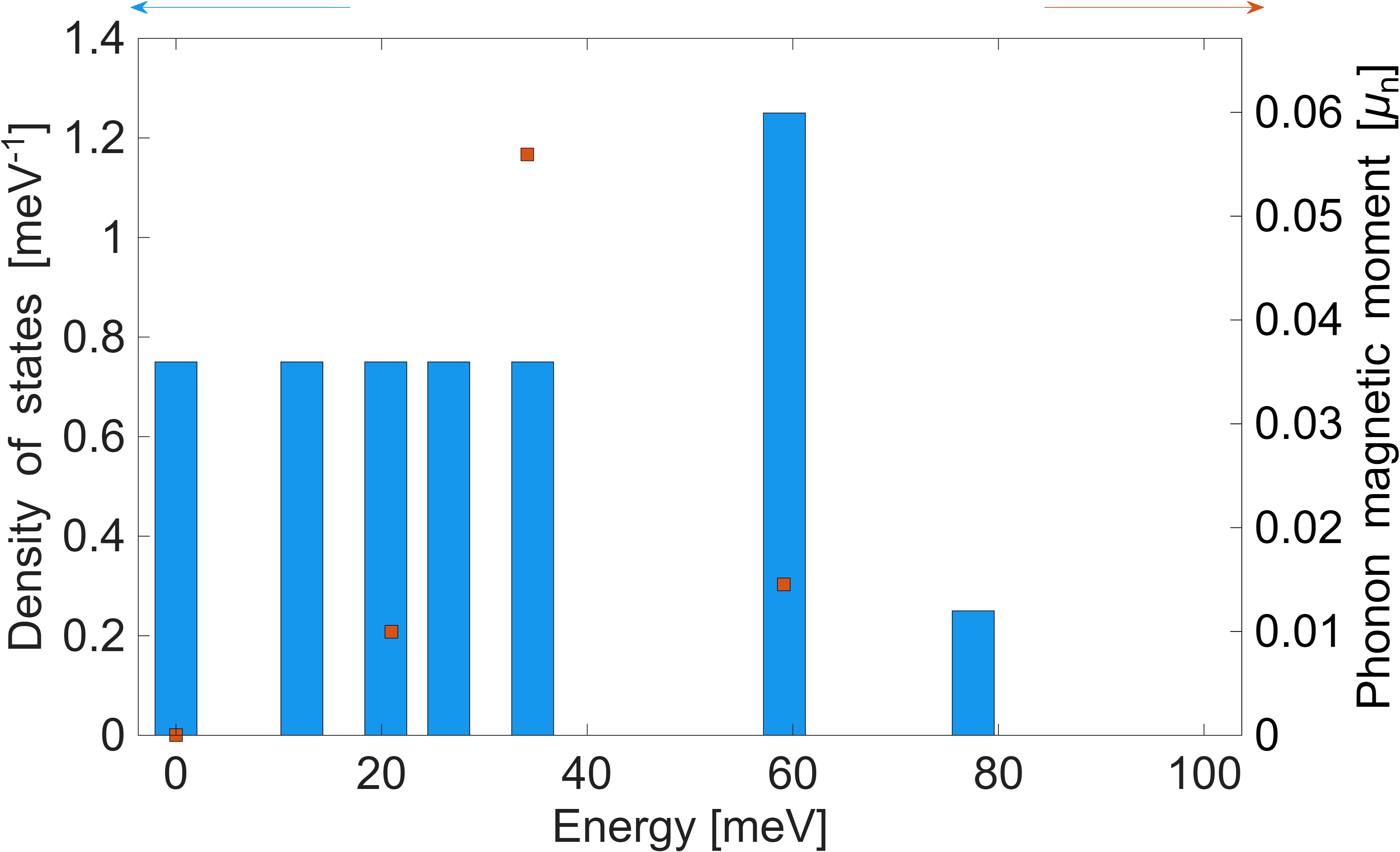}
    \put(3,63){\small\textbf{c)}}
  \end{overpic}
  \caption{Phonon magnetism in CaZrF$_6$. \textbf{a)} Crystal structure of the primitive unit cell, with atoms colored by element (Zr green, F purple, Ca blue). \textbf{b)} Projection onto the $yz$ plane of the atomic trajectories  following the model pulse. \textbf{c)} Phonon density of states and phonon magnetic moments as a function of energy.}
  \label{fig:cazrf6}
\end{figure*}

\begin{figure*}
  \vspace{1em}       
  \centering
  \begin{overpic}[height=\panelheight]{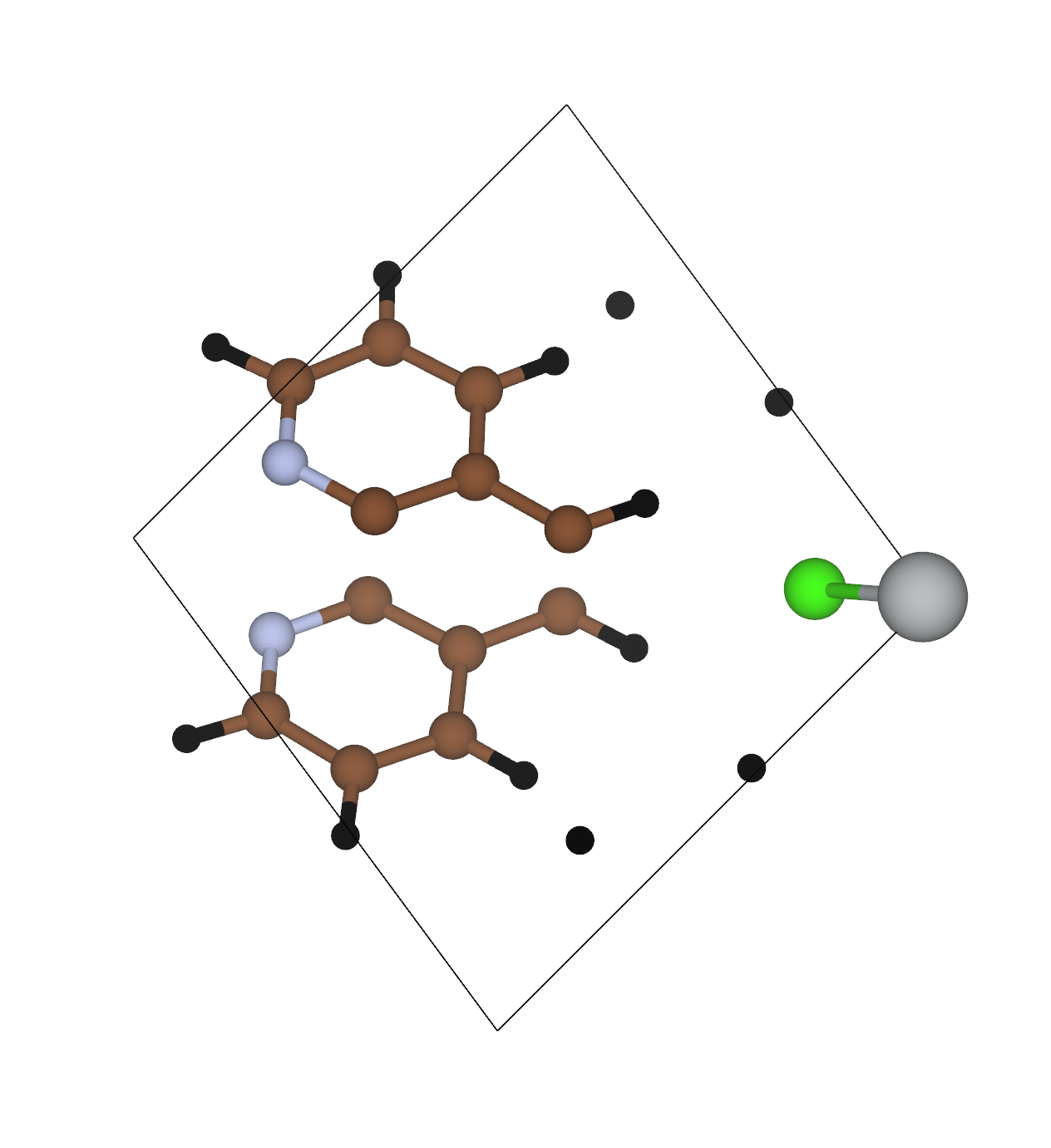}
    \put(3,104){\small\textbf{a)}}
  \end{overpic}
  \hfill
  \begin{overpic}[height=\panelheight]{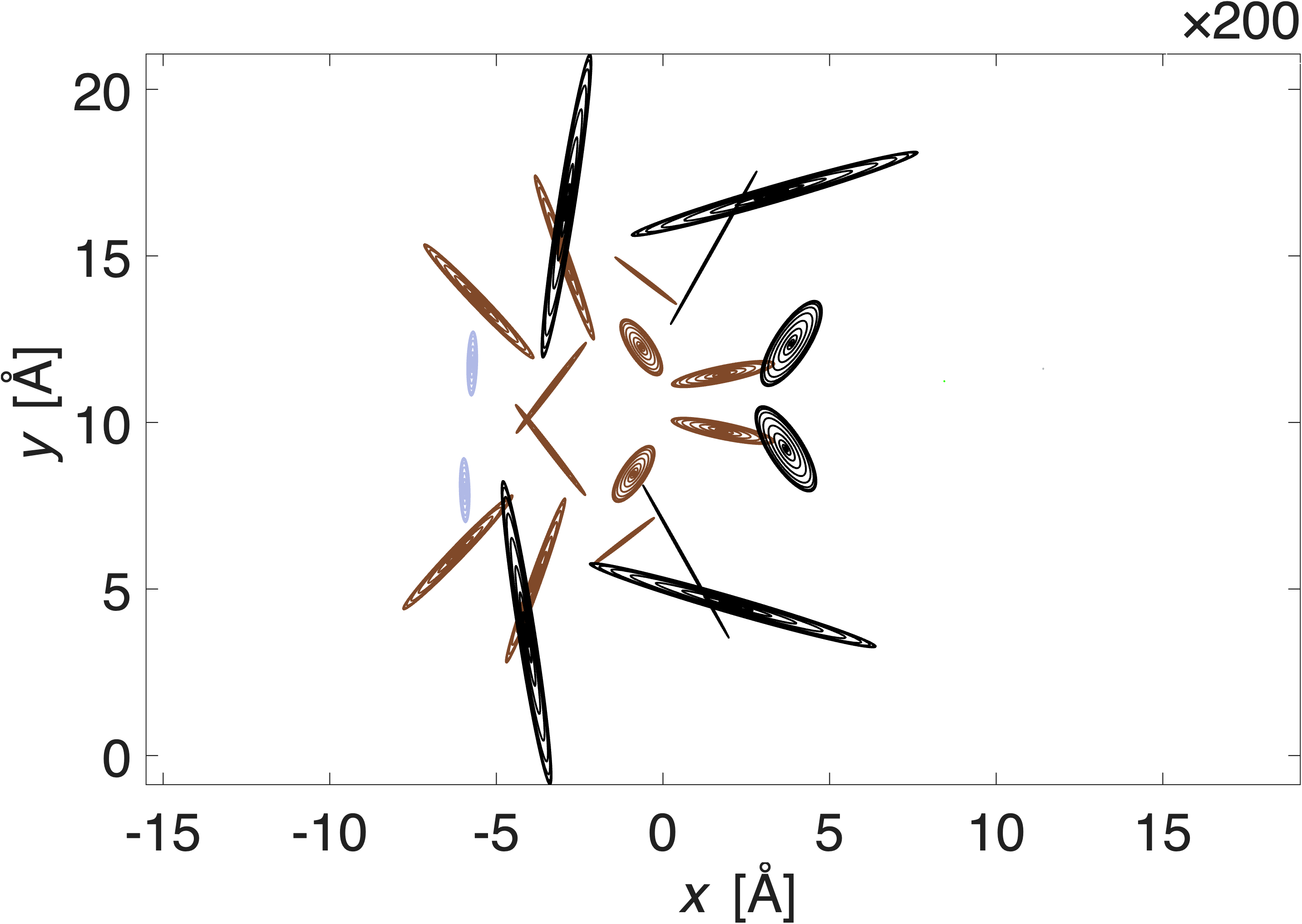}
    \put(3,75){\small\textbf{b)}}
            \put(87,69){\colorbox{white}{\small x141}}

  \end{overpic}
  \hfill
  \begin{overpic}[height=\panelheight]{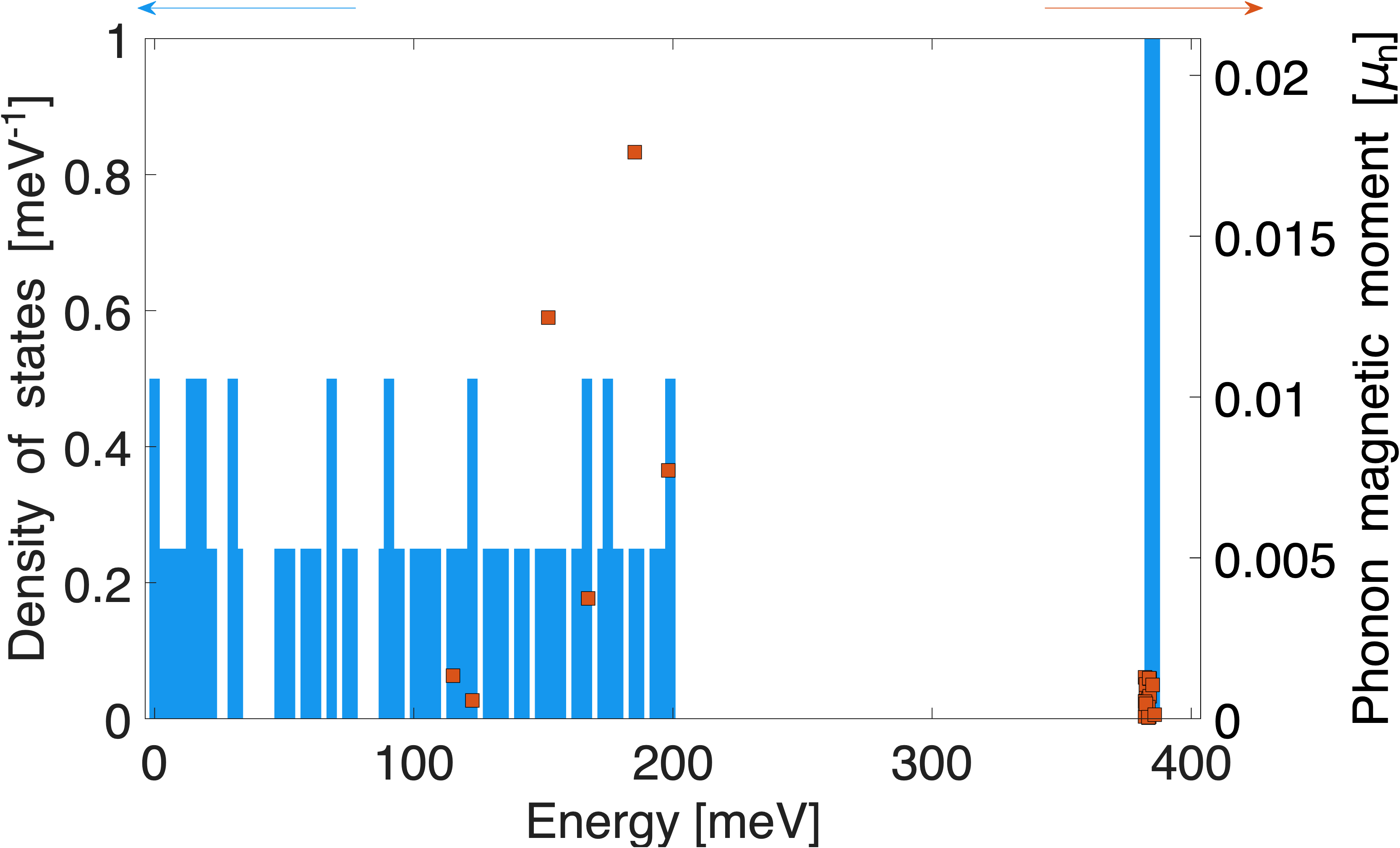}
    \put(3,63){\small\textbf{c)}}
  \end{overpic}
    \caption{Phonon magnetism in AgCl(phen). \textbf{a)} Crystal structure of the primitive unit cell, with atoms colored by element (Cl green, Ag grey, N light blue, C brown, H black). \textbf{b)} Projection onto the $xy$ plane of the atomic trajectories  following the model pulse. \textbf{c)} Phonon density of states and phonon magnetic moments as a function of energy.}
  \label{fig:mahtul}
\end{figure*}

\begin{figure*}
  \vspace{1em}      
  \centering
  \begin{overpic}[height=\panelheight]{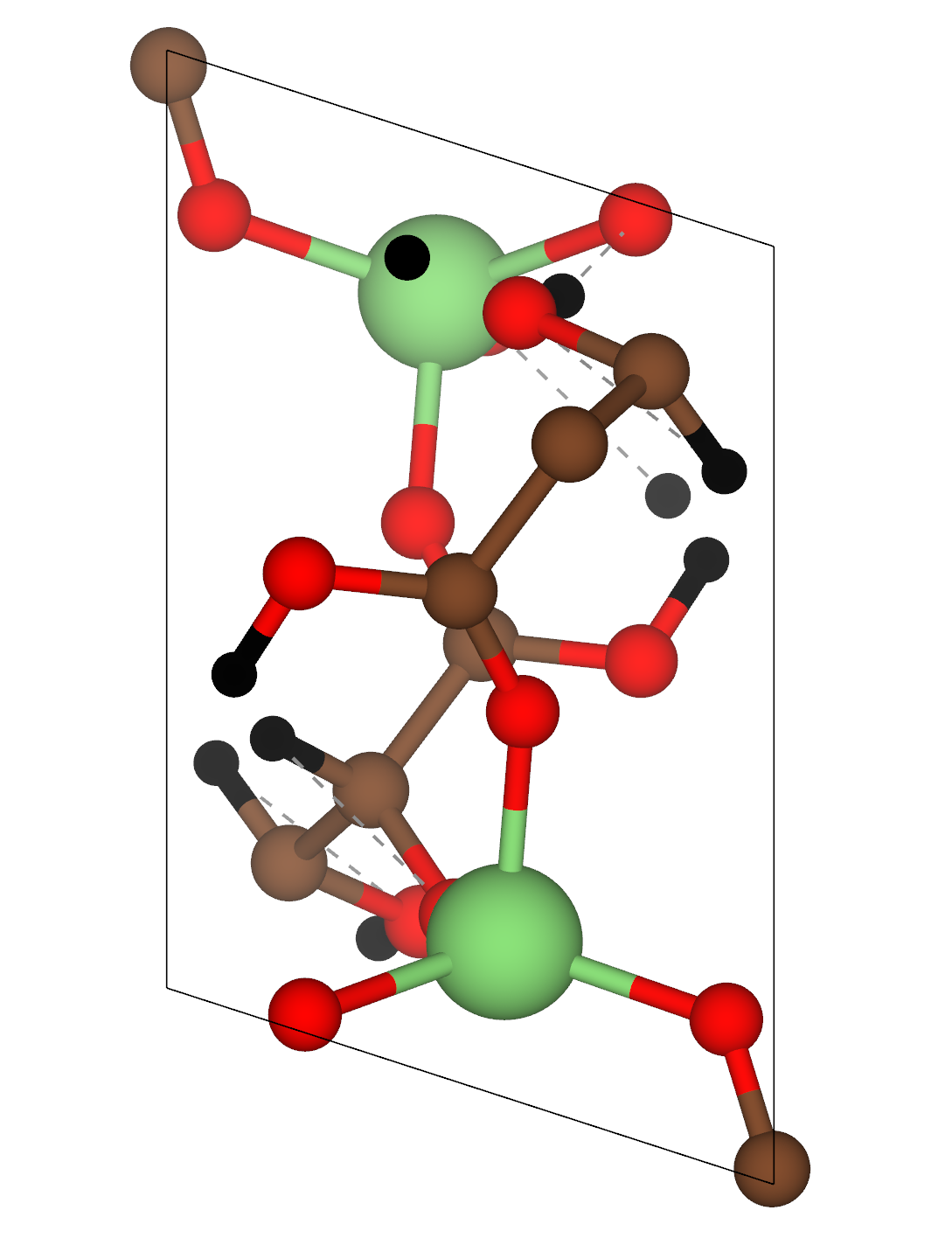}
    \put(3,105){\small\textbf{a)}}
  \end{overpic}
  \hfill
  \begin{overpic}[height=\panelheight]{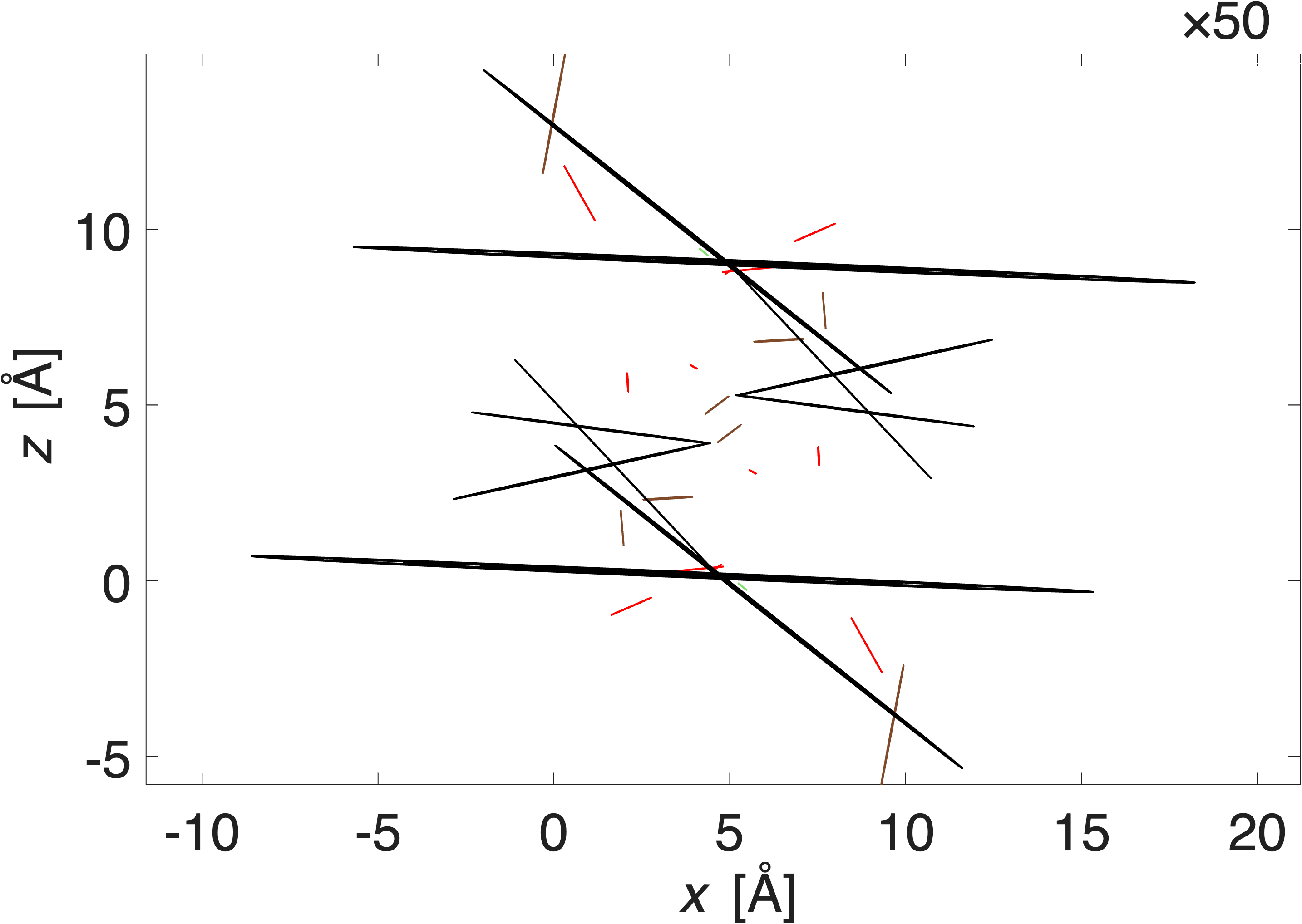}
    \put(3,75){\small\textbf{b)}}
    \put(87,69){\colorbox{white}{\small x35}}

  \end{overpic}
  \hfill
  \begin{overpic}[height=\panelheight]{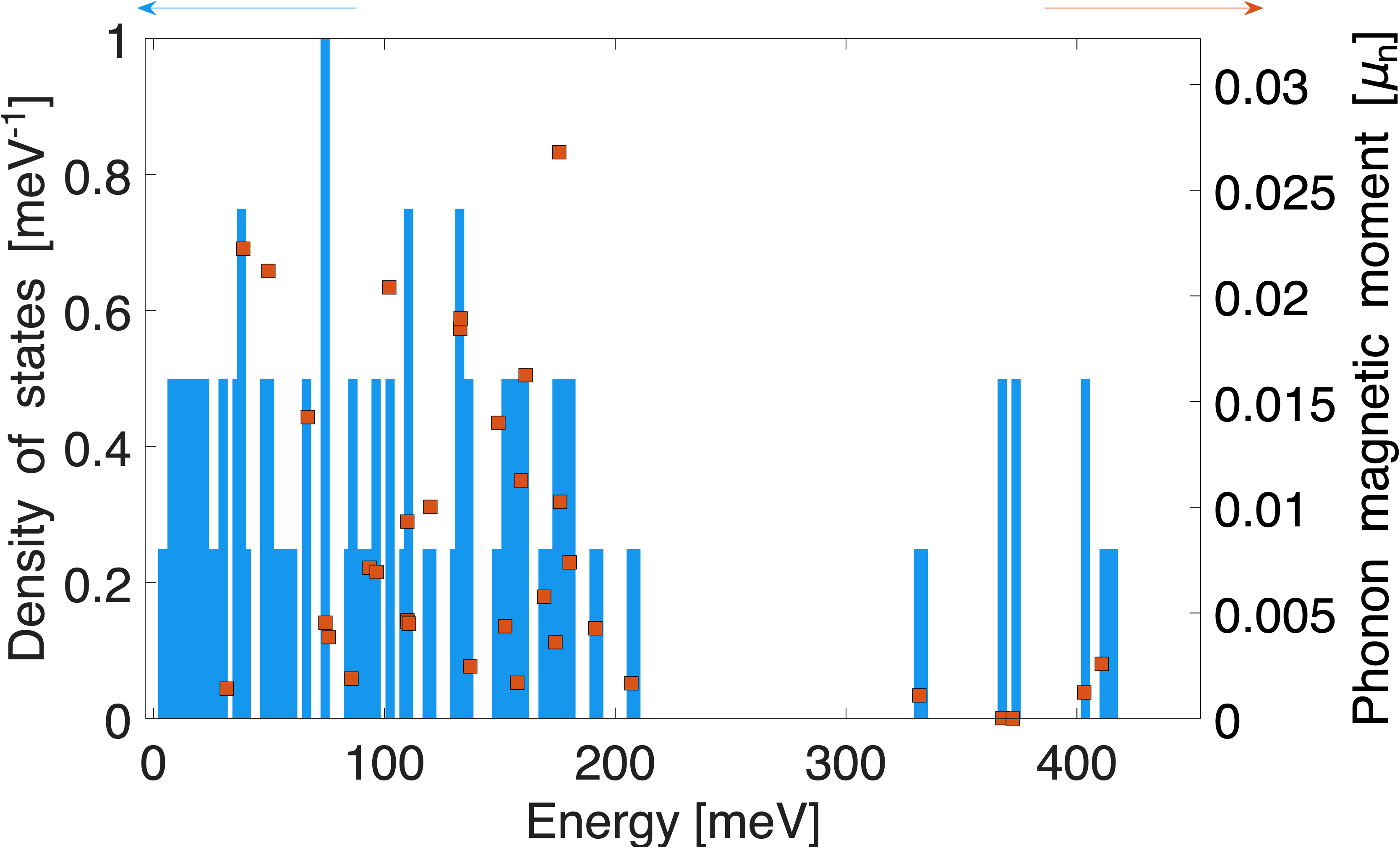}
    \put(3,64){\small\textbf{c)}}
  \end{overpic}
  \caption{Phonon magnetism in Li$_2$(tar). \textbf{a)} Crystal structure of the primitive unit cell, with atoms colored by element (Li light green, O red, C brown, H black). \textbf{b)} Projection onto the $xz$ plane of the atomic trajectories  following the model pulse. \textbf{c)} Phonon density of states and phonon magnetic moments as a function of energy.}
  \label{fig:litar}
\end{figure*}

\begin{figure*}
  \vspace{1em}       
  \centering
  \begin{overpic}[height=\panelheight]{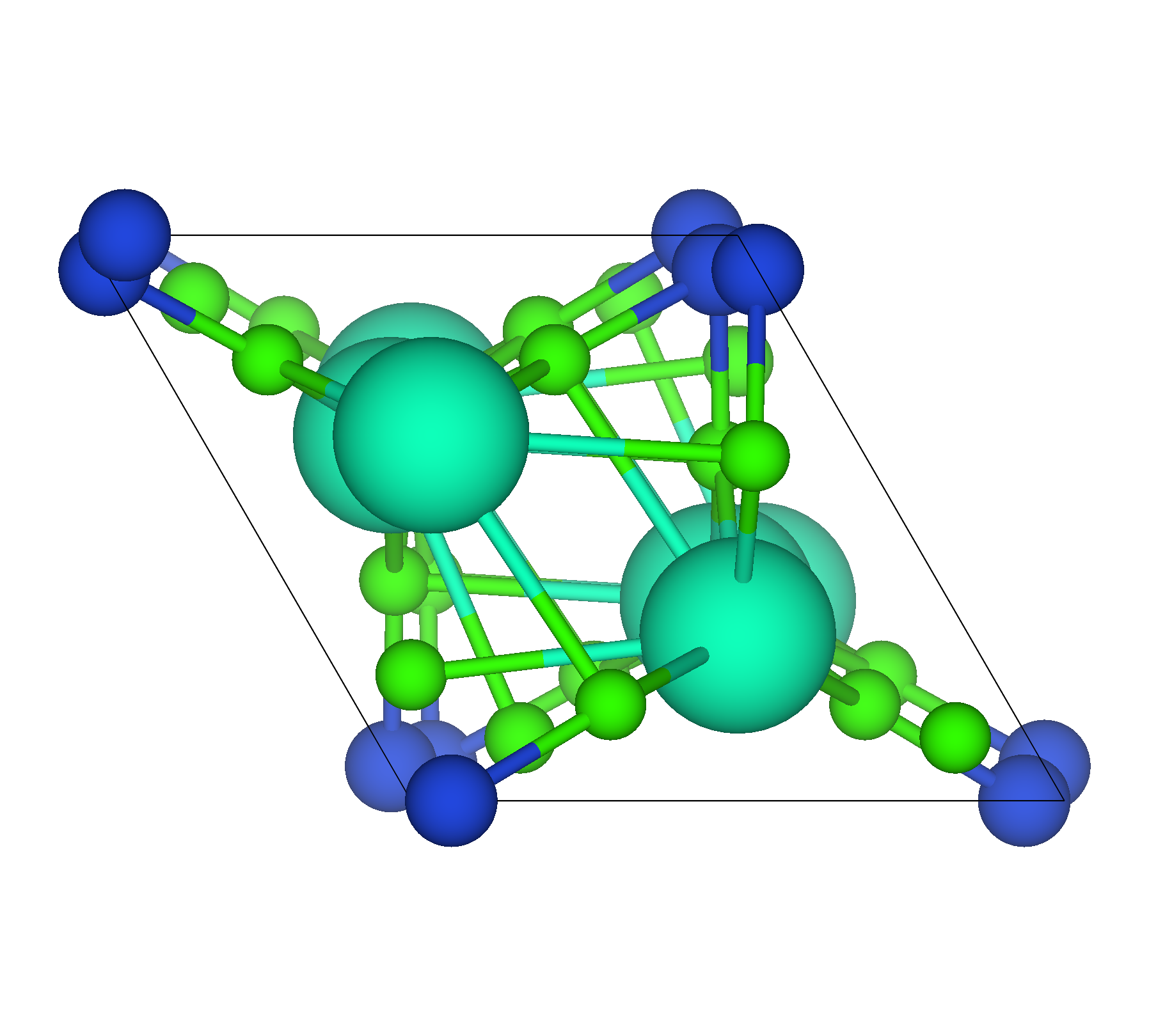}
    \put(3,95){\small\textbf{a)}}
  \end{overpic}
  \hfill
  \begin{overpic}[height=\panelheight]{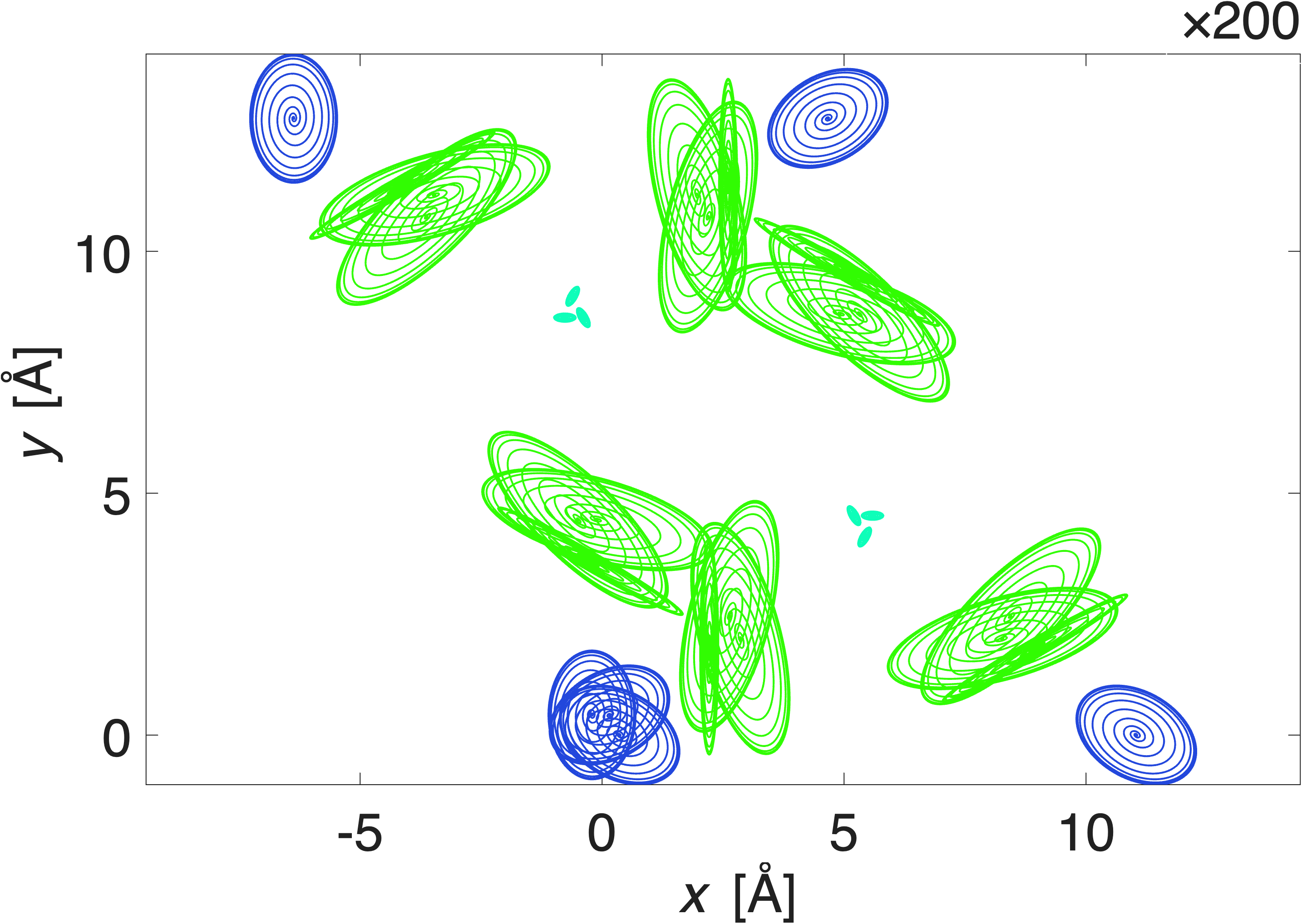}
    \put(3,75){\small\textbf{b)}}
    \put(87,69){\colorbox{white}{\small x141}}
  \end{overpic}
  \hfill
  \begin{overpic}[height=\panelheight]{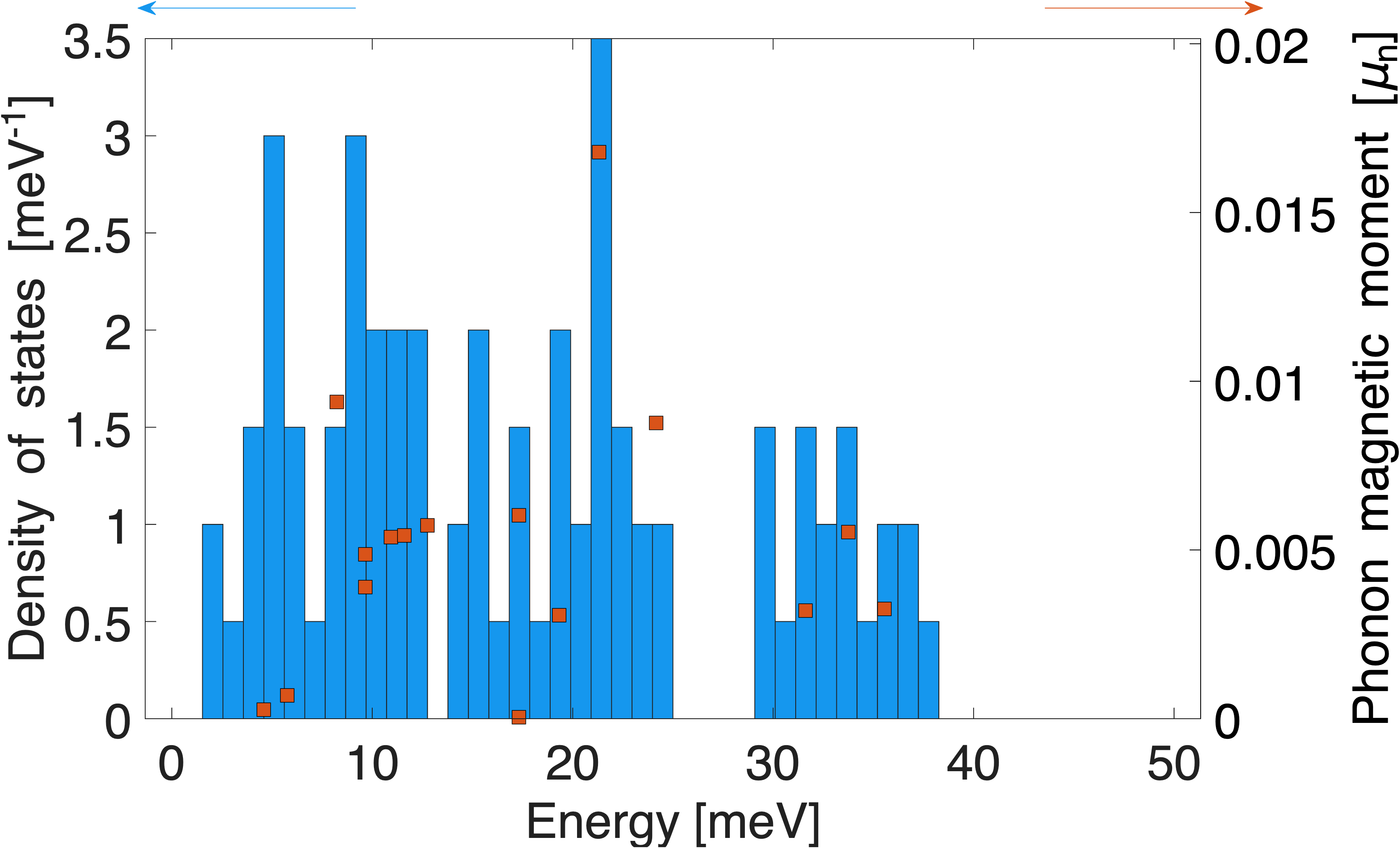}
    \put(3,63){\small\textbf{c)}}
  \end{overpic}
  \caption{Phonon magnetism in CsCuCl$_3$. \textbf{a)} Crystal structure of the primitive unit cell, with atoms colored by element (Cl green, Cs teal, Cu blue). \textbf{b)} Projection onto the $xy$ plane of the atomic trajectories  following the model pulse. \textbf{c)} Phonon density of states and phonon magnetic moments as a function of energy.}
  \label{fig:CsCuCl3}
\end{figure*}

\begin{figure*}
  \vspace{1em}       
  \centering
  \begin{overpic}[height=\panelheight]{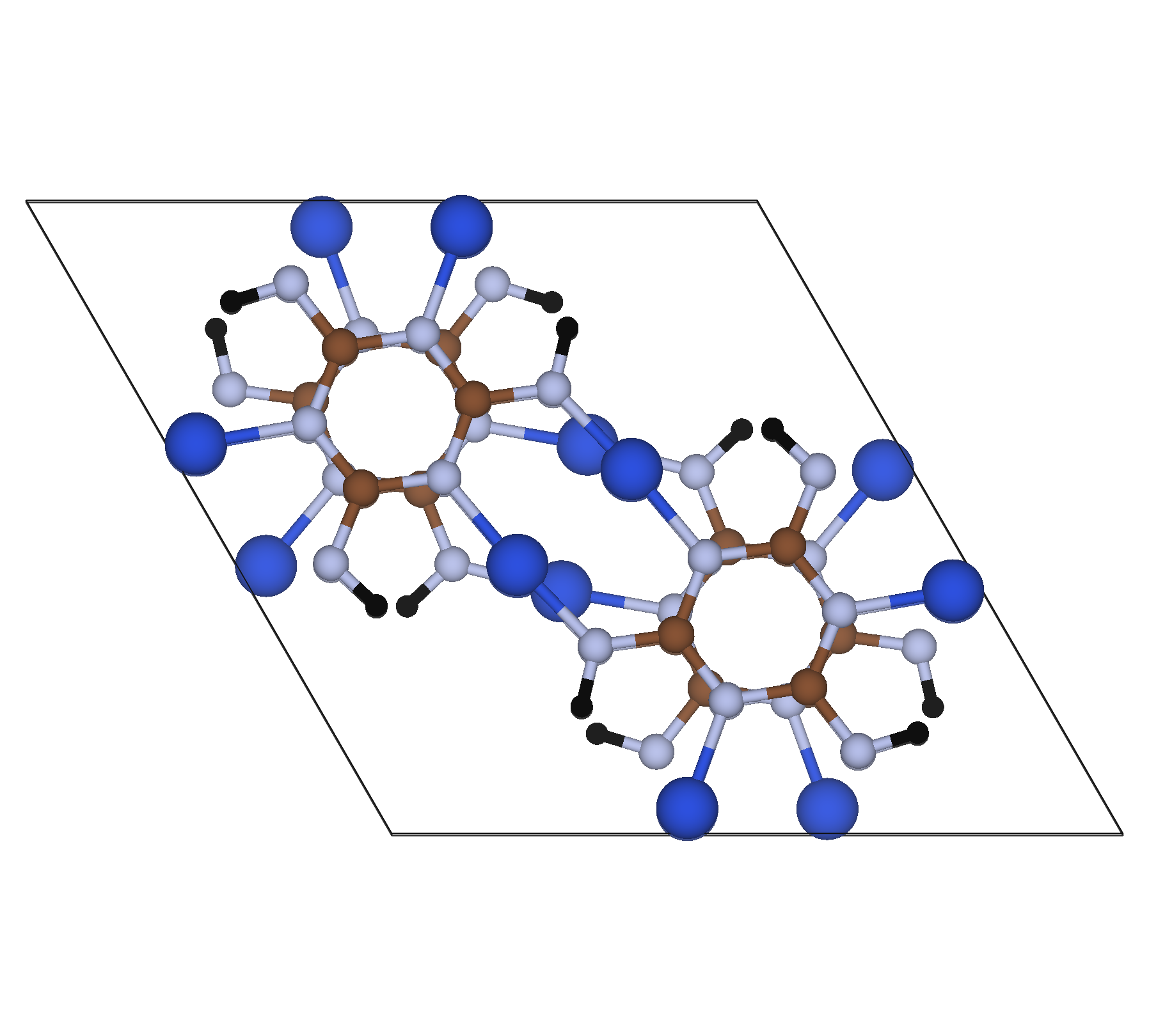}
    \put(3,95){\small\textbf{a)}}
  \end{overpic}
  \hfill
  \begin{overpic}[height=\panelheight]{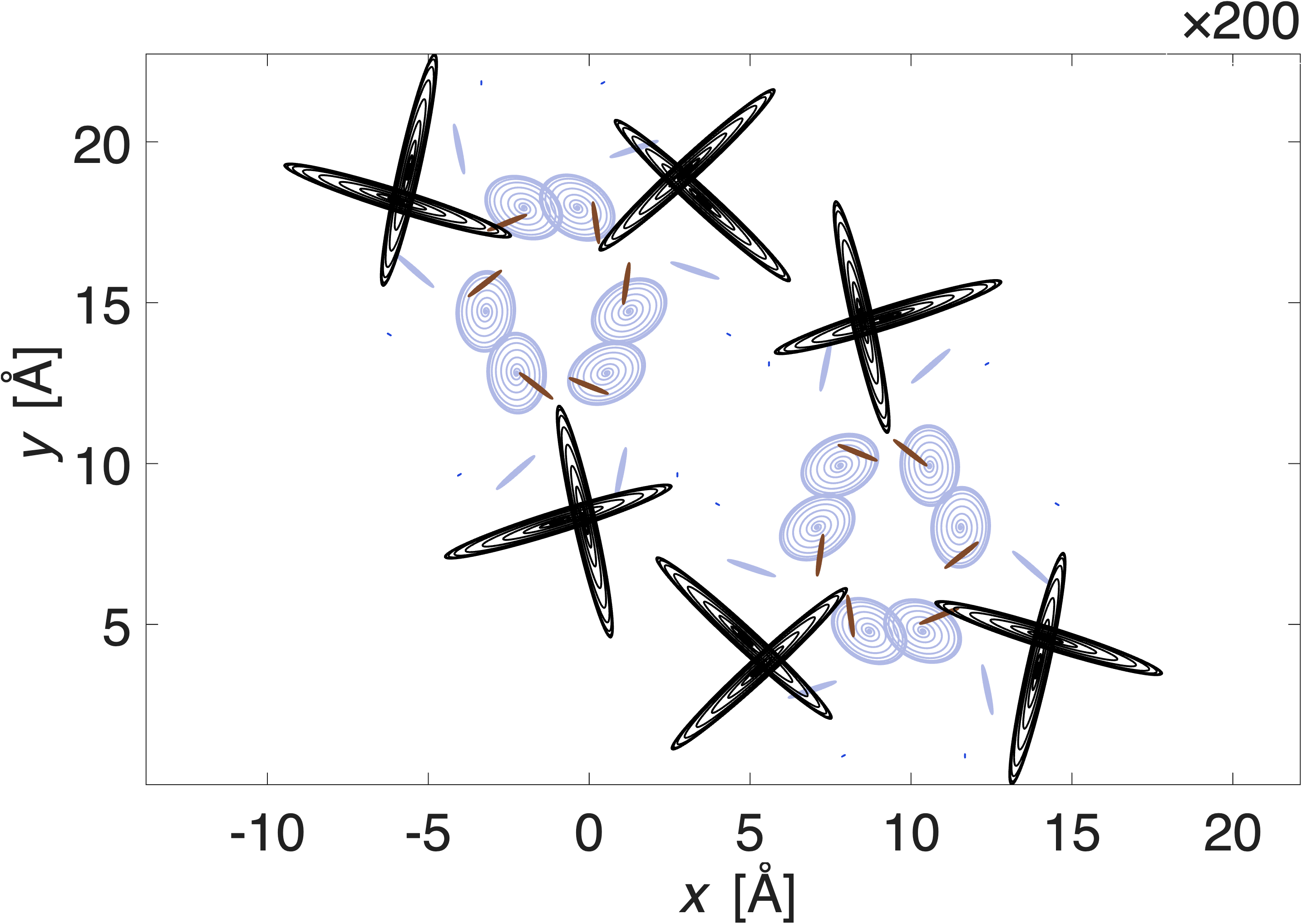}
    \put(3,75){\small\textbf{b)}}
    \put(87,69){\colorbox{white}{\small x141}}

  \end{overpic}
  \hfill
  \begin{overpic}[height=\panelheight]{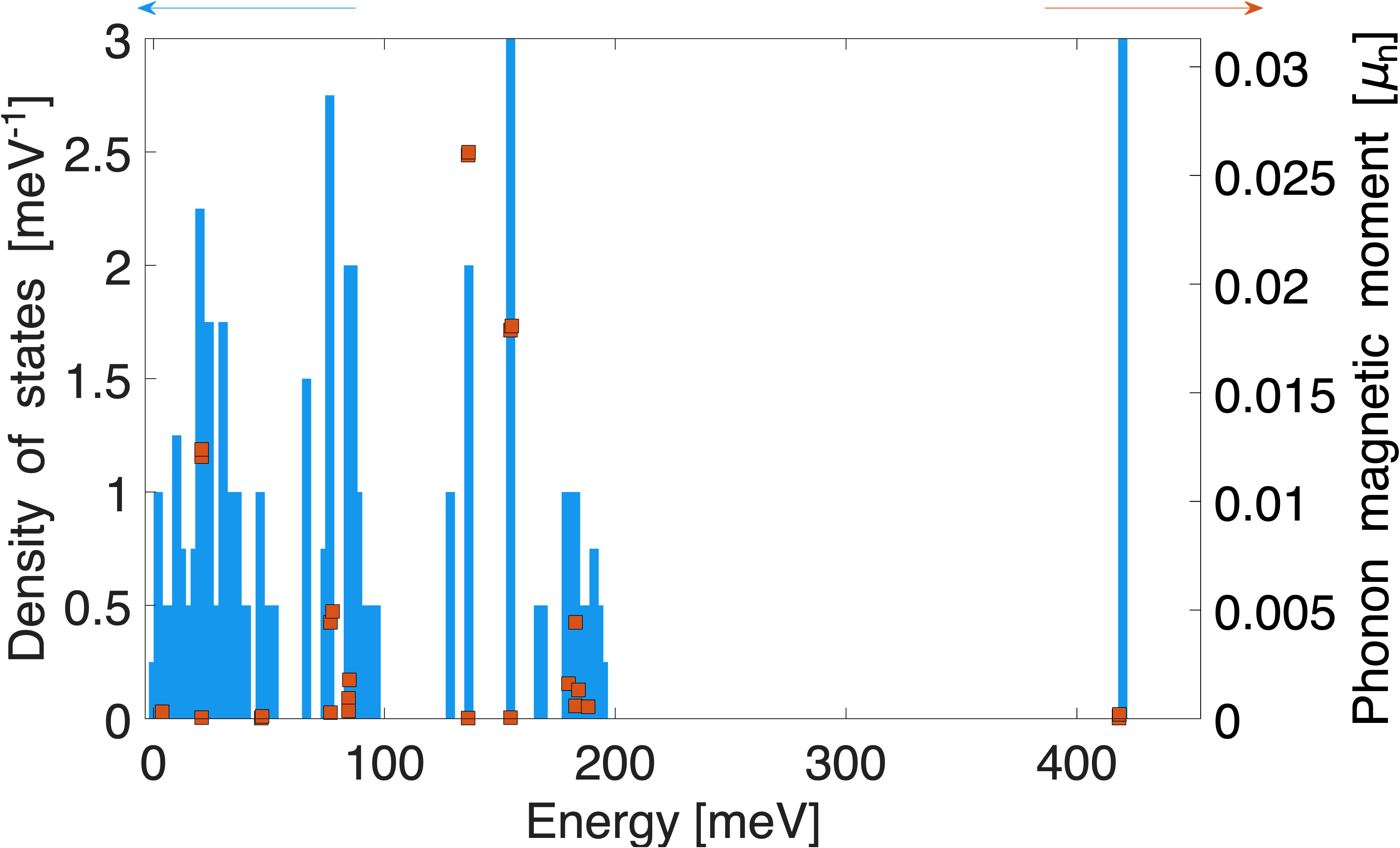}
    \put(3,63){\small\textbf{c)}}
  \end{overpic}
  \caption{Phonon magnetism in Cu$_3$(mel). \textbf{a)} Crystal structure of the primitive unit cell, with atoms colored by element (Cu blue, N light blue, C brown, H black). \textbf{b)} Projection onto the $xy$ plane of the atomic trajectories  following the model pulse. \textbf{c)} Phonon density of states and phonon magnetic moments as a function of energy.}
  \label{fig:cumel}
\end{figure*}

\begin{figure*}
  \vspace{1em}       
  \centering
  \begin{overpic}[height=\panelheight]{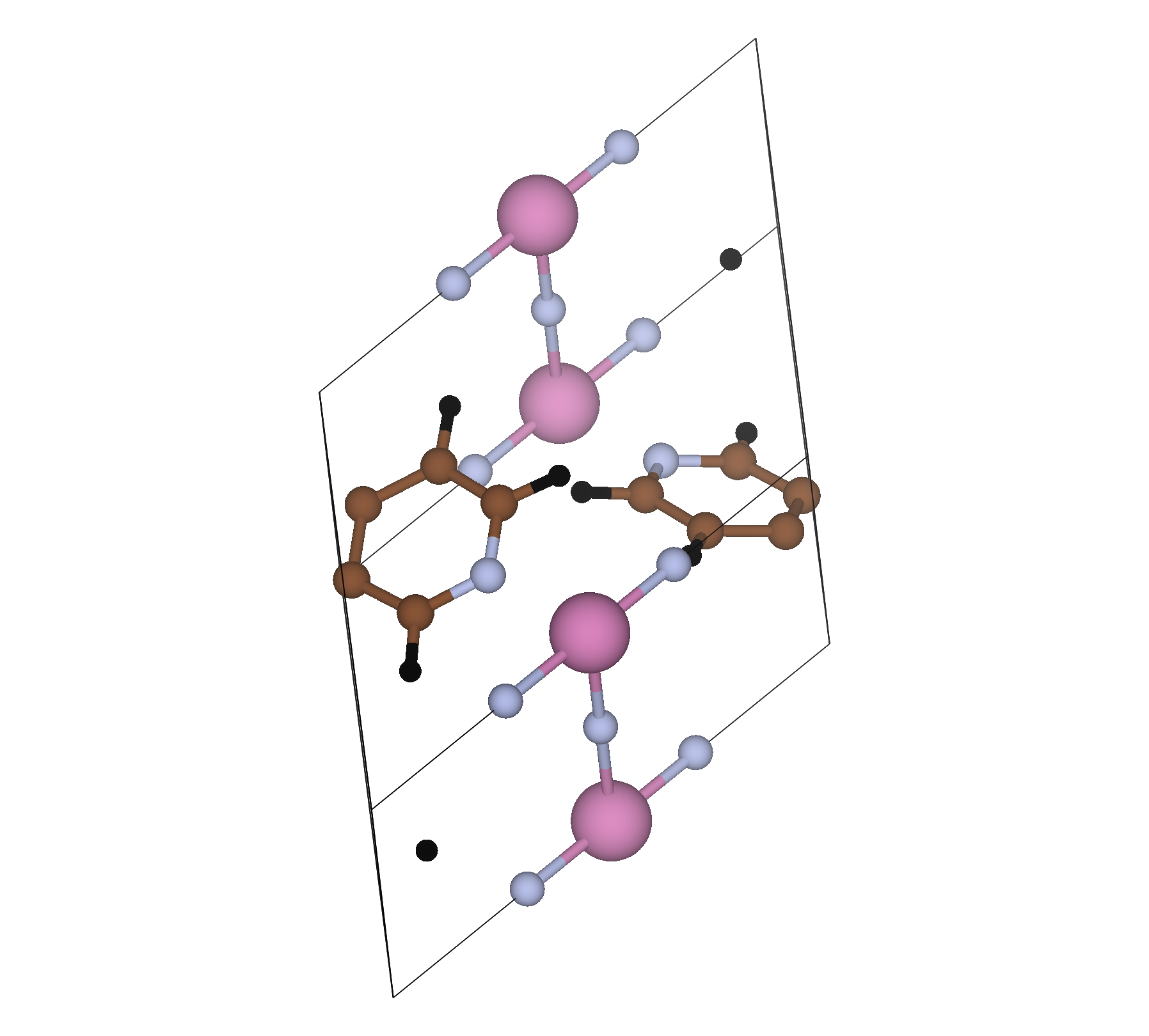}
    \put(3,95){\small\textbf{a)}}
  \end{overpic}
  \hfill
  \begin{overpic}[height=\panelheight]{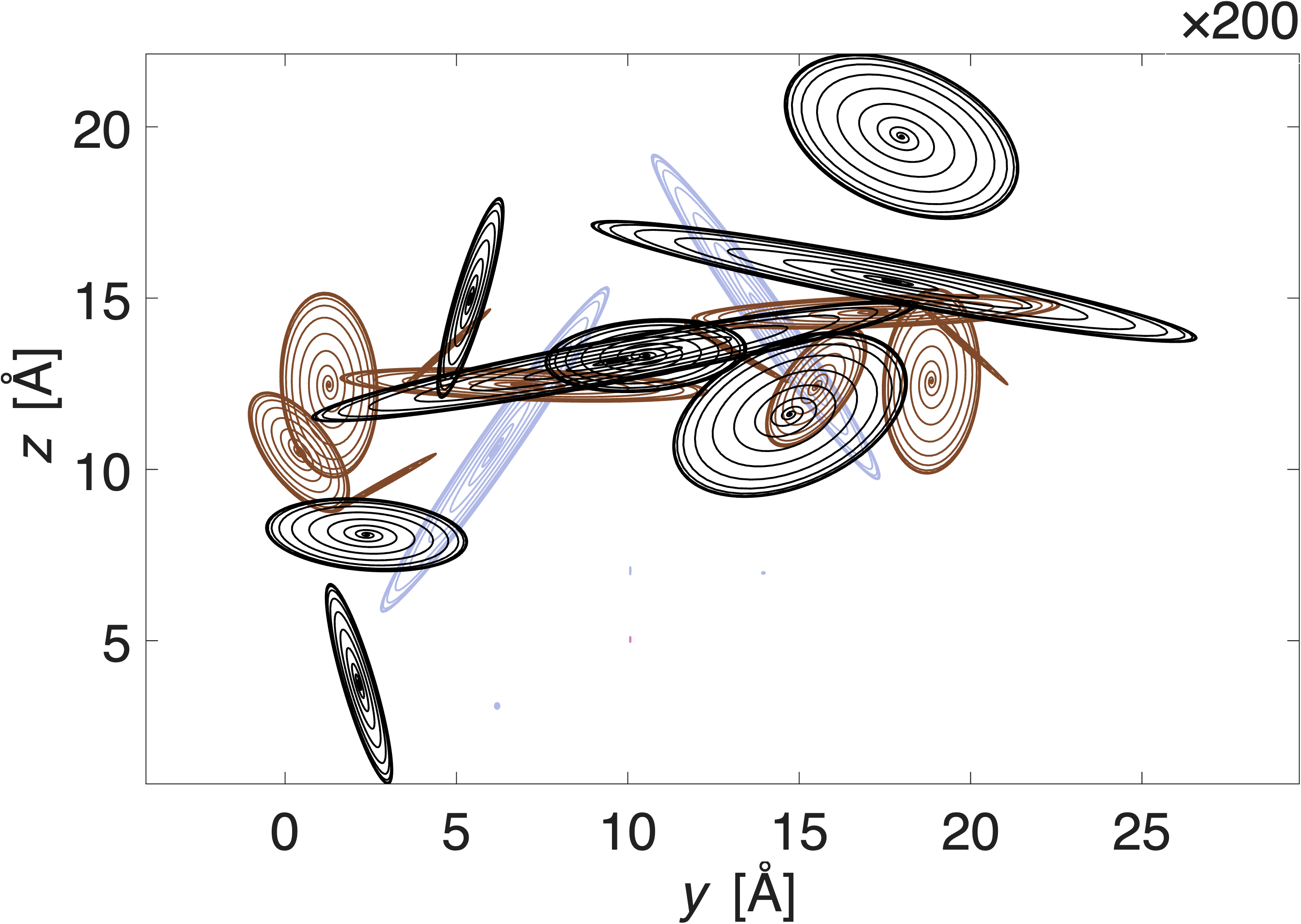}
    \put(3,75){\small\textbf{b)}}
        \put(87,69){\colorbox{white}{\small x141}}
  \end{overpic}
  \hfill
  \begin{overpic}[height=\panelheight]{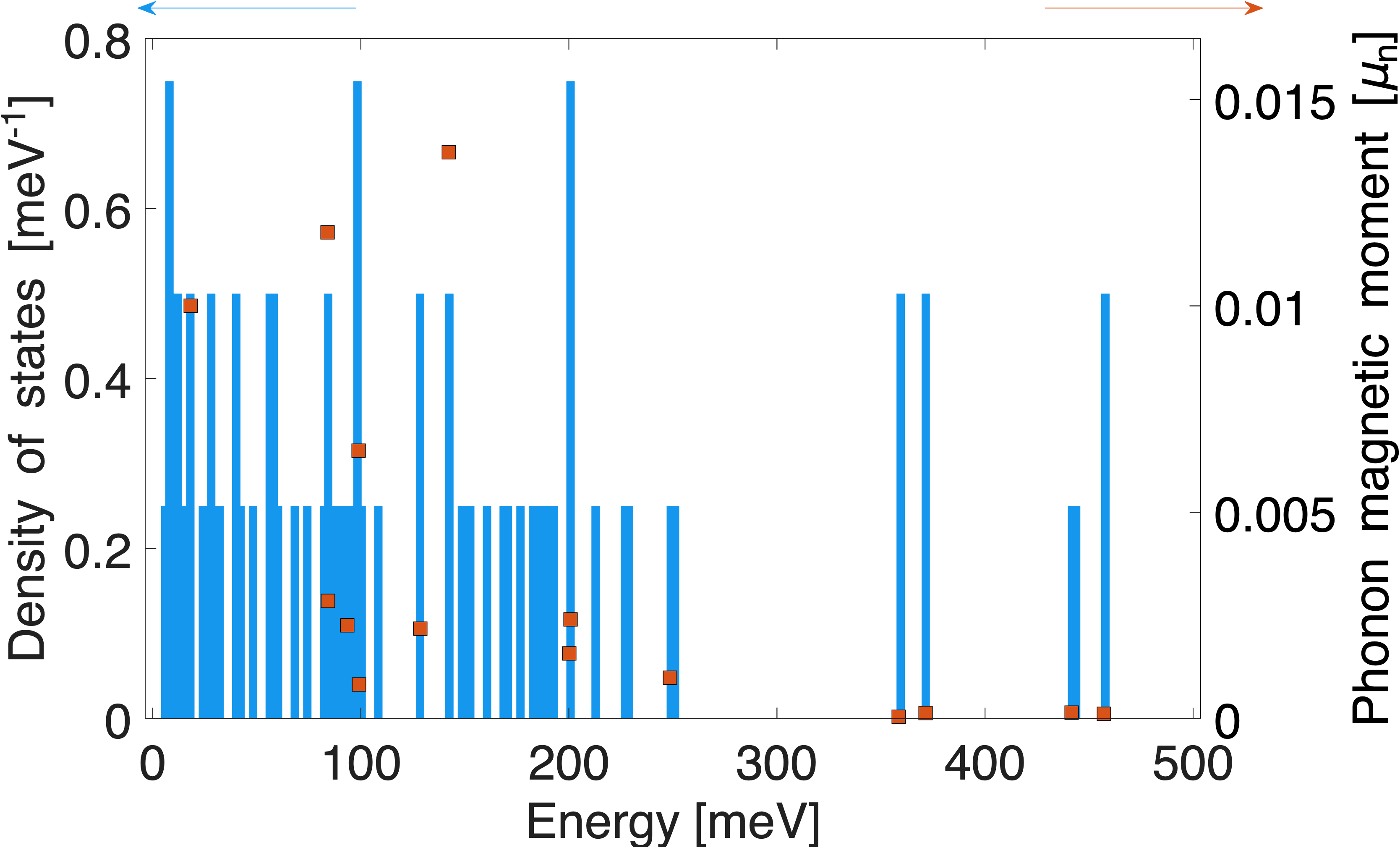}
    \put(3,63){\small\textbf{c)}}
  \end{overpic}
  \caption{Phonon magnetism in InF$_3$(bpy). \textbf{a)} Crystal structure of the primitive unit cell, with atoms colored by element (In light red, F purple, C brown, H black). \textbf{b)} Projection onto the $yz$ plane of the atomic trajectories  following the model pulse. \textbf{c)} Phonon density of states and phonon magnetic moments as a function of energy.}
  \label{fig:infb}
\end{figure*}

\begin{figure*}
  \vspace{1em}       
  \centering
  \begin{overpic}[height=\panelheight]{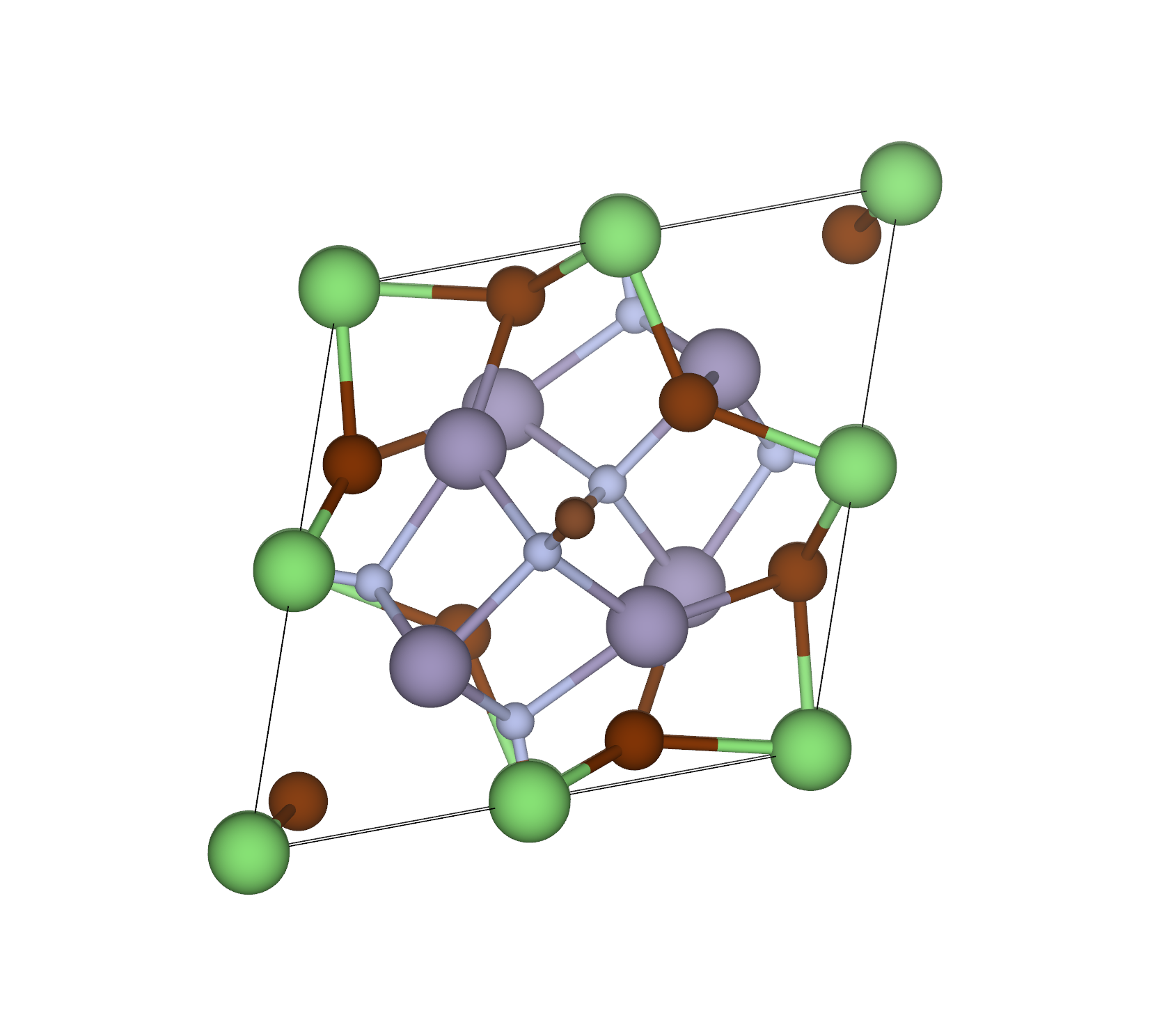}
    \put(3,94){\small\textbf{a)}}
  \end{overpic}
  \hfill
  \begin{overpic}[height=\panelheight]{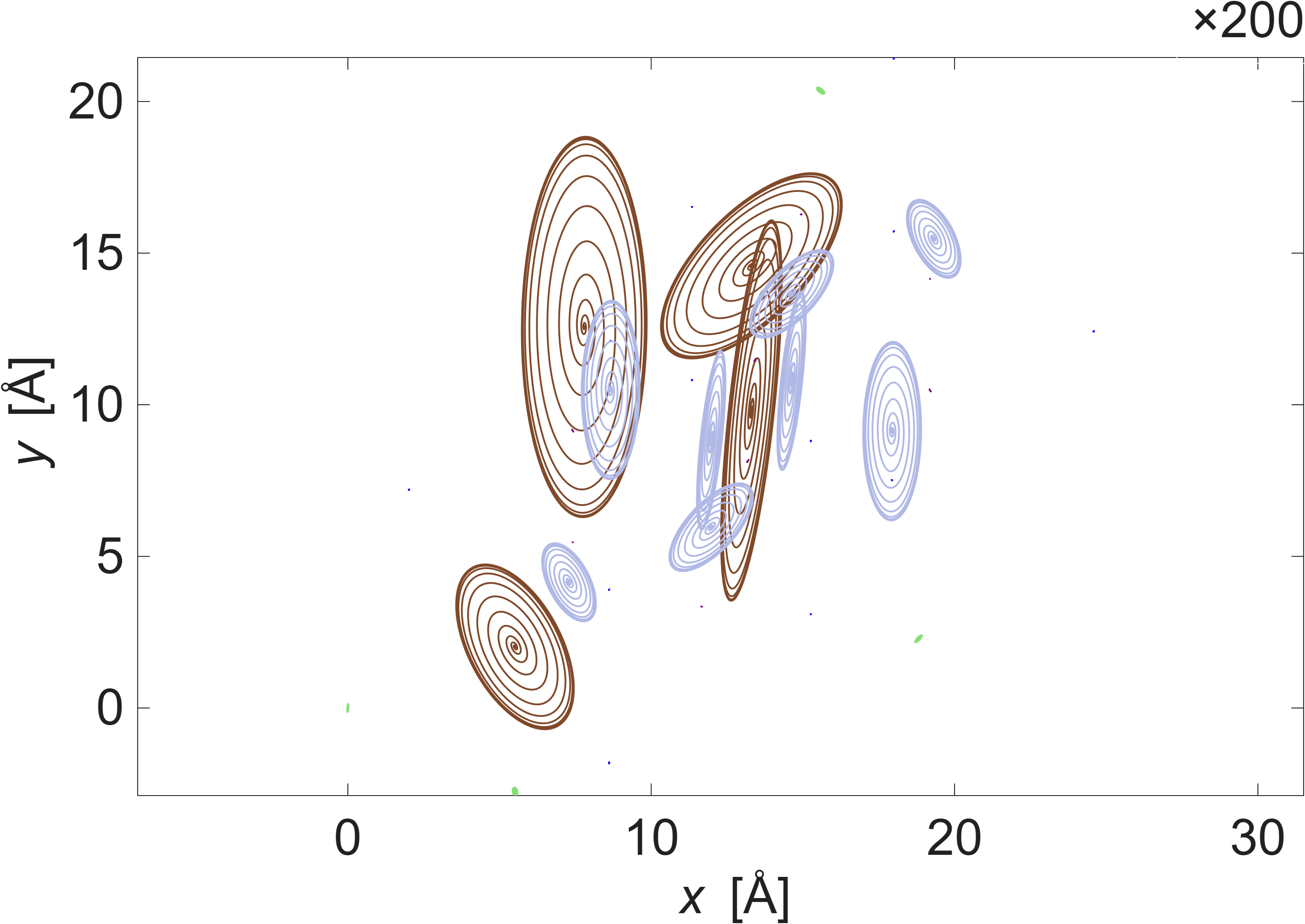}
    \put(3,75){\small\textbf{b)}}
        \put(87,69){\colorbox{white}{\small x141}}
  \end{overpic}
  \hfill
  \begin{overpic}[height=\panelheight]{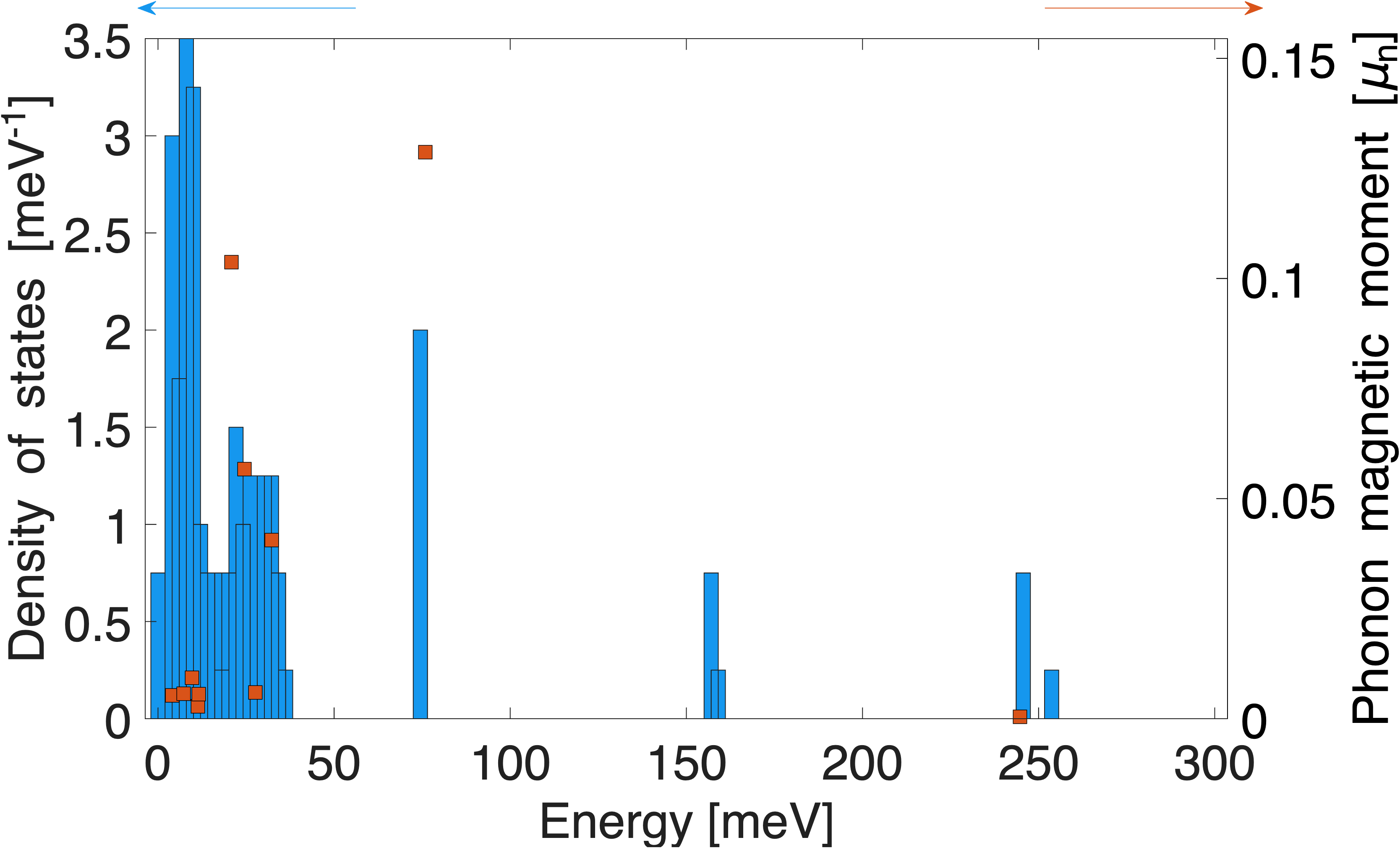}
    \put(3,63){\small\textbf{c)}}
  \end{overpic}
  \caption{Phonon magnetism in LiSn$_2$Br$_3$(CN$_2$). \textbf{a)} Crystal structure of the primitive unit cell, with atoms colored by element (Li green, Sn purple, N light blue, C brown). \textbf{b)} Projection onto the $xy$ plane of the atomic trajectories  following the model pulse. \textbf{c)} Phonon density of states and phonon magnetic moments as a function of energy.}
  \label{fig:lisnbrcn2}
\end{figure*}

\begin{figure*}
  \vspace{1em}       
  \centering
  \begin{overpic}[height=\panelheight]{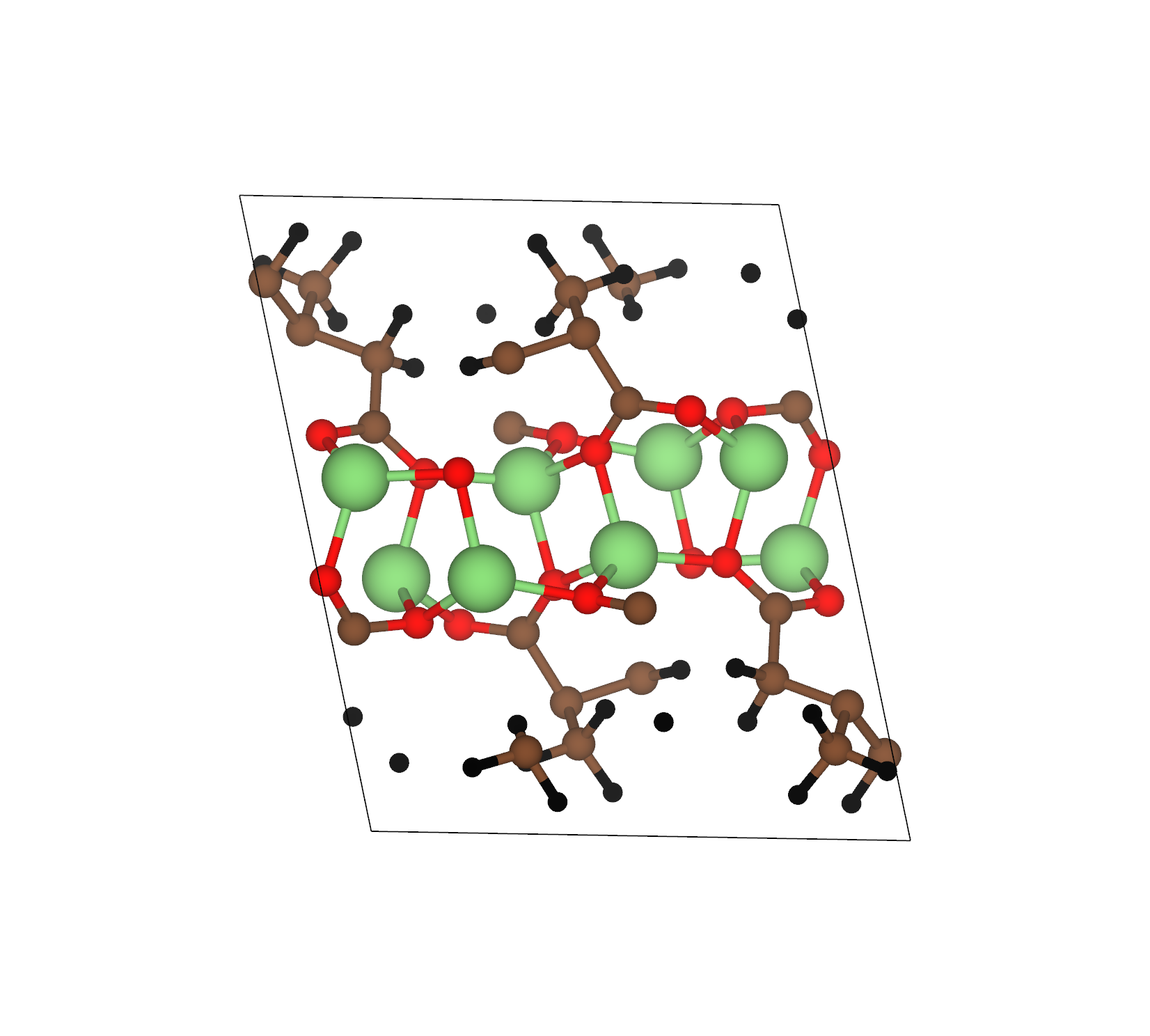}
    \put(3,95){\small\textbf{a)}}
  \end{overpic}
  \hfill
  \begin{overpic}[height=\panelheight]{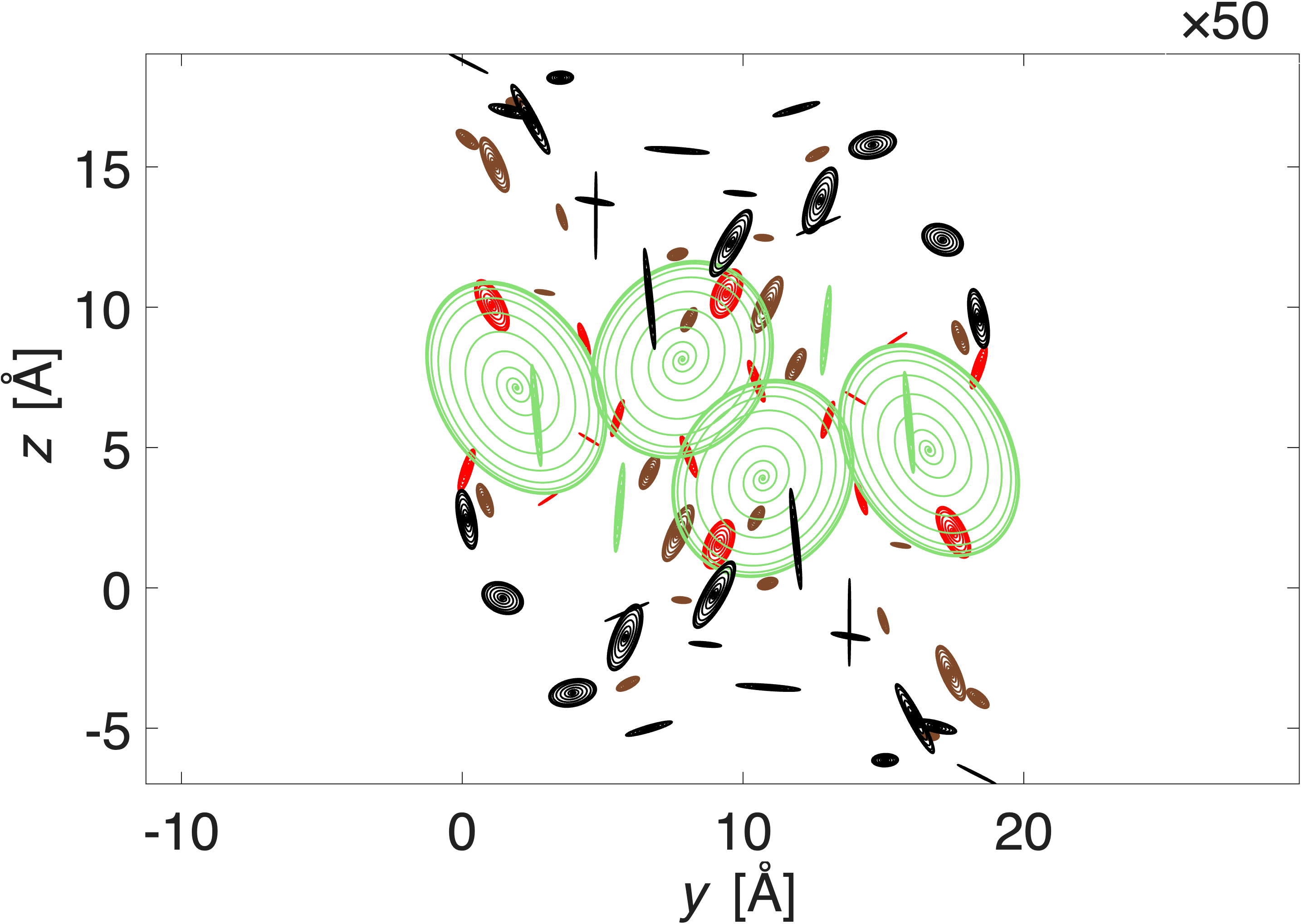}
    \put(3,75){\small\textbf{b)}}
        \put(87,69){\colorbox{white}{\small x35}}

  \end{overpic}
  \hfill
  \begin{overpic}[height=\panelheight]{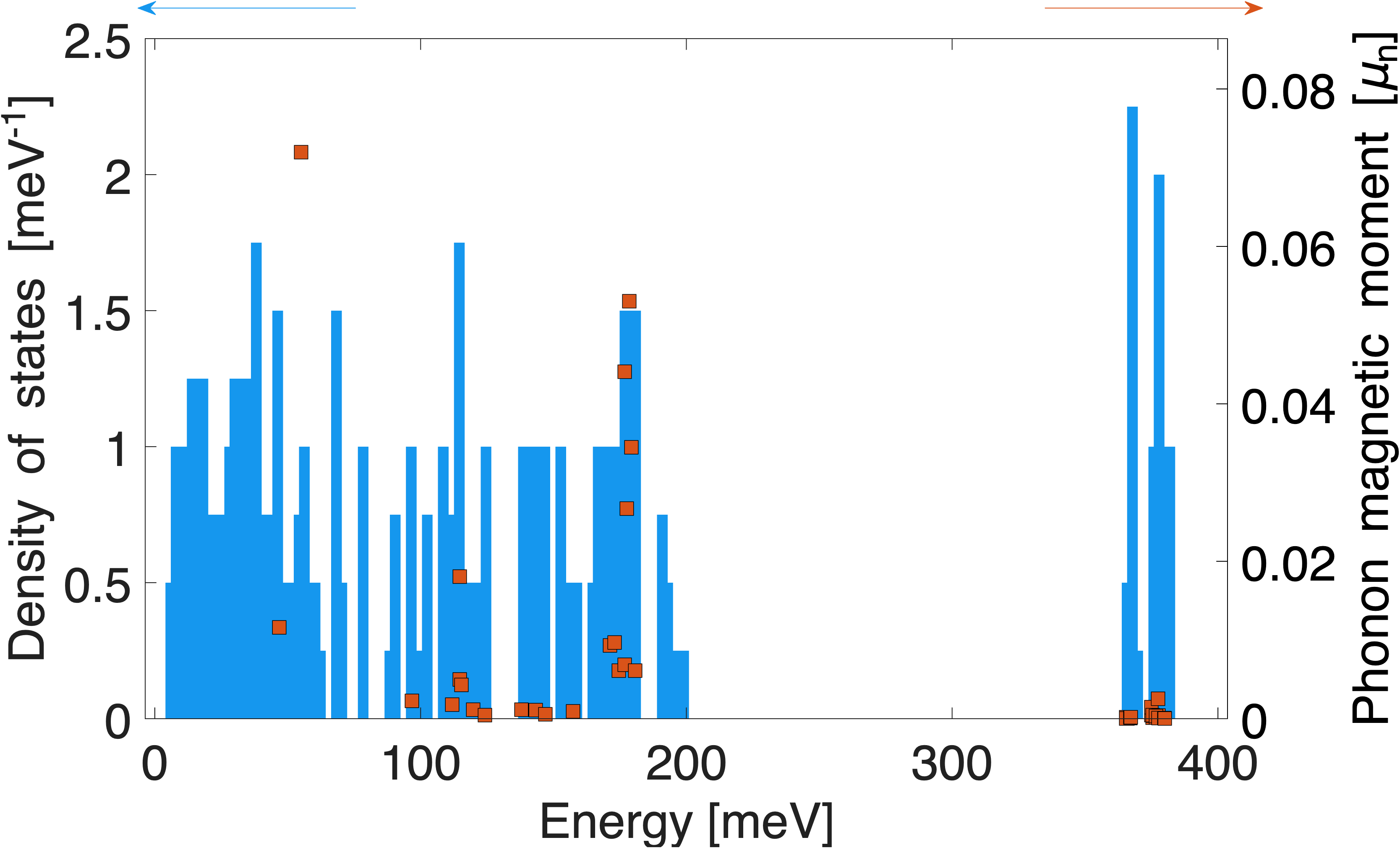}
    \put(3,63){\small\textbf{c)}}
  \end{overpic}
  \caption{Phonon magnetism in Li$_2$(dmsu). \textbf{a)} Crystal structure of the primitive unit cell, with atoms colored by element (Li green, O red, C brown, H black). \textbf{b)} Projection onto the $yz$ plane of the atomic trajectories  following the model pulse. \textbf{c)} Phonon density of states and phonon magnetic moments as a function of energy.}
  \label{fig:lisucc}
\end{figure*}

\begin{figure*}
  \vspace{1em}       
  \centering
  \begin{overpic}[height=\panelheight]{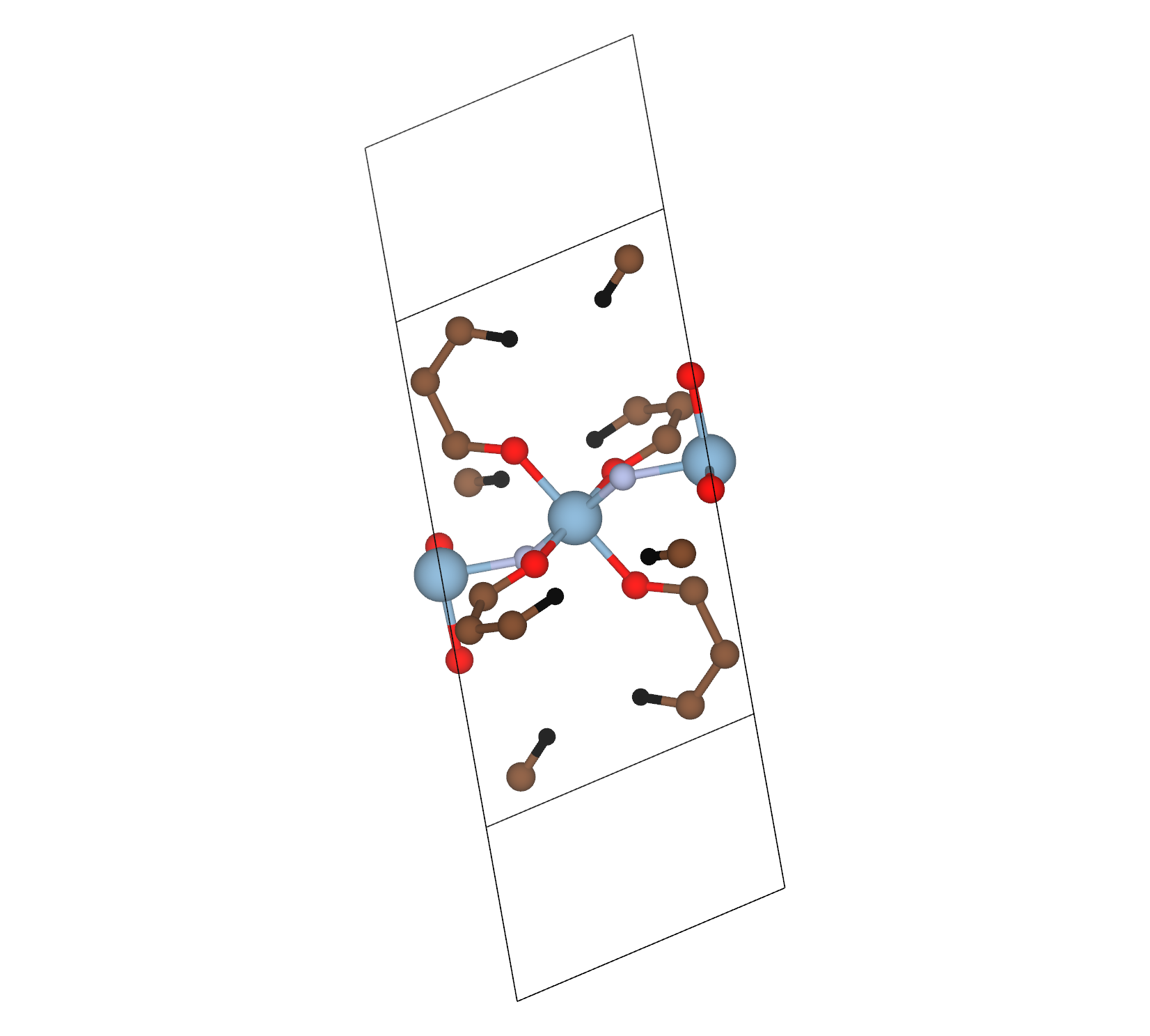}
    \put(3,96){\small\textbf{a)}}
  \end{overpic}
  \hfill
  \begin{overpic}[height=\panelheight]{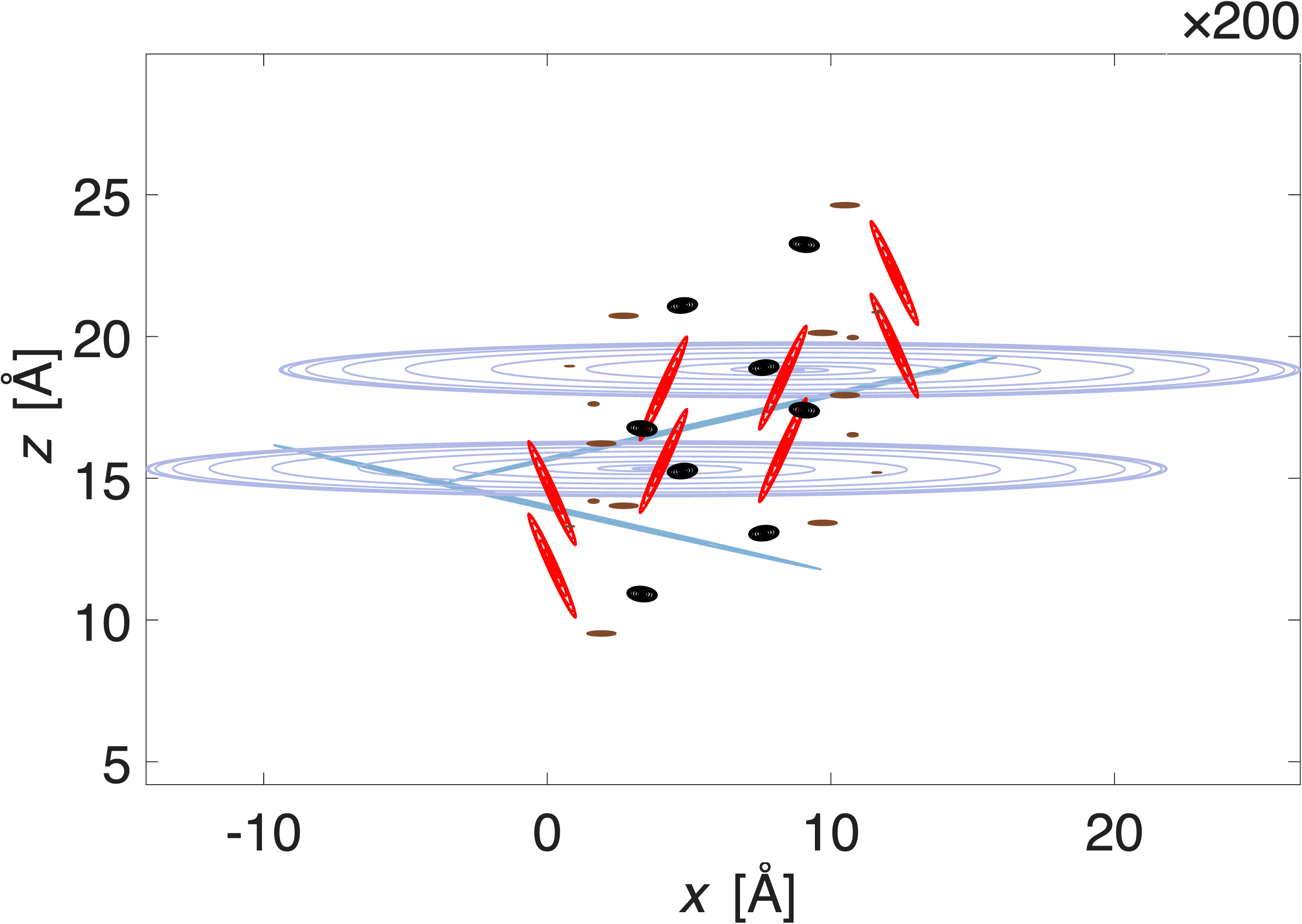}
    \put(3,75){\small\textbf{b)}}
        \put(87,69){\colorbox{white}{\small x141}}

  \end{overpic}
  \hfill
  \begin{overpic}[height=\panelheight]{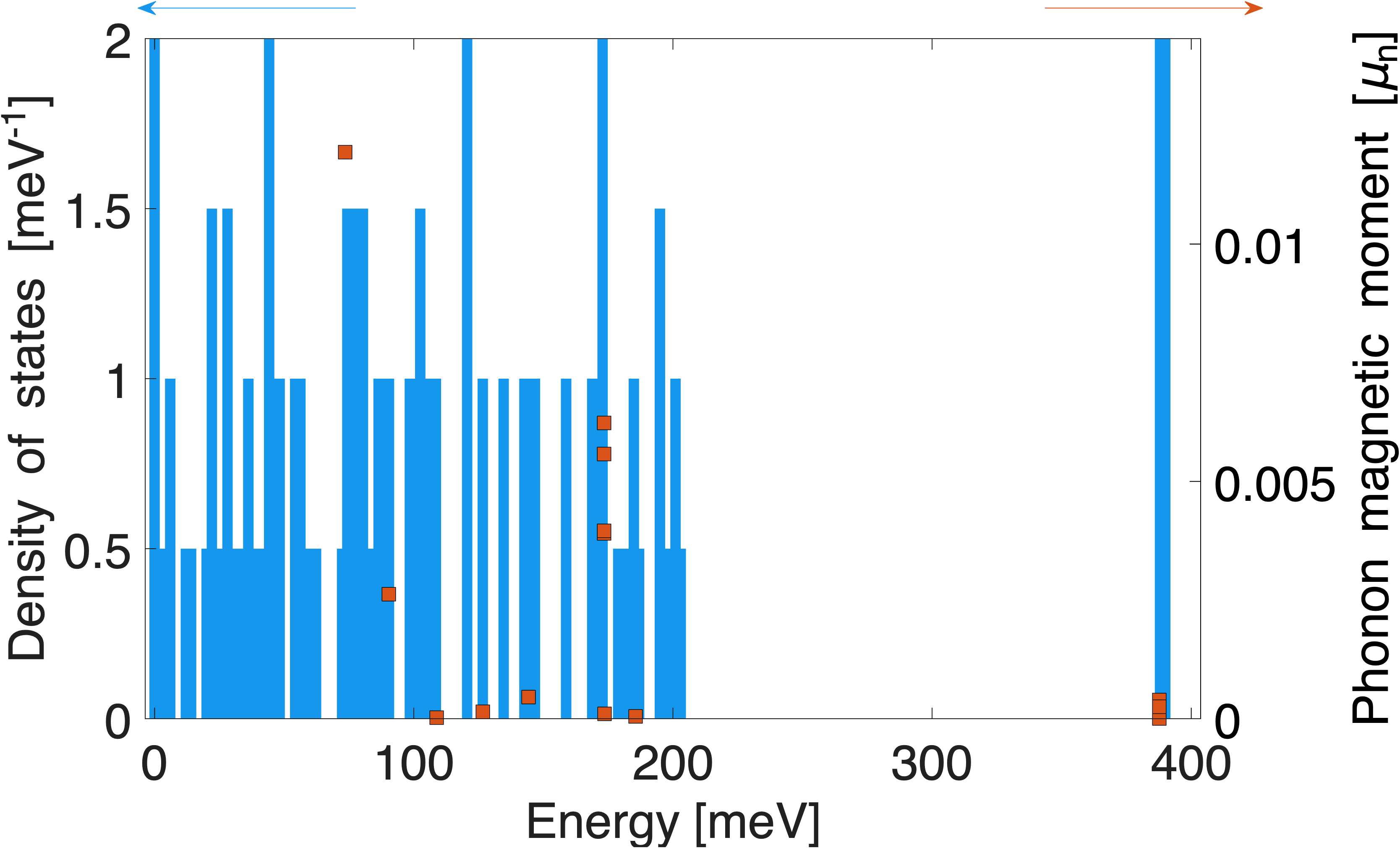}
    \put(3,63){\small\textbf{c)}}
  \end{overpic}
  \caption{Phonon magnetism in AlF(bdc). \textbf{a)} Crystal structure of the primitive unit cell, with atoms colored by element (Al blue, F purple, O red, C brown, H black). \textbf{b)} Projection onto the $xz$ plane of the atomic trajectories  following the model pulse. \textbf{c)} Phonon density of states and phonon magnetic moments as a function of energy.}
  \label{fig:mil53f}
\end{figure*}

\begin{figure*}
  \vspace{1em}       
  \centering
  \begin{overpic}[height=\panelheight]{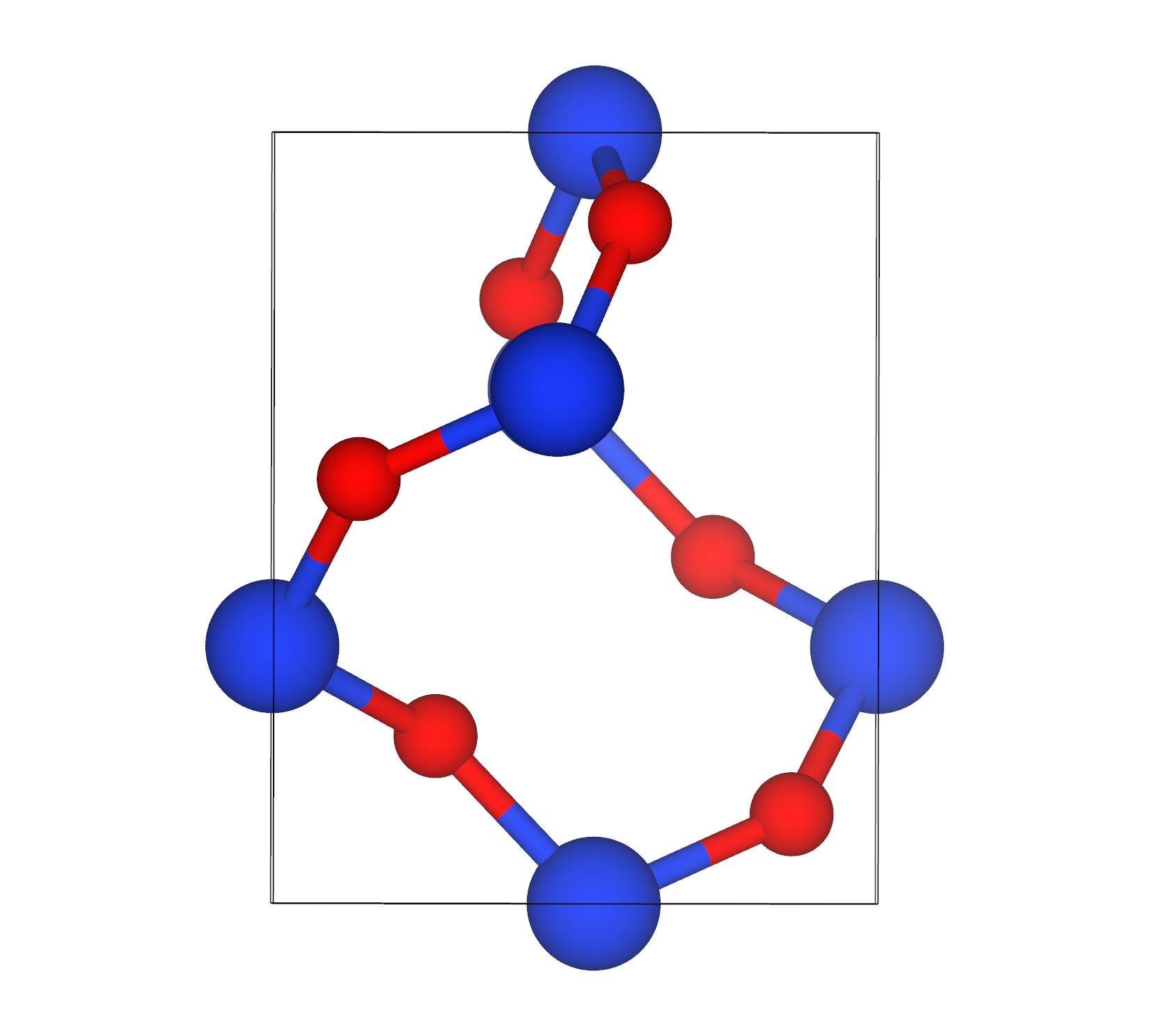}
    \put(3,95){\small\textbf{a)}}
  \end{overpic}
  \hfill
  \begin{overpic}[height=\panelheight]{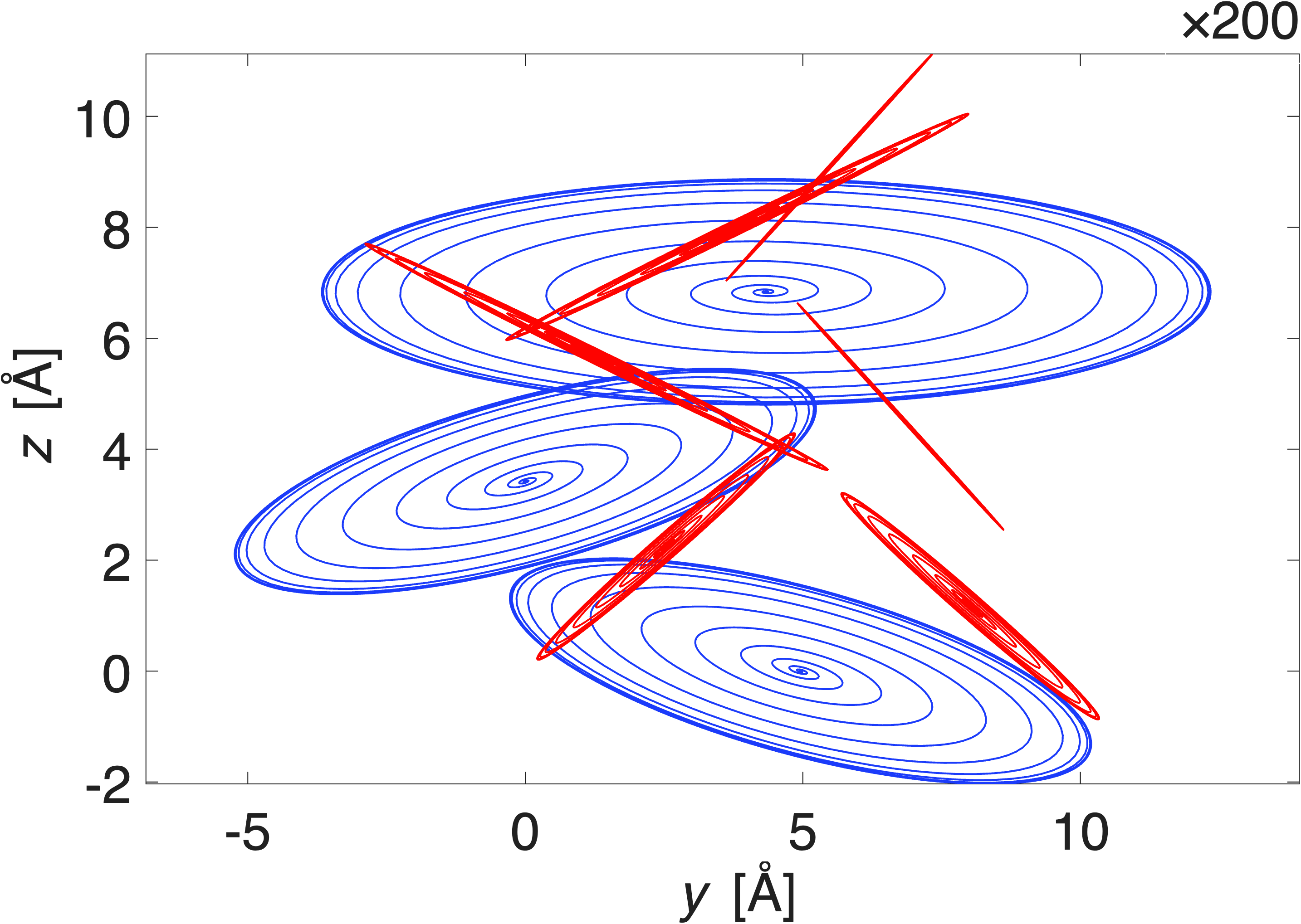}
    \put(3,75){\small\textbf{b)}}
        \put(87,69){\colorbox{white}{\small x141}}

  \end{overpic}
  \hfill
  \begin{overpic}[height=\panelheight]{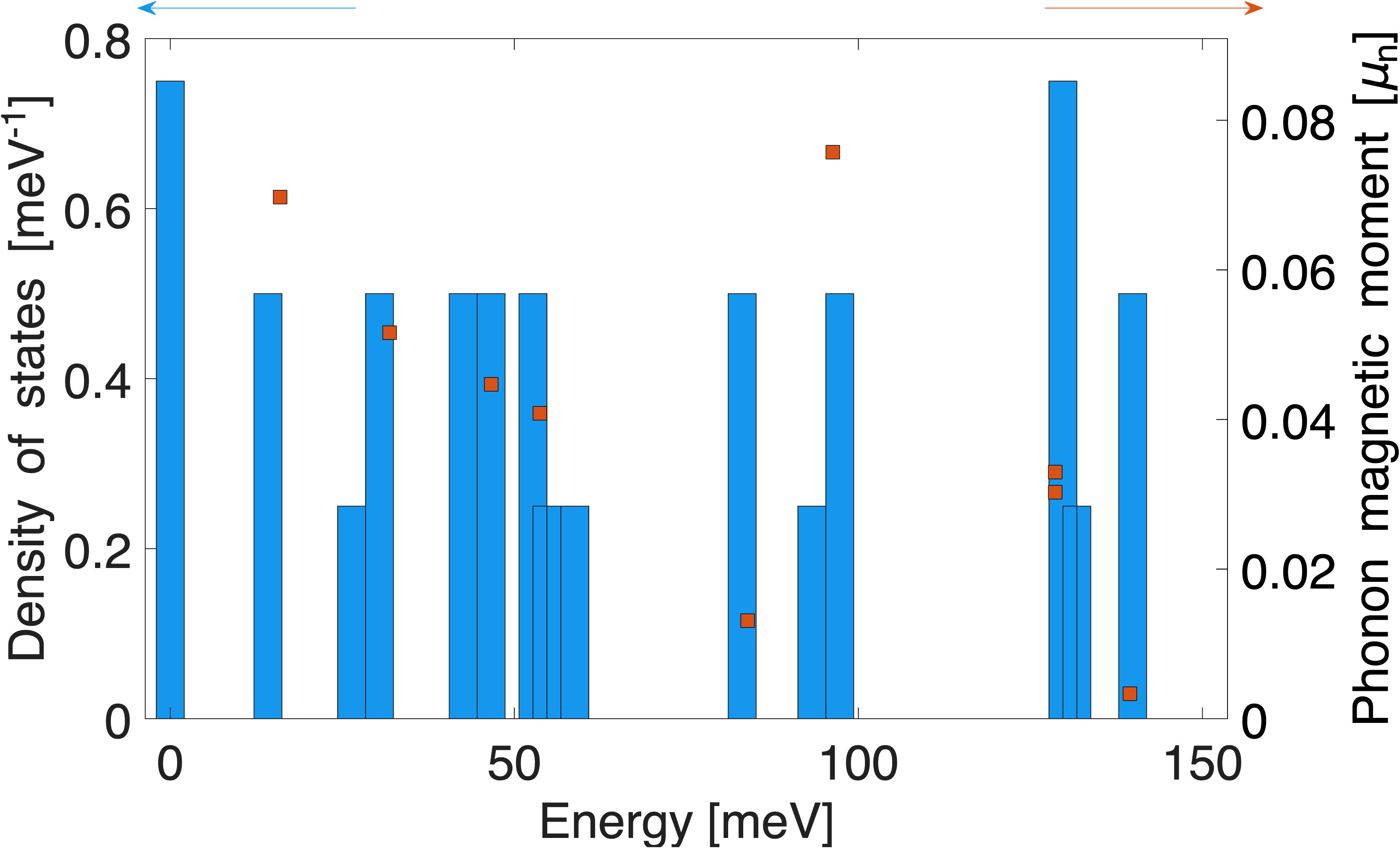}
    \put(3,63){\small\textbf{c)}}
  \end{overpic}
  \caption{Phonon magnetism in quartz. \textbf{a)} Crystal structure of the primitive unit cell, with atoms colored by element (Si blue, O red). \textbf{b)} Projection onto the $yz$ plane of the atomic trajectories  following the model pulse. \textbf{c)} Phonon density of states and phonon magnetic moments as a function of energy.}
  \label{fig:quartz}
\end{figure*}

\begin{figure*}
  \vspace{1em}       
  \centering
  \begin{overpic}[height=\panelheight]{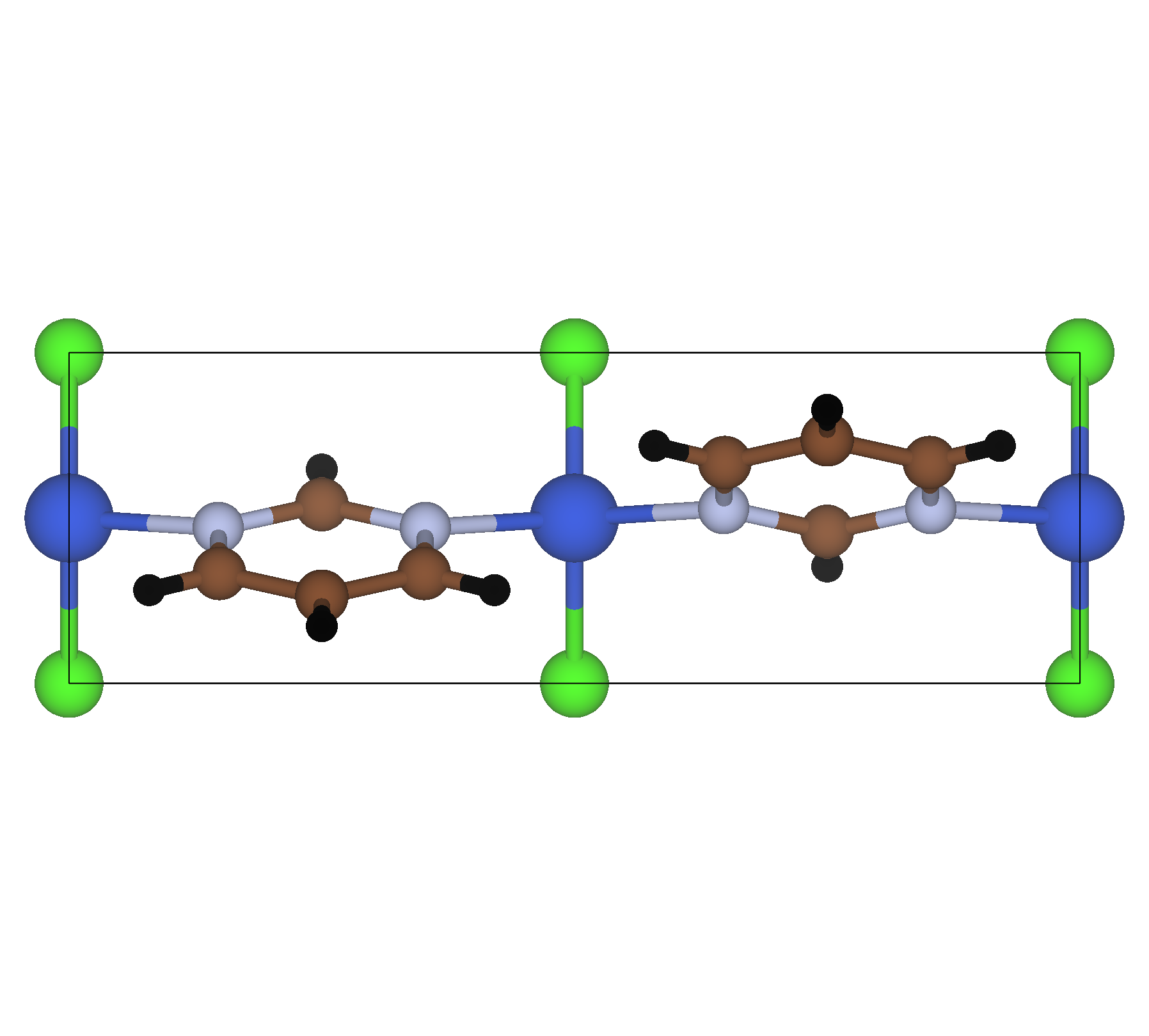}
    \put(3,97){\small\textbf{a)}}
  \end{overpic}
  \hfill
  \begin{overpic}[height=\panelheight]{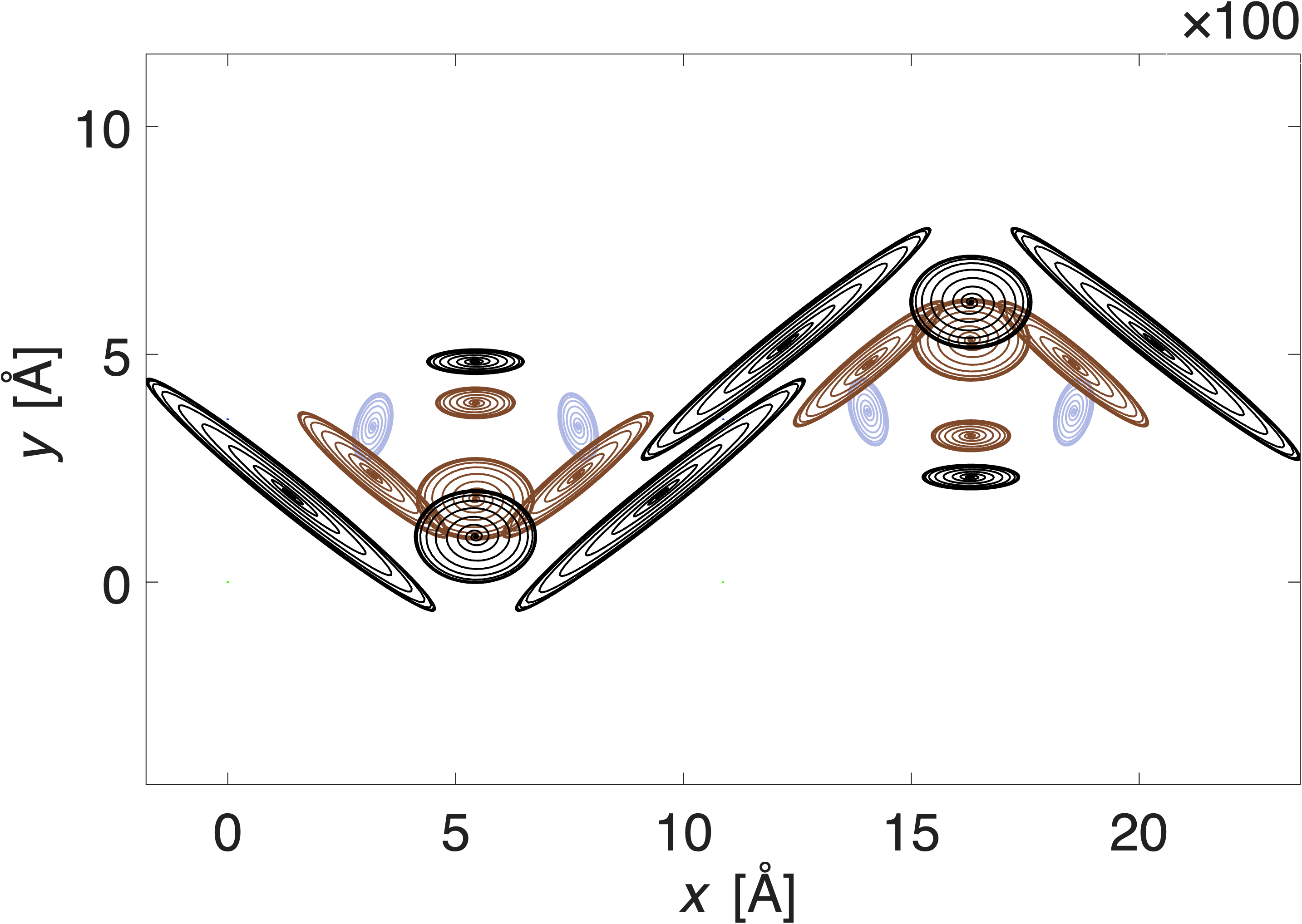}
    \put(3,75){\small\textbf{b)}}
            \put(88,69){\colorbox{white}{\small x70}}
  \end{overpic}
  \hfill
  \begin{overpic}[height=\panelheight]{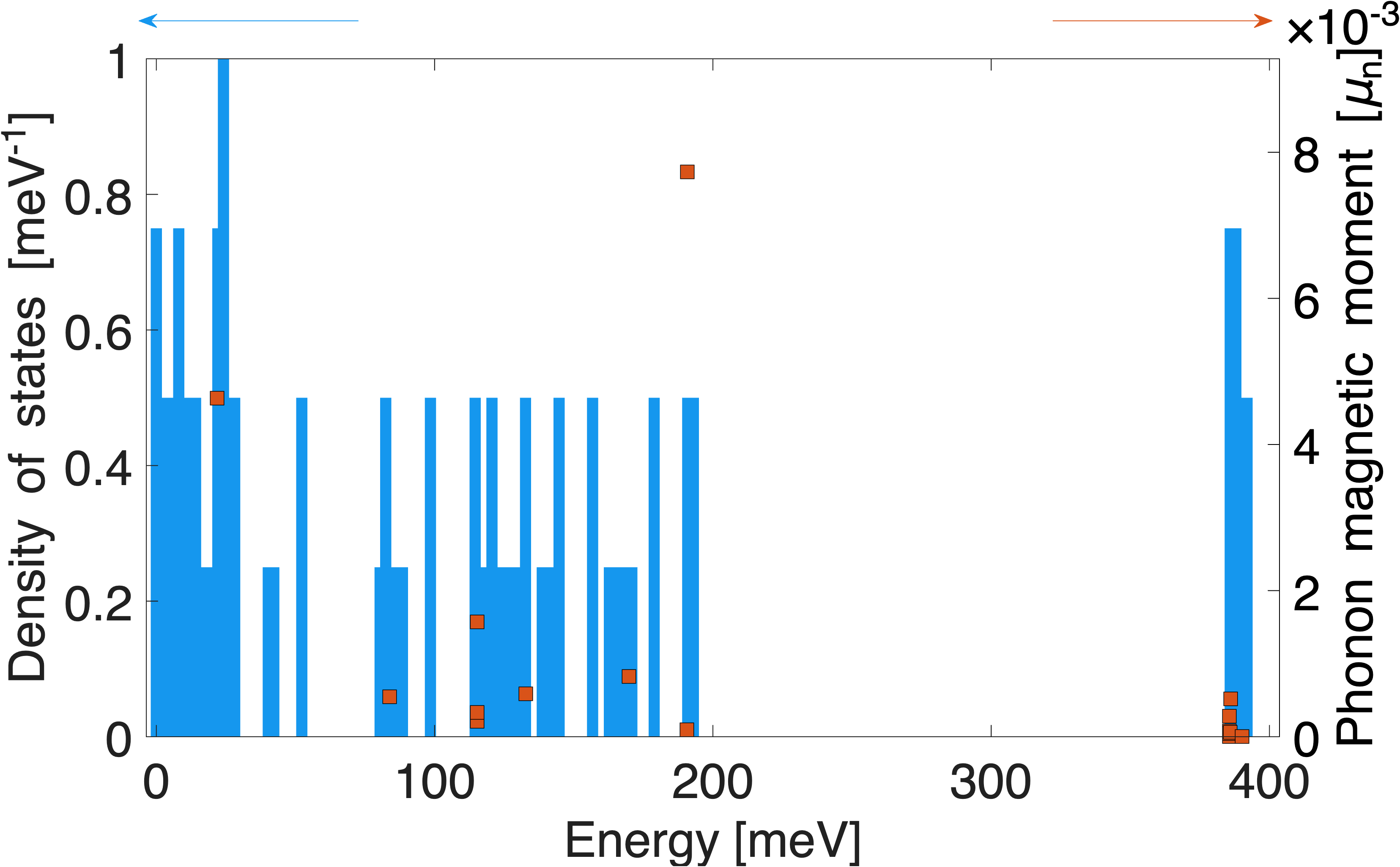}
    \put(3,64){\small\textbf{c)}}
  \end{overpic}
  \caption{Phonon magnetism in CuCl(pyr). \textbf{a)} Crystal structure of the primitive unit cell, with atoms colored by element (Cu blue, Cl green, N light blue, C brown, H black). \textbf{b)} Projection onto the $xy$ plane of the atomic trajectories  following the model pulse. \textbf{c)} Phonon density of states and phonon magnetic moments as a function of energy.}
  \label{fig:sacsev}
\end{figure*}

\begin{figure*}
  \vspace{1em}       
  \centering
  \begin{overpic}[height=\panelheight]{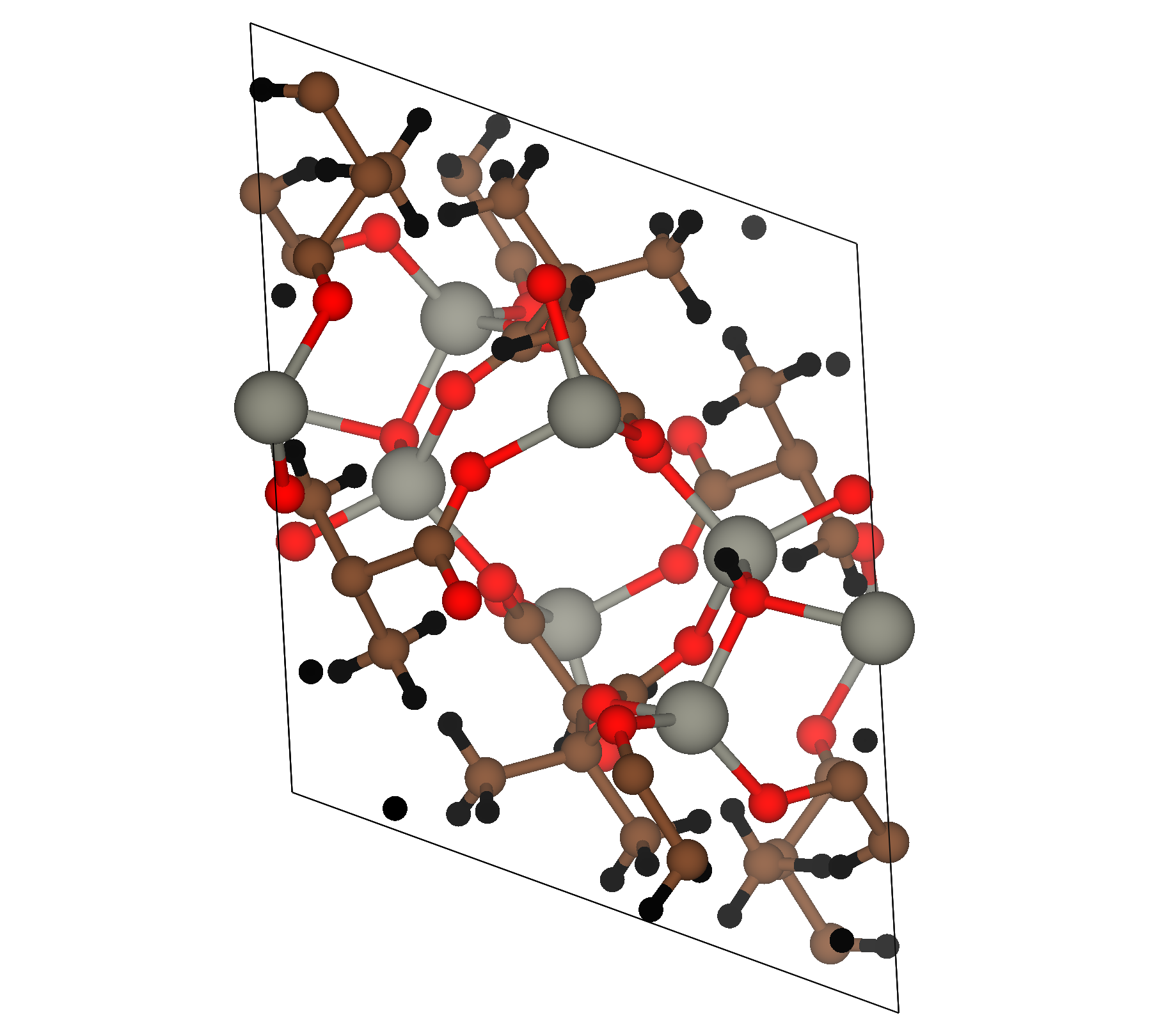}
    \put(3,95){\small\textbf{a)}}
  \end{overpic}
  \hfill
  \begin{overpic}[height=\panelheight]{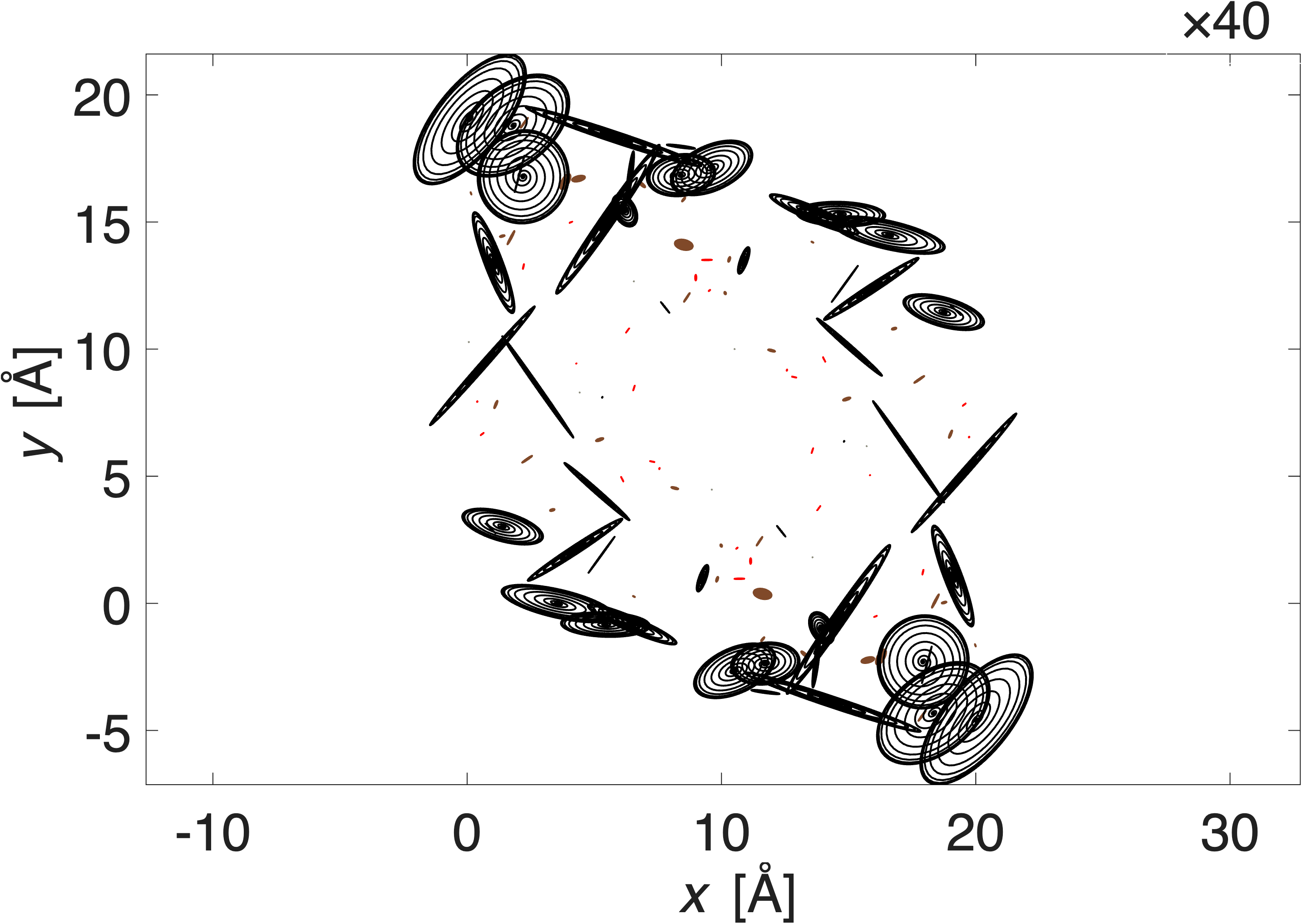}
    \put(3,75){\small\textbf{b)}}
                \put(88,69){\colorbox{white}{\small x28}}

  \end{overpic}
  \hfill
  \begin{overpic}[height=\panelheight]{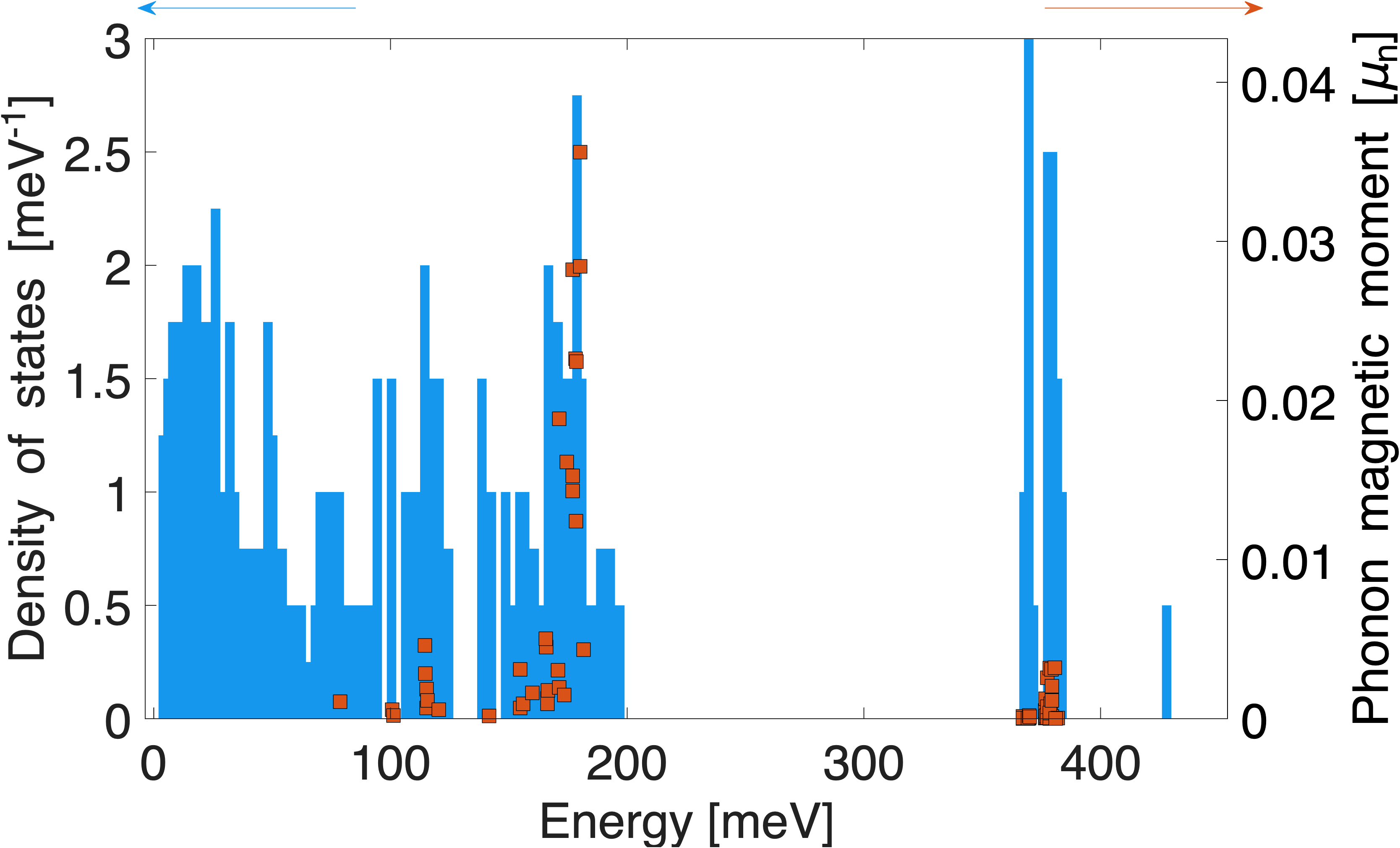}
    \put(3,63){\small\textbf{c)}}
  \end{overpic}
  \caption{Phonon magnetism in Zn$_7$(dmsu)$_6$(OH)$_2$. \textbf{a)} Crystal structure of the primitive unit cell, with atoms colored by element (Zn gray, O red, C brown, H black). \textbf{b)} Projection onto the $xy$ plane of the atomic trajectories  following the model pulse. \textbf{c)} Phonon density of states and phonon magnetic moments as a function of energy.}
  \label{fig:znsucc}
\end{figure*}

The maximum phonon magnetic moments and related properties for all nineteen materials are collected in Table~\ref{tab:phonon_moments}. Figs.~\ref{fig:znform}--\ref{fig:znsucc} show the atomic trajectories of these modes, as well as the moments of each IR-active mode as a function of energy. As expressed in Eq.~(\ref{eq:magmoment}), the contribution of an atom to the phonon magnetic moment of a given mode is the product of two factors: its gyromagnetic ratio and the circular polarization of the atomic motions $S$. Additionally, when the angular momenta of anionic and cationic sublattices are co-aligned, their magnetic moments will partially cancel~\cite{juraschek2017dynamical, juraschek2019orbital}. Therefore, it is desirable for the mode to only excite one sublattice, or to excite orthogonal motions of different sublattices~\cite{chaudharyAnomalous2026a}.

The magnitudes of the magnetic moments are $< 1~\upmu_\mathrm{n}$ as these are the Maxwellian moments generated directly by the ions, not the much larger effective magnetic moments which can reach values $> 1~\upmu_\mathrm{B}$~\cite{luo2023large}. However, at comparable pulse intensities, the magnetization measured for CeF$_3$ (${\sim}1~\upmu_\mathrm{n}/\mathrm{nm}^3$)~\cite{luo2023large} is of the same order as some of our predicted values, indicating that the bare phonon orbital contribution alone can already reach experimentally relevant magnetization densities.

\begin{table*}[htbp]
\centering
\caption{Phonon magnetic moments and related properties for selected materials, sorted by increasing $\mu_\mathrm{ph}$, the magnetic moment of one phonon. $S$ is the dimensionless circular polarization of the selected phonon, $\gamma$ is the phonon gyromagnetic ratio, $\omega/2\pi$ is the phonon frequency, $N_\mathrm{at}$ is the number of atoms in the primitive unit cell, and $M_\mathrm{pulse}$ is the magnetization density induced by the model pump pulse. $M_\mathrm{melt}^{(\mathrm{A})}$ and $M_\mathrm{melt}^{(\mathrm{M})}$ are the estimated maximum magnetization densities allowed by the Lindemann criterion when evaluated using all interatomic distances and only metal--ligand distances, respectively.}
\label{tab:phonon_moments}
\renewcommand{\arraystretch}{1.2}
\small
\begin{tabular*}{\textwidth}{@{\extracolsep{\fill}} l r r r r r r r r}
\toprule
\addlinespace[4pt]
Material &
\multicolumn{1}{c}{$\mu_\mathrm{ph}$ [$\upmu_\mathrm{n}$]} &
\multicolumn{1}{c}{$S$} &
\multicolumn{1}{c}{$\gamma$ [$\upmu_\mathrm{n}/\hbar$]} &
\multicolumn{1}{c}{$\omega/2\pi$ [THz]} &
\multicolumn{1}{c}{$N_\mathrm{at}$} &
\multicolumn{1}{c}{$M_\mathrm{pulse}$ [$\upmu_\mathrm{n}$/nm$^3$]} &
\multicolumn{1}{c}{$M_\mathrm{melt}^{(\mathrm{A})}$ [$\upmu_\mathrm{n}$/nm$^3$]} &
\multicolumn{1}{c}{$M_\mathrm{melt}^{(\mathrm{M})}$ [$\upmu_\mathrm{n}$/nm$^3$]} \\
\addlinespace[4pt]
\midrule
CuCl(pyr) & 0.0077 & 0.495 & 0.0156 & 46.1 & 24 & 0.292 & 2.43 & 11.3 \\
AlF(bdc) & 0.0119 & 0.331 & 0.0360 & 17.8 & 36 & 0.244 & 2.03 & 2.03 \\
InF$_3$(bpy) & 0.0137 & 0.013 & 1.03 & 34.4 & 24 & 0.229 & 2.12 & 3.30 \\
CsCuCl$_3$ & 0.0168 & 0.158 & 0.106 & 5.16 & 30 & 0.0326 & 1.43 & 1.43 \\
AgCl(phen) & 0.0176 & 0.263 & 0.0671 & 44.8 & 24 & 0.0908 & 2.00 & 2.05 \\
ScF$_3$ & 0.0218 & 0.071 & 0.306 & 12.7 & 4 & 3.44 & 37.0 & 37.0 \\
Cu$_3$(mel) & 0.0261 & 0.442 & 0.0590 & 33.0 & 60 & 0.0362 & 0.0284 & 0.596 \\
Li$_2$(tar) & 0.0268 & 0.023 & 1.15 & 42.5 & 32 & 2.52 & 5.24 & 5.24 \\
Zn$_7$(dmsu)$_6$(OH)$_2$ & 0.0356 & 0.131 & 0.272 & 43.6 & 119 & 0.468 & 2.77 & 6.30 \\
Sr(tar) & 0.0421 & 0.013 & 3.26 & 6.23 & 30 & 0.0682 & 2.82 & 2.82 \\
Cd(Gua)(fmt)$_3$ & 0.0427 & 0.121 & 0.354 & 35.8 & 46 & 0.00756 & 0.424 & 0.762 \\
CuPt(CN)$_6$ & 0.0544 & 0.980 & 0.0555 & 3.93 & 14 & 0.0612 & 0.325 & 0.325 \\
CaZrF$_6$ & 0.0559 & 0.852 & 0.0656 & 8.27 & 8 & 0.743 & 19.2 & 19.2 \\
Li$_2$(dmsu) & 0.0720 & 0.519 & 0.139 & 13.3 & 80 & 1.38 & 12.2 & 12.2 \\
Quartz & 0.0758 & 0.561 & 0.135 & 23.3 & 9 & 6.23 & 80.5 & 80.5 \\
LiSn$_2$Br$_3$(CN$_2$) & 0.129 & 0.501 & 0.257 & 18.3 & 36 & 1.11 & 11.5 & 11.5 \\
SrTiO$_3$ & 0.179 & 0.506 & 0.354 & 7.33 & 10 & 0.0685 & 3.65 & 3.65 \\
BPO$_4$ & 0.204 & 0.815 & 0.250 & 26.5 & 6 & 161 & 386 & 386 \\
Zn(NH$_4$)(fmt)$_3$ & 0.316 & 0.117 & 2.70 & 51.0 & 108 & 0.0400 & 1.79 & 197 \\
\addlinespace[4pt]
\bottomrule
\end{tabular*}
\end{table*}

The largest moment belongs to Zn(NH$_4$)(fmt)$_3$ ($0.316~\upmu_\mathrm{n}$), specifically to a mode at 51 THz involving displacements of hydrogen atoms within the NH$_4$ cation (Fig.~\ref{fig:znform}). This mode has low circular polarization ($S = 0.117$), but a high phonon gyromagnetic ratio ($\gamma = 2.7~\upmu_\mathrm{n}/\hbar$), as the NH$_4^+$ hydrogens in Zn(NH$_4$)(fmt)$_3$ have a scalar Born effective charge $Z_\mathrm{H}^* = \mathrm{tr}(\mathbf{Z}_\mathrm{H}^*)/3  \approx 1/3~\mathrm{e}$. In the other materials studied, the hydrogens generally have lower effective charges (\textit{e.g.} $Z_\mathrm{H}^* \approx 0.05~\mathrm{e}$ in Zn$_7$(dmsu)$_6$(OH)$_2$). Cd(Gua)(fmt)$_3$, which contains the polyatomic guanidinium ($[\mathrm{C}(\mathrm{NH}_2)_3]^+$) cation, shows even higher atomic gyromagnetic ratios with $Z_\mathrm{H}^* \approx 1/2~\mathrm{e}$; however it lacks a mode with a high degree of angular momentum localization on the hydrogen atoms (Fig.~\ref{fig:cdgform}). 

These gyromagnetic ratios and magnetic moments are smaller than those reported for the metal hydrides CsH and CuH, where $Z_\mathrm{H}^* \approx -1~\mathrm{e}$~\cite{juraschek2019orbital}; however such materials are difficult to study experimentally due to their chemical reactivity. We are therefore able to suggest the use of hydrogen-containing polyatomic ions as a practical method to harness the high gyromagnetic ratio of the hydrogen atom in dynamical multiferroicity. For example, hybrid lead halide perovskites (\textit{e.g.} (CH$_3$NH$_3$)PbI$_3$) are known to possess chiral phonons~\cite{polsChiral2025} and strong electron-phonon coupling~\cite{duanPhotoinduced2020}, and are therefore promising candidates for future research.

The second largest magnetic moment ($0.204~\upmu_\mathrm{n}$) is found in BPO$_4$, due to angular momentum localization on its boron sublattice (Fig.~\ref{fig:BPO4}), which has a significant gyromagnetic ratio due to boron's small mass and $Z^*_\mathrm{B} \approx 2.75~\mathrm{e}$. BPO$_4$ also has, by a wide margin, the largest magnetization per pulse ($M_\mathrm{pulse}$) due to its combination of high magnetic moment with high mode effective charge. The strong light--matter interaction indicated by the large mode effective charge has been used to transiently induce structural chirality in BPO$_4$ \textit{via} nonlinear phononics~\cite{zengPhotoinduced2025, romaoChirality2025}. As the photoinduced state has both chiral and polar symmetry (point group $C_2$~\cite{romaoPhononInduced2024}), phonon driving by an elliptical pulse could induce both a magnetization and ferroelectric polarization, leading to a light-induced multiferroic state~\cite{paivaDynamically2025}.

SrTiO$_3$ has the third-largest moment, $0.179~\upmu_\mathrm{n}$, matching the value previously reported in Ref.~\cite{juraschek2019orbital}. The large phonon magnetic moment arises primarily from the existence of a mode involving only motion of the oxygen atoms (Fig.~\ref{fig:STO}), with one oxygen sublattice having a circular trajectory and one having a linear trajectory, leading to an overall circular polarization of 0.5. The phonon magnetism is further enhanced by the anomalous effective charge of the oxygen atoms $Z_\mathrm{O}^* \approx -3.25~\mathrm{e}$, which is a consequence of incipient ferroelectricity~\cite{mullerSrTiO31979}. Experiments have found a much larger effective phonon magnetic moment in SrTiO$_3$~\cite{basiniTerahertz2024}, due to enhancement of the non-Maxwellian effective magnetic field by electron--phonon coupling to the oxygen $p$ orbitals~\cite{shabalaPhonon2024, merlinUnraveling2024}.

An impact of crystallographic symmetry can be seen by comparing the atomic displacements in SrTiO$_3$ to those in ScF$_3$. Both materials adopt the perovskite structure with different modifications: ScF$_3$ has no A-site cation and is stable in the cubic $Pm\bar{3}m$ space group down to 0~K~\cite{grevePronounced2010}, whereas SrTiO$_3$ undergoes a structural phase transition into a $I4/mcm$ polymorph upon cooling~\cite{mccallaUnified2016}. The higher symmetry of ScF$_3$ reduces the number of phonon modes, mixing the Sc and F displacements (Fig.~\ref{fig:scf3}), and we do not see the localization of angular momentum onto a single site that we see in the $I4/mcm$ polymorph of SrTiO$_3$ (Fig.~\ref{fig:STO}). 

Several materials with large $\gamma$ have small magnetic moments because they lack modes which also have high circular polarization. Sr(tar), for example, has the largest $\gamma$ in the dataset ($3.26~\upmu_\mathrm{n}/\hbar$) but $S = 0.013$, giving only $0.042~\upmu_\mathrm{n}$. As Fig.~\ref{fig:cugson} shows, this is not due to a lack of circular motion of the atoms. Rather, two pairs of oxygen sublattices circle in opposite directions, as a consequence of the pair of linear modes both preserving the 2-fold rotational symmetry along $c$. This effect also reduces the circular polarization of the mode shown in Fig.~\ref{fig:znform}, although to a lesser degree. On the other hand, modes with high circular polarization can have low magnetic moments as a consequence of low effective charges (CuPt(CN)$_6$, Fig.~\ref{fig:cuptcn6}, Sec.~\ref{sec:appendix}) or large atomic mass (CaZrF$_6$, Fig.~\ref{fig:cazrf6}. Sec.~\ref{sec:appendix}).

While the phonon magnetic moment controls the magnetic field generated by an individual phonon, in ultrafast experiments other parameters also contribute to the generated magnetization, \textit{e.g.}, the maximum phonon amplitude that can be reached before the sample melts. We estimate the maximum magnetization at this limiting amplitude using the Lindemann criterion, as described above (Eq.~(\ref{eq:lindemann})). We evaluate two variants of this criterion, one based on only the metal--ligand distances (M), and one including all interatomic distances (A). In this way, we consider the potential for complex frameworks to accommodate large light-atom vibrational amplitudes away from the structurally vital metal--ligand bonds~\cite{gaillacMelting2018}.

In Zn(NH$_4$)(fmt)$_3$, the magnetization at the metal--ligand limit ($M_\mathrm{melt}^{(\mathrm{M})} = 197~\upmu_\mathrm{n}/\mathrm{nm}^3$) exceeds the all-atom one ($M_\mathrm{melt}^{(\mathrm{A})} = 1.79~\upmu_\mathrm{n}/\mathrm{nm}^3$) by two orders of magnitude, while significant enhancements also appear in CuCl(pyr), Cu$_3$(mel), and Zn$_7$(dmsu)$_6$(OH)$_2$. Materials with monoatomic ligands (such as SrTiO$_3$) show no such enhancement, and therefore MOFs with lower per-phonon moments could match or exceed the melting-point magnetization of SrTiO$_3$ ($3.65~\upmu_\mathrm{n}/\mathrm{nm}^3$). We note that our values of $M_\mathrm{melt}^{(\mathrm{M})}$ only provide a limiting case; in practice it is, of course, not possible to create arbitrarily large displacements away from the metal--ligand bonds. However, especially in cases suchas Zn(NH$_4$)(fmt)$_3$, where the displacement is mostly localized outside of the framework itself, enhancement of $M_\mathrm{melt}$ is an interesting possibility.

The phonon frequency is another important practical factor, as lab-scale THz laser setups generally rely on photon downconversion, decreasing laser power significantly~\cite{songReview2026}; generation of circularly polarized THz light poses further challenges~\cite{jiaEfficient2019}. Many of the studied MOFs have magnetic phonons with $\omega/2\pi\geq12$~THz (\textit{i.e.}, in the mid-IR regime), making pumping to high phonon numbers more accessible than for non-framework materials such as SrTiO$_3$ (7~THz), which require THz optics. 

\section{Conclusions}
We have extended the study of dynamical multiferroicity to complex, low-symmetry crystals; surveying the light-induced phonon magnetic moments of nineteen materials ranging from perovskites to metal--organic frameworks. Zn(NH$_4$)(fmt)$_3$ has the largest phonon magnetic moment out of the studied materials ($0.316~\upmu_\mathrm{n}$), arising from the electrically charged hydrogens of its NH$_4^+$ cation, despite a low overall circular polarization. We suggest that such hydrogen-containing polyatomic ions are a practical route to harness the high gyromagnetic ratio of hydrogen, especially in materials with large electron--phonon coupling, such as hybrid lead-halide perovskites. The material with the second largest moment (BPO$_4$, $\mu_\mathrm{ph} = 0.204~\upmu_\mathrm{n}$) is a promising candidate for light-induced multiferroicity through a combination of nonlinear phononic rectification and phonon magnetism.

Metal--organic frameworks offer a further, structural advantage over materials with monoatomic ligands do not: because their light atoms can be displaced far from the metal--ligand bonds, the magnetization attainable before melting could greatly exceed that of SrTiO$_3$ or similar materials. Many of these frameworks additionally host magnetic modes in the mid-infrared, making them accessible with standard IR laser sources. These results establish framework materials as a promising and largely unexplored platform for light-driven magnetism, and motivate their experimental study.

\section{Computational Methods}\label{sec:methods}

\textbf{Density functional theory:} DFT and DFPT calculations of the phonon energies, eigendisplacements, and Born effective charges of AlF(bdc), ScF$_3$, Li$_2$(tar), Zn$_7$(dmsu)$_6$(OH)$_2$, Cd(Gua)(fmt)$_3$, CaZrF$_6$, Li$_2$(dmsu), BPO$_4$, SrTiO$_3$, and Zn(NH$_4$)(fmt)$_3$ were performed using the \textsc{Abinit} software package \cite{gonze2020abinit, verstraeteAbinit2025, gonze1997dynamical, bottin2008large, bjorkman2011cif2cell}. DFT data for the remaining materials were taken from the literature~\cite{romaoChiral2023, matasChiral2025, kallenbachTricopper2021, uedaTuning2025, uedaChiral2023, loeberSynthesis2021}. The Perdew--Burke--Ernzerhof exchange--correlation functional \cite{perdew1996generalized} was used with the dispersion correction of Grimme \cite{grimmeSemiempirical2006, grimme2010consistent}, except in the case of SrTiO$_3$, where the dispersion correction was omitted for consistency with Ref.~\cite{juraschek2019orbital}. A plane-wave basis set was used with PseudoDojo~\cite{vansettenPseudoDojo2018} and ONCV~\cite{hamannOptimized2013} pseudopotentials. The energy cutoffs used for the basis sets were: 30 Ha (Li$_2$(tar), Li$_2$(dmsu)), 38 Ha (Zn$_7$(dmsu)$_6$(OH)$_2$, Cd(Gua)(fmt)$_3$), 40 Ha (CaZrF$_6$), 45 Ha (AlF(bdc), BPO$_4$), and 50 Ha (ScF$_3$). Calculations were performed on Monkhorst--Pack grids \cite{monkhorst1976special} of $\mathbf{k}$-points with real-space basis vectors: $[3~0~0],~[0~3~0],~[0~0~3]$ (Zn$_7$(dmsu)$_6$(OH)$_2$); $[3~0~0],~[0~3~0],~[0~0~4]$ (Zn(NH$_4$)(fmt)$_3$); $[4~0~0],~[0~4~0],~[0~0~4]$ (AlF(bdc), BPO$_4$); $[{-2}~2~1],~[2~{-2}~1],~[2~2~{-1}]$ (Li$_2$(dmsu)); $[4~4~{-4}],~[{-4}~4~{-4}],~[{-4}~4~4]$ (CaZrF$_6$); $[4~5~{-4}],~[4~5~4],~[8~0~0]$ (Cd(Gua)(fmt)$_3$); and  $[-6~4~4],~[6~{-4}~4],~[6~4~{-4}]$ (Li$_2$(tar)); $[16~0~0],~[0~16~0],~[0~0~16]$ (ScF$_3$). These parameters were chosen following convergence studies, with the convergence criterion being 1\% of the pressure. Phonon calculations were performed at $\Gamma~(\mathbf{q} = 0)$. Crystal structures were relaxed prior to DFPT calculations. All calculations were performed with an insulating, diamagnetic electronic structure. Postprocessing to obtain the light-induced phonon magnetism was performed using a MATLAB script available from Ref.~\cite{github}.

\textbf{Data availability:} All computational data are publicly available from Ref. \cite{sdata}.

\acknowledgements{MM acknowledges support by the CTU Mobility Project MSCA-F-CZ-III under the number \texttt{CZ.02.01.01/00/22\_010/0008601} and the support of the Ministry of Education, Youth and Sports of the Czech Republic through e-INFRA CZ (ID:90254). CPR acknowledges support from the project FerrMion of the Ministry of Education, Youth and Sports, Czech Republic, co-funded by the European Union (\texttt{CZ.02.01.01/00/22\_008/0004591}), the European Union and Horizon 2020 through Grant No. 101030352, ETH Zurich, and the Swiss National Supercomputing Center (CSCS) under project IDs s1128 and eth3.}

\bibliography{ref}

\section{Appendix}\label{sec:appendix}

The Appendix contains the corresponding crystal structures, circularly polarized atomic trajectories, phonon densities of states, and phonon magnetic moments for the additional materials included in the survey but not shown in the main text. These figures follow the same format as the main-text examples and provide the mode-specific information underlying the values reported in Table~\ref{tab:phonon_moments}.

\end{document}